\documentstyle[12pt, epsf]{article}

\def \a{\alpha}
\def \b{\beta}
\def \g{\gamma}
\def \d{\delta}
\def \ep{\epsilon}

\def \l{\lambda}
\def \m{\mu}
\def \n{\nu}
\def \x{\xi}
\def \vp{\varpi}
\def \r{\rho}

\def \s{\sigma}

\def \t{\tau}
\def \u{\upsilon}
\def \ph{\phi}
\def \vp{\varphi}
\def \c{\chi}

\def \L{\Lambda}
\def \X{\Xi}

\def \Ph{\Phi}
\def \Ps{\Psi}

\def \re{{\rm e}}
\def \im{{\rm i}}

\def \Tr{{\rm Tr}}
\def \ma{{\rm matrix}}
\def \hatleftix{{\hat{\mathit L}_{\Lambda}}}
\def \leftix{{{\mathit L}_{\Lambda}}}
\def \hatrightix{{\hat{\mathit R}_{\Lambda}}}

\def \hatmultix{{\hat{\mathit M}_{\Lambda}}}
\def \multix{{{\mathit M}_{\Lambda}}}
\def \hatcentrix{{\hat{\mathit \Sigma}_{\Lambda}}}

\def \vectrix{{\cal V}_{\Lambda}}

\def \hatheterix{{\hat{\Gamma}_{\L | \L}}}
\def \cyclix{{\hat{\mathit C}_{\Lambda}}}

\def \salt{{\mathit F}_{\Lambda}}

\def \ignore#1{}
\def \gl{gl_{+\infty}}

\def \la#1{\label{#1}}
\def \ift{\infty}
\def \le{\left}
\def \ri{\right}
\def \da{\dagger}
\def \ti#1{\tilde{#1}}
\def \lb{\lbrack}
\def \rb{\rbrack}

\def \rar{\rightarrow}

\def \lrar{\leftrightarrow}
\def \ld{\ldots}
\def \cd{\cdots}
\def \nn{\nonumber}
\def \pa{\partial}

\newcommand \beq{\begin{eqnarray}}
\newcommand \eeq{\end{eqnarray}}
\newcommand \beqs{\begin{eqnarray}}
\newcommand \eeqs{\begin{eqnarray}}
\newcommand \ba{\begin{array}}
\newcommand \ea{\end{array}}

\newtheorem{lemma}{Lemma}

\begin{document}

\begin{flushright}
hep-th/9906060 \\
\end{flushright}

\begin{center}
   {\LARGE\bf A Review of Symmetry Algebras of Quantum Matrix Models in the Large-$N$ Limit} \\
   \vspace{2cm}
   {\large\bf C.-W. H. Lee and S. G. Rajeev} \\
   {\it Department of Physics and Astronomy, P.O. Box 270171, University of Rochester, 
    Rochester, New York 14627} \\
   \vspace{.5cm}
   {June 8, 1999} \\
   \vspace{2cm}
   {\Large\bf Abstract}
\end{center}

This is a review article in which we will introduce, in a unifying fashion and with more intermediate steps in some
difficult calculations, two infinite-dimensional Lie algebras of quantum matrix models, one for the open string 
sector and one for the closed string sector.  Physical observables of quantum matrix models in the large-$N$ limit 
can be expressed as elements of these Lie algebras.  We will see that both algebras arise as quotient algebras of a
larger Lie algebra.  We will also discuss some properties of these Lie algebras not published elsewhere yet, and 
briefly review their relationship with well-known algebras like the Cuntz algebra, the Witt algebra and the Virasoro
algebra.  We will also review how Yang--Mills theory, various low energy effective models of string theory, quantum 
gravity, string-bit models, and quantum spin chain models can be formulated as quantum matrix models.  Studying 
these algebras thus help us understand the common symmetry of these physical systems.

\pagebreak

\section{Introduction}
\la{c1}

As physicists, there are a number of phenomena in strong interaction, quantum gravity and condensed matter physics 
which we want to understand.

Quantum chromodynamics (QCD) is the widely accepted theory of strong interaction.  It postulates that the basic 
entities participating strong interaction are quarks, antiquarks and gluons.  In the high-energy regime, the theory
displays asymptotic freedom.  The coupling among these entities becomes so weak that we can use perturbative means
to calculate experimentally measurable quantities like differential cross sections in particle reactions.  Indeed, 
the excellent agreement between perturbative QCD and high energy particle phenomena form the experimental basis of 
the theory.

Nevertheless, strong interaction manifests itself not only in the high-energy regime but also in the low-energy one.
Here, the strong coupling constant becomes large.  Quarks, anti-quarks and gluons are permanently confined to form 
bound states called hadrons, like protons and neutrons.  Hadrons can be observed in laboratories.  We can measure 
their charges, spins, masses and other physical quantities.  One challenging but important problem in physics is to 
understand the structures of hadrons within the framework of QCD; this serves as an experimental verification of
QCD in the low-energy regime.  Hadronic structure can be described by something called a structure function which 
tells us the (fractional) numbers of constituent quarks, antiquarks or gluons carrying a certain fraction of the 
total momentum of the hadron.  The structure function of a proton has been measured carefully \cite{cteq}.  There 
has been no systematic theoretical attempt to explain the structure function until very recently 
\cite{krra}\footnote{See Ref.\cite{rajeev99} for a more pedagogical and updated account.}.  In this work, the 
number of colors $N$ is taken to be infinitely large as an approximation.  The resulting model can be treated as a 
classical mechanics \cite{berezin, yaffe, rajeev94}; i.e., the space of observables form a phase space of position 
and momentum, and the dynamics of a point on this phase space is governed by the Hamiltonian of this classical 
system and a Poisson bracket\footnote{Ref.\cite{arnold} provides an excellent discussion for such a geometric 
formulation of classical mechanics.}.  This is because quantum fluctuations abate in the large-$N$ limit --- the 
Green function of a product of color singlets is dominated by the product of the Green functions of these color 
singlets, and other terms are of subleading order\footnote{For an introductory discussion on this point, see 
Refs.\cite{witten78} and \cite{coleman}.}.  It is possible to derive a Poisson bracket for Yang--Mills theory in 
the large-$N$ limit.  This Poisson bracket can be incorporated into a commutative algebra of dynamical variables to 
form something called a Poisson algebra \cite{chpr}.  

As an initial attempt, only quarks and anti-quarks in the QCD model in Ref.\cite{krra} are dynamical.  A more 
realistic model should have dynamical gluons in addition to quarks.  Gluons carry a sizeable portion of the total
momentum of a proton \cite{cteq} and are thus significant entities.  One notable feature of a gluon field is that it
is in the adjoint representation of the gauge group and carries two color indices.  These can be treated as row and
column indices of a {\em matrix}.  This suggests gluon dynamics can be described by an abstract model of matrices.  

Besides strong interaction phenomena, another fundamental question in physics is how one quantizes gravity.  The 
most promising solution to this problem is superstring theory \cite{polchinski}.  Here we postulate that the basic 
dynamical entities are one-dimensional objects called strings.  Quarks, gluons, gravitons and photons all arise as 
excitations of string states.  If the theory consists of bosonic strings only, the ground states will be tachyons.
To remove tachyons, fermions are introduced into the theory in such a way that there exists a symmetry between 
bosons and fermions.  This boson--fermion symmetry is called supersymmetry\footnote{Supersymmetry could also be
viewed as a symmetry which unifies, in a non-trivial manner, the space-time symmetry described by the Poincar\'{e}
group, and the local gauge symmetry at each point of space-time.  The symmetry between bosons and fermions then 
come as a corollary.  See Refs.\cite{weba} or \cite{buku} for further details.}  A superstring theory which is free 
of quantum anomaly must be ten-dimensional.  Since we see only four macroscopic dimensions, the extra ones have to 
be compactified.  

A partially non-perturbative treatment of superstring theory is through entities called D$p$-branes 
\cite{polchinski}.  They are extended objects spanning $p$ dimensions.  Open strings stretch between D$p$-branes in 
the remaining dimensions, which are all compactified.  There is a non-trivial background gauge field permeating the 
whole ten-dimensional space-time.  The dynamics of the end points of open strings can be regarded as the dynamics 
of D$p$-branes themselves, each of which behaves like space-time of $p$ dimensions.  

Different versions of string theory were put forward in the 80's.  Lately, evidence suggests that there exist 
duality relationship among these different versions of string theory and so there is actually only one theory for
strings.  The most fundamental formulation of string theory is called M(atrix)-theory \cite{bfss}.  Currently, there
is a widely-believed M-theory conjecture which states that in the infinite momentum frame (a frame in which the 
momentum of a physical entity in one dimension is very large), M-theory \cite{bfss} can be described by the quantum 
mechanics of an infinite number $N$ of point-like D0-branes, the dynamics of which is in turn described by a {\em 
matrix model with supersymmetry}.

Besides the M-theory conjecture, there are a number of different ways of formulating the low-energy dynamics of
superstring theory as supersymmetric matrix models \cite{polchinski}.

We can use matrix models to describe condensed matter phenomena, too.  One major approach condensed matter 
physicists use to understand high-$T_c$ superconductivity, quantum Hall effect and superfluidity is to mimic them 
by integrable models like the Hubbard model \cite{hubbard}.  (This is a model for strongly correlated electron 
systems.  Its Hamiltonian consists of some terms describing electron hopping from site to site, and a term which 
suppresses the tendency of two electrons to occupy the same site.  We will write down the one-dimensional version 
of this model in a later section.)  It turns out that the Hubbard model and many other integrable models can 
actually be formulated as matrix models with or without supersymmetry.

Thus matrix models provide us a unifying formalism for a vast variety of physical phenomena.  Now, we would like to 
propose an {\em algebraic} approach to matrix models.  The centerpiece of the classical mechanical model of QCD in 
Refs.\cite{rajeev94} and \cite{krra} is a Poisson algebra.  We can write the Hamiltonian as an element of this 
Poisson algebra, and can describe the dynamics of hadrons and do calculations through it.  In string theory, the 
string is a one-dimensional object and so it sweeps out a two-dimensional surface called a worldsheet as time goes 
by.  Worldsheet dynamics possesses a remarkable symmetry called conformal symmetry --- the Lagrangian is invariant 
under an invertible mapping of worldsheet coordinates $x \rar x'$ which leaves the worldsheet metric tensor 
$g_{\m\n}(x)$ invariant up to a scale, i.e., $g'_{\m\n}(x') = \L (x) g_{\m\n}(x)$, where $\L (x)$ is a non-zero 
function of worldsheet coordinates.  This invertible mapping is called the conformal transformation.  It turns out 
that the conformal charges, i.e., the conserved charges associated with conformal symmetry, in particular the 
Hamiltonian of bosonic string theory, can be written as elements of a Lie algebra called the Virasoro algebra 
\cite{virasoro, gool}.  Through the Virasoro algebra, we learn a lot about string theory like the mass spectrum and 
the S-matrix elements.  

The relationship between conformal symmetry and the Virasoro algebra illustrates one powerful approach to physics
--- identify the symmetry of a physical system, express the symmetry in terms of an algebra, and use the properties
of the algebra to work out the physical behavior of the system.  Sometimes, the symmetry of the physical system is
so perfect that it completely determines the key properties of the system.

The above argument suggests that we may get fruitful discovery in gluon dynamics, M-theory and superconductivity if 
there is a Lie algebra for a generic matrix model, and we are able to write its Hamiltonian in terms of this Lie 
algebra, which expresses a new symmetry in physics.

\section{Formalism and Examples of Quantum Matrix Models in the Large-$N$ Limit}
\la{c2}

Examples of quantum matrix models fall into three broad classes: Yang--Mills theory, string theory and 
one-dimensional quantum spin systems.   

It should be obvious why Yang--Mills theory can be expressed as a quantum matrix model.  Identify the trivial 
vacuum state $|0\rangle$ with respect to the annihilation operators characterized by Eq.(\ref{2.2.1}) below, i.e., 
the action of any of these annihilation operators on this vacuum state yields 0.  A typical {\em physical state} is 
a linear combination of traces of products of creation operators acting on this vacuum state.  (In the context of
Yang--Mills theory, this is a color-singlet state of a single meson or glueball.)  A typical {\em observable} is a 
linear combination of the trace of a product of creation and annihilation operators.  Naively, we may think that 
the ordering of these operators is arbitrary, depending on the details of that physical model only.  Actually, the 
planar property of the large-$N$ limit \cite{thooft74a} comes into play here, and only those traces in which all 
creation operators come to the left of all annihilation operators survive the large-$N$ limit.  Moreover, the 
large-$N$ limit brings about a dramatic simplification of the action of this collective operator on a physical 
state --- the unique trace of a product of creation operators acting on the vacuum will not be broken up into a 
product of more than one trace of products of creation operators acting on the vacuum, and it is not possible for a 
product of more than one trace of products of creation operators to merge into just one trace of products of 
creation operators.  In other words, {\em an observable propagates a color-singlet state into a linear combination 
of color-singlet states} \cite{thorn79}.

This explains why we can mimic a string as a collection of string bits.  In this context, a physical state is a
string, open or closed, and the concluding statement of the previous paragraph implies that an observable replaces
a segment of the string with another segment, without breaking it and without joining several strings together.  
This is a desirable simplifying feature for some string models.  An important feature of this string-bit model is
that wherever this segment lies on the string, it will be replaced by the new segment at exactly the same location.
In other words, the action of an observable involves neighbouring string bits only, and is translationally 
invariant with respect to the string.

This brings us to quantum spin chain models.  Since the two most important properties of a quantum spin chain model 
is that a spin interacts with neighbouring spins (not necessarily nearest neighbouring spins though) only, and
the interaction is translationally invariant, we can write down a typical quantum spin chain model as a quantum
matrix model in the large-$N$ limit, so long as the boundary conditions match.

We are going to present these ideas systematically in this section.

First of all, we will present an abstract formalism of quantum matrix models.  We will then use a Yang--Mills 
theory with only bosonic adjoint matter fields to serve as a generic example to show how Yang--Mills theories with
different matter contents are expressed as quantum matrix models.  Next, we will turn to string theory, and see the
various ways quantum matrix models are related to it.  Finally, we will present a formal way of transcribing a 
quantum spin chain system into a quantum matrix model in the large-$N$ limit, and give some examples from bosonic 
spin systems.  Since many of them are known to be exactly integrable, and have even been exactly solved, we thus 
give here some examples of quantum matrix models, with or without supersymmetry, which are exactly integrable or 
even exactly solved.  These models either involve cyclically symmetric spin chains satisfying the periodic boundary 
condition, or involve open spin chains satisfying open boundary conditions.

\subsection{Formulation of Quantum Matrix Models in the Large-$N$ Limit}
\la{s2.2}

Let us give a unifying and more comprehensive review of the formulation of quantum matrix models in our previous 
papers \cite{opstal, clstal}.  The multi-index notations introduced in Appendix~A of Ref.\cite{opstal} and that of
Ref.\cite{clstal} will be used extensively.  The reader can understand these two appendices without reading the main
texts of those two references.

Think of the row and column indices of those annihilation and creation operators as the row and column indices of 
an element in $U(N)$.  We will call them {\em color indices}, as this is the case in Yang--Mills theory 
(Subsection~\ref{s2.3}).  Let $a^{\mu_1}_{\mu_2}(k)$ be an annihilation operator of a boson in the adjoint 
representation for $1 \leq k \leq \L$.  (Here $\L$ is a positive integer.)  Let $\c^{\mu}(\l)$ be an annihilation 
operator of a fermion in the fundamental representation for $1 \leq \l \leq \L_F$.  (Here $\L_F$ is also a positive 
integer.)  Lastly, let $\bar{\c}_{\mu}(\l)$ be an annihilation operator of a fermion in the conjugate 
representation.  We will call $k$ and $\l$ {\em quantum states other than color}.  The corresponding creation 
operators are $a^{\da\mu_1}_{\mu_2}(k)$, $\c^{\da}_{\mu}(\l)$ and $\bar{\c}^{\da\mu}(\l)$ with appropriate values 
for $k$ and $\l$.  We will say that these operators create an {\em adjoint parton}, a {\em fundamental parton} and 
a {\em conjugate parton}, respectively.  Most of these operators commute with one another except the following 
non-trivial cases:
\beq
   \le[ a^{\mu_1}_{\mu_2}(k_1), a^{\da\mu_3}_{\mu_4}(k_2) \ri] & = &
   \d_{k_1 k_2} \d^{\mu_3}_{\mu_2} \d^{\mu_1}_{\mu_4}; 
\la{2.2.1} \\
   \le[ \bar{\c}_{\mu_1}(\l_1), \bar{\c}^{\da\mu_2}(\l_2) \ri]_+ & = & \d_{\l_1 \l_2} \d^{\mu_2}_{\mu_1}; 
\la{2.2.3}
\eeq
and
\beq
   \le[ \c^{\mu_1}(\l_1), \c^{\da}_{\mu_2}(\l_2) \ri]_+ = \d_{\l_1 \l_2} \d^{\mu_1}_{\mu_2}.
\la{2.2.4}
\eeq

\begin{figure}
\epsfxsize=5in
\centerline{\epsfbox{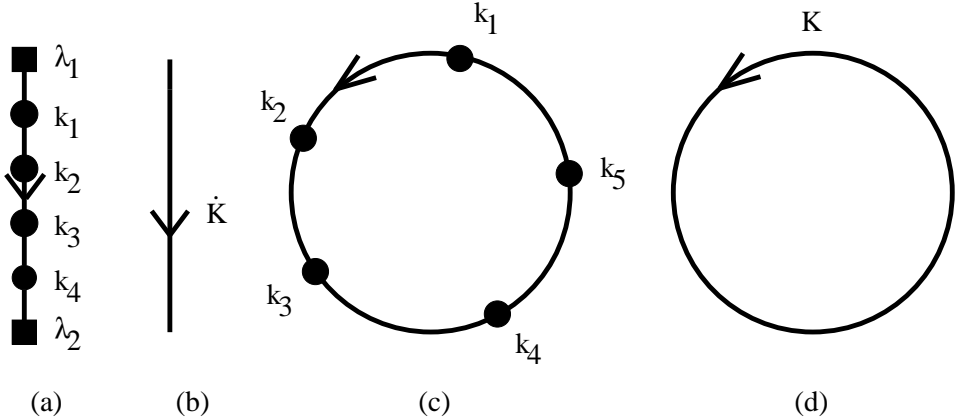}}
\caption{\em (a) A typical open singlet state in detail.  It consists of the creation operators of a conjugate 
parton of the quantum state $\l_1$, four adjoint partons of the quantum states $k_1$, $k_2$, $k_3$ and $k_4$, and a 
fundamental parton of the quantum state $\l_2$.  These creation operators are represented by a solid square at the 
top, four solid circles in the middle, and a solid square at the bottom, respectively.  All these operators carry 
color indices.  If two circles, or a circle and a square, are joined by a solid line, then the two corresponding 
operators share a color index, all possible values of which we sum over, regardless of how thick the solid line is.
The arrow indicates the direction of the integer sequence $\dot{K}$.  (b) A simplified diagrammatic representation 
of an open singlet state.  Here conjugate and fundamental partons are ignored.  We will ignore them in all future 
simplified diagrams.  The adjoint partons in between are represented by the integer sequence $\dot{K}$.  There is 
no relationship between the length of the thick line and the number of adjoint partons it carries.  (c) A typical 
closed singlet state $\Psi^K$ in detail.  This state consists of a series of adjoint partons of the quantum states 
$k_1$, $k_2$, \ldots, and $k_5$, and is cyclically symmetric.  (d) A simplified diagrammatic representation of the 
closed singlet state.  We use the integer sequence $K$ to represent it.  Cyclic symmetry is manifest.  The size of 
this big circle does not tell us the number of indices the circle carries.}     
\label{f2.1}
\end{figure}

There are two families of physical states (or color-invariant states in the context of gauge theory).  One family
consists of linear combinations of states of the form
\beq
   \bar{\ph}^{\l_1} \otimes s^K \otimes \ph^{\l_2} & \equiv & N^{-(c+1)/2} \bar{\c}^{\da\u_1}(\l_1) 
   \a^{\da\u_2}_{\u_1}(k_1) \a^{\da\u_3}_{\u_2}(k_2) \cd \nn \\
   & & \a^{\da\u_{c+1}}_{\u_c}(k_c) \c^{\da}_{\u_{c+1}}(\l_2) |0 \rangle.
\la{2.2.5}
\eeq
Here we use the capital letter $K$ to denote the integer sequence $k_1$, $k_2$, \ld, $k_c$.  {\em Unless otherwise
specified, the summation convention applies to all repeated color indices throughout this whole review.}  We will
justify the use of the notation $\otimes$ later.  This term carries a factor of $N$ raised to a certain power to 
make its norm finite in the large-$N$ limit.  (The proof that this is the correct power is similar to that given in 
Appendix~\ref{sa1.3} in which we will prove a closely related statement.)  We will call these {\em open singlet 
states}.  These are open string states in string-bit model.  The other family consists of linear combinations of 
states of the form
\beq
   \Ps^K \equiv N^{-c/2} \a^{\da\u_2}_{\u_1}(k_1) \a^{\da\u_3}_{\u_2}(k_2) \cd
   \a^{\da\u_1}_{\u_c}(k_c) |0 \rangle.
\la{2.2.6}
\eeq
These are {\em closed singlet states}.  These are closed string states in string-bit model.  This state is 
manifestly cyclic:
\beq
   \Ps^{K_1 K_2} = \Ps^{K_2 K_1}.
\la{2.2.8}
\eeq 
Figs.\ref{f2.1}(a) and (b) show a typical open singlet state, whereas Figs.\ref{f2.1}(c) and (d) show a typical 
closed singlet state.

\begin{figure}
\epsfxsize=5.5in
\centerline{\epsfbox{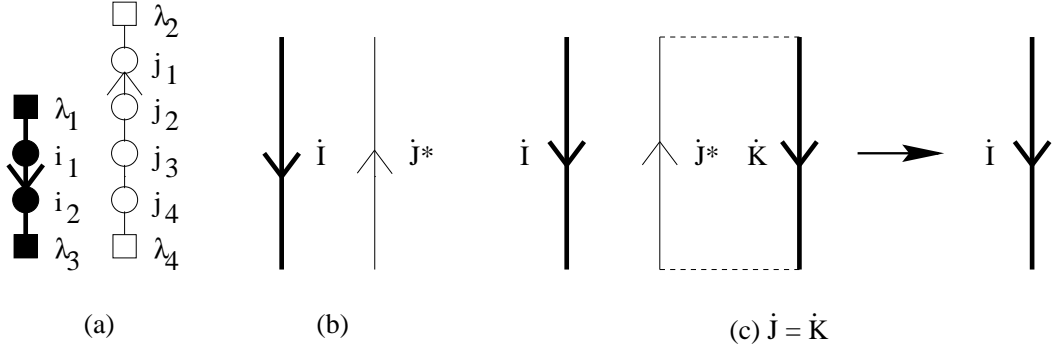}}
\caption{\em (a) A typical operator of the first kind.  On the left, there is a solid square representing the
creation operator of a conjugate parton with the quantum state $\l_1$.  Below it are two solid circles representing
the creation operators of adjoint partons with the quantum states $i_1$ and $i_2$.  At the bottom is another solid 
circle representing the creation operator of a fundamental parton with the quantum state $\l_3$; on the right, there
is a hollow square representing the annihilation operator of a conjugate parton with the quantum state $\l_2$.  
Below it are four hollow cirlces representing the annihilation of adjoint partons with the quantum states $j_1$, 
$j_2$, $j_3$ and $j_4$.  At the bottom is another hollow square representing the annihilation operator of a 
fundamental parton with the quantum state $\l_4$.  Again two circles, or a circle and a square, are joined by a 
solid or thin line whenever the corresponding operators share a common color index with all possible values.  The
arrows indicate the directions of the integer sequences $I$ and $\dot{J}$.  Notice that $\dot{J}$ is put in 
reverse.  (b)  An operator of the first kind in brief.  The sequence of creation operators is simplified to a thick 
line, and the sequence of annihilation operators is simplified to a thin line.  $\dot{J}^*$ is the reverse of 
$\dot{J}$.  The lengths of the two lines have no relationship with the numbers of creation or annihilation 
operators they carry.  (c) The action of an operator of the first kind on the adjoint parton portion of an open 
singlet state (Eq.(\ref{2.2.10})).  The dotted lines connect the line segments to be `annihilated' together.  This 
action produces an open singlet state as shown on the right of the arrow.}
\la{f2.2}
\end{figure}

Now let us construct physical operators acting on these singlet states.  It turns out that there are {\em five} 
families of them.  The first family consists of {\em finite} linear combinations of operators of the form
\beq
   \bar{\X}^{\l_1}_{\l_2} \otimes f^{\dot{I}}_{\dot{J}} \otimes \X^{\l_3}_{\l_4} & \equiv &
   N^{-(\dot{a} + \dot{b} + 2)/2} \bar{\c}^{\da\mu_1}(\l_1) a^{\da\mu_2}_{\mu_1}(i_1) \cd 
   a^{\da\mu_{\dot{a}+1}}_{\mu_{\dot{a}}}(i_{\dot{a}}) \c^{\da}_{\mu_{\dot{a}+1}}(\l_3) \cdot \nn \\
   & & \bar{\c}_{\n_1}(\l_2) a^{\n_1}_{\n_2}(j_1) \cd a^{\n_{\dot{b}}}_{\n_{\dot{b}+1}}(j_{\dot{b}}) 
   \c^{\n_{\dot{b}+1}}(\l_4),
\la{2.2.9}
\eeq
where $\dot{a} = \#(\dot{I})$ and $\dot{b} = \#(\dot{J})$.  We say that this is an {\em operator of the first 
kind}.  Figs.~\ref{f2.2}(a) and (b) show the diagrammatic representations of this family of operators.  In the 
planar large-$N$ limit, a typical term of an operator of the first kind propagates an open singlet state to another 
open singlet state:
\beq
   \bar{\X}^{\l_1}_{\l_2} \otimes f^{\dot{I}}_{\dot{J}} \otimes \X^{\l_3}_{\l_4} 
   \le( \bar{\ph}^{\l_5} \otimes s^{\dot{K}} \otimes \ph^{\l_6} \ri) = 
   \d^{\l_5}_{\l_2} \d^{\dot{K}}_{\dot{J}} \d^{\l_6}_{\l_4} 
   \bar{\ph}^{\l_1} \otimes s^{\dot{I}} \otimes \ph^{\l_3}.
\la{2.2.10}
\eeq
However, it annihilates a closed singlet state:
\beq
   \bar{\X}^{\l_1}_{\l_2} \otimes f^{\dot{I}}_{\dot{J}} \otimes \X^{\l_3}_{\l_4} \le( \Ps^K \ri) = 0.
\la{2.2.11}
\eeq
We can visualize Eq.(\ref{2.2.10}) in Fig.~\ref{f2.2}(c).  The fact that this operator does not split a singlet 
state into a product of singlet states, and it does not combine a product of singlet states together to form a 
singlet state has been known for a long time \cite{thorn79}, and is ultimately related to the planarity of the 
large-$N$ limit \cite{thooft74a}.  We will provide a non-rigorous diagrammatic proof of this fact in 
Appendix~\ref{sa1.3}, where we work on the action of an operator of the second kind (to be defined below) on an 
open singlet state.  The reader can easily work out the actions of operators of other kinds by the same reasoning.

Eq.(\ref{2.2.10}) justifies the use of the direct product symbol $\otimes$.  This equation shows that 
$\bar{\X}^{\l_1}_{\l_2}$ acts as a $\L_F \times \L_F$ matrix on the vector $\bar{\ph}^{\l_5}$ in a 
$\L_F$-dimensional space, $f^{\dot{I}}_{\dot{J}}$ acts as an infinite-dimensional matrix on the vector $s^{\dot{K}}$
in an infinite-dimensional space, and $\X^{l_3}_{l_4}$ acts as another $\L_F \times \L_F$ matrix on the vector
$\ph^{\l_6}$ in another $\L_F$-dimensional space.  Thus an operator of the first kind is an element of the direct 
product $gl(\Lambda_F) \otimes \salt \otimes gl(\Lambda_F)$.  Here, $gl(\Lambda_F)$ is the Lie algebra of the 
general linear group $GL(2\Lambda_F)$, and the infinite-dimensional Lie algebra $\salt$ spanned by 
$f^{\dot{I}}_{\dot{J}}$ is isomorphic to the inductive limit $gl_{+\infty}$ of the $gl(n)$'s as $n\to \infty$.

\begin{figure}
\epsfxsize=3.8in
\epsfysize=3.8in
\centerline{\epsfbox{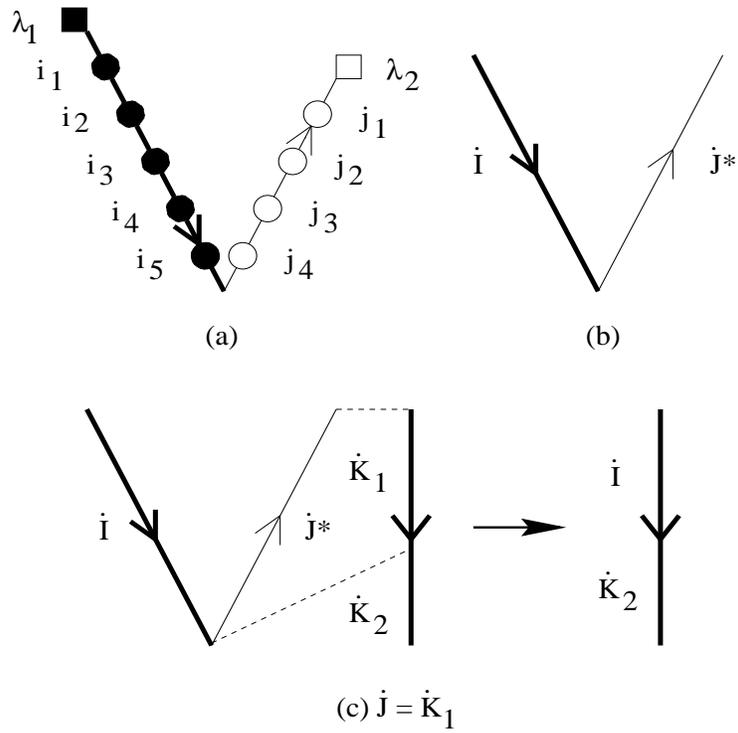}}
\caption{\em  (a) A typical operator of the second kind with five creation and four annihilation operators in 
detail.  The solid and hollow squares represent creation and annihilation operators of conjugate partons, 
respectively.  (b) An operator of the second kind in brief. (c) The action of an operator of the second kind on the 
adjoint parton portion of an open singlet state (Eq.(\ref{2.2.13})).}
\label{f2.3}
\end{figure}

{\em Operators of the second kind} are {\em finite} linear combinations of operators of the form
\beq
   \bar{\X}^{\l_1}_{\l_2} \otimes l^{\dot{I}}_{\dot{J}} & \equiv & N^{-(\dot{a} + \dot{b})/2}
   \bar{\c}^{\da\mu_1}(\l_1) a^{\da\mu_2}_{\mu_1}(i_1) a^{\da\mu_3}_{\mu_2}(i_2) \cd
   a^{\da\mu_{\dot{a} + 1}}_{\mu_{\dot{a}}}(i_{\dot{a}}) \cdot \nn \\
   & & a^{\n_{\dot{b}}}_{\mu_{\dot{a} + 1}}(j_{\dot{b}}) a^{\n_{\dot{b} - 1}}_{\n_{\dot{b}}}(j_{\dot{b} - 1}) \cd
   a^{\n_1}_{\n_2}(j_1) \bar{\c}_{\n_1}(\l_2).
\la{2.2.12}
\eeq
A typical operator of this kind is depicted in Fig.~\ref{f2.3}(a) and (b).  An operator of the second kind acts on 
the end with a conjugate parton and propagates an open singlet state to a linear combination of open singlet states:
\beq
   \bar{\X}^{\l_1}_{\l_2} \otimes l^{\dot{I}}_{\dot{J}} 
   \le( \bar{\ph}^{\l_3} \otimes s^{\dot{K}} \otimes \ph^{\l_4} \ri) = 
   \d^{\l_3}_{\l_2} \sum_{\dot{K_1} \dot{K_2} = \dot{K}} \d^{\dot{K}_1}_{\dot{J}} 
   \bar{\ph}^{\l_1} \otimes s^{\dot{I} \dot{K}_2} \otimes \ph^{\l_4}.
\la{2.2.13}
\eeq
On the R.H.S. of this equation, there will only be a finite number of non-zero terms in the sum (bounded by the 
number of ways of splitting $\dot{K}$ into subsequence), so there is no problem of convergence.  For example, if 
$\dot{K}$ is shorter than $\dot{J}$, the right hand side will vanish.  We can visualize this equation in 
Fig.~\ref{f2.3}(c).  This equation shows why we can treat this operator as a direct product of the operators 
$\bar{\Xi}^{\lambda_1}_{\lambda_2}$, $l^{\dot{I}}_{\dot{J}}$ and the identity operator.  The first operator acts as 
a $\Lambda_F \times \Lambda_F$ matrix on $\bar{\phi}^{\l_3}$'s, the second one acts on $s^{\dot{K}}$, whereas the 
last one acts trivially on $\phi^{\l_4}$'s.  This operator annihilates a closed singlet state:
\beq
   \bar{\X}^{\l_1}_{\l_2} \otimes l^{\dot{I}}_{\dot{J}} \le( \Ps^K \ri) = 0.
\la{2.2.14}
\eeq

\begin{figure}
\epsfxsize=3.8in
\epsfysize=3.8in
\centerline{\epsfbox{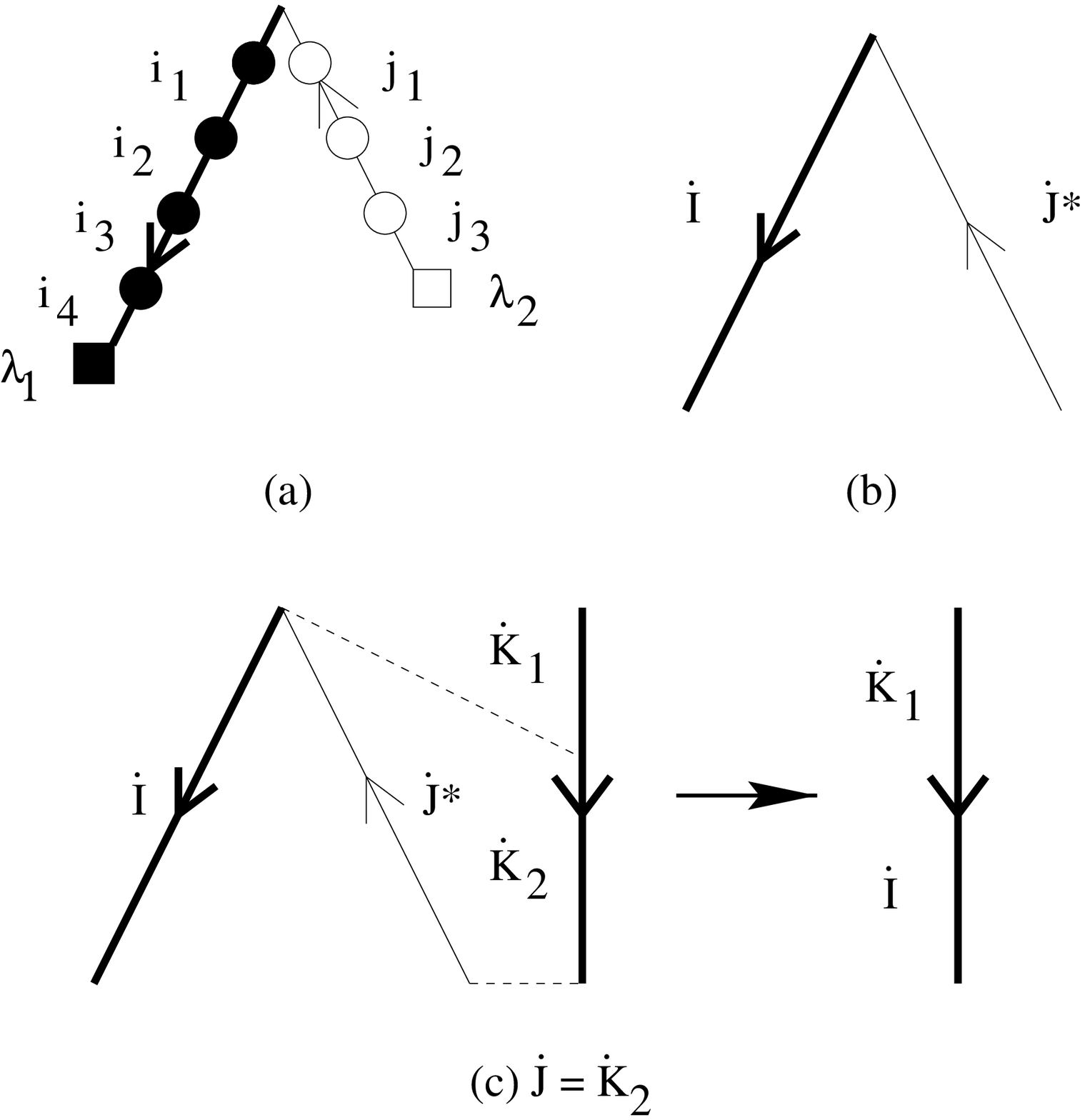}}
\caption{\em  (a) A typical operator of the third kind with four creation and three annihilation operators in 
detail.  This time the solid and hollow squares are creation and annihilation operators of fundamental partons.  
Notice the difference between the orientations of the arrows here and those in Fig.~\ref{f2.3}(a).  This reflects 
the difference in the manner the color indices are contracted in Eqs.(\ref{2.2.12}) and (\ref{2.2.15}).  (b) An 
operator of the third kind in brief.  (c)  The action of an operator of the third kind on the adjoint parton 
portion of an open singlet state (Eq.(\ref{2.2.16})).}
\label{f2.4}
\end{figure}

{\em Operators of the third kind} are very similar to those of the second kind.  They are linear combinations of
operators of the form
\beq
   r^{\dot{I}}_{\dot{J}} \otimes \X^{\l_1}_{\l_2} & \equiv & N^{-(\dot{a} + \dot{b})/2}
   \c^{\da}_{\mu_{\dot{a} + 1}}(\l_1) a^{\da\mu_{\dot{a} + 1}}_{\mu_{\dot{a}}}(i_{\dot{a}})
   a^{\da\mu_{\dot{a}}}_{\mu_{\dot{a} - 1}}(i_{\dot{a} - 1}) \cd \nn \\
   & & a^{\da\mu_2}_{\mu_1}(i_1) a^{\mu_1}_{\n_1}(j_1) a^{\n_1}_{\n_2}(j_2) \cd
   a^{\n_{\dot{b} - 1}}_{\n_{\dot{b}}}(j_{\dot{b}}) \c^{\n_{\dot{b}}}(\l_2).
\la{2.2.15}
\eeq
A typical operator of this kind is depicted in Fig.~\ref{f2.4}(a) and (b).  They act on the end with a fundamental 
parton instead of a conjugate parton as shown below:
\beq
   r^{\dot{I}}_{\dot{J}} \otimes \X^{\l_1}_{\l_2} \le( \bar{\ph}^{\l_3} \otimes s^{\dot{K}} \otimes \ph^{\l_4} \ri) 
   = \d^{\l_4}_{\l_2} \sum_{\dot{K}_1 \dot{K}_2 = \dot{K}} 
   \bar{\ph}^{\l_3} \otimes s^{\dot{K}_1 \dot{I}} \otimes \ph^{\l_1}.
\la{2.2.16}
\eeq
Fig.~\ref{f2.4}(c) shows this action diagrammatically. This equation shows that a term of an operator of the third
kind is a direct product of the identity operator, the operator $r^{\dot{I}}_{\dot{J}}$ and the operator 
$\Xi^{\lambda_1}_{\lambda_2}$.  Like the previous two kinds of operators, an operator of the third kind also 
annihilates a closed singlet state:
\beq
   r^{\dot{I}}_{\dot{J}} \otimes \X^{\l_3}_{\l_4} \le( \Ps^K \ri) = 0.
\la{2.2.17}
\eeq

\begin{figure}
\epsfxsize=5.5in
\centerline{\epsfbox{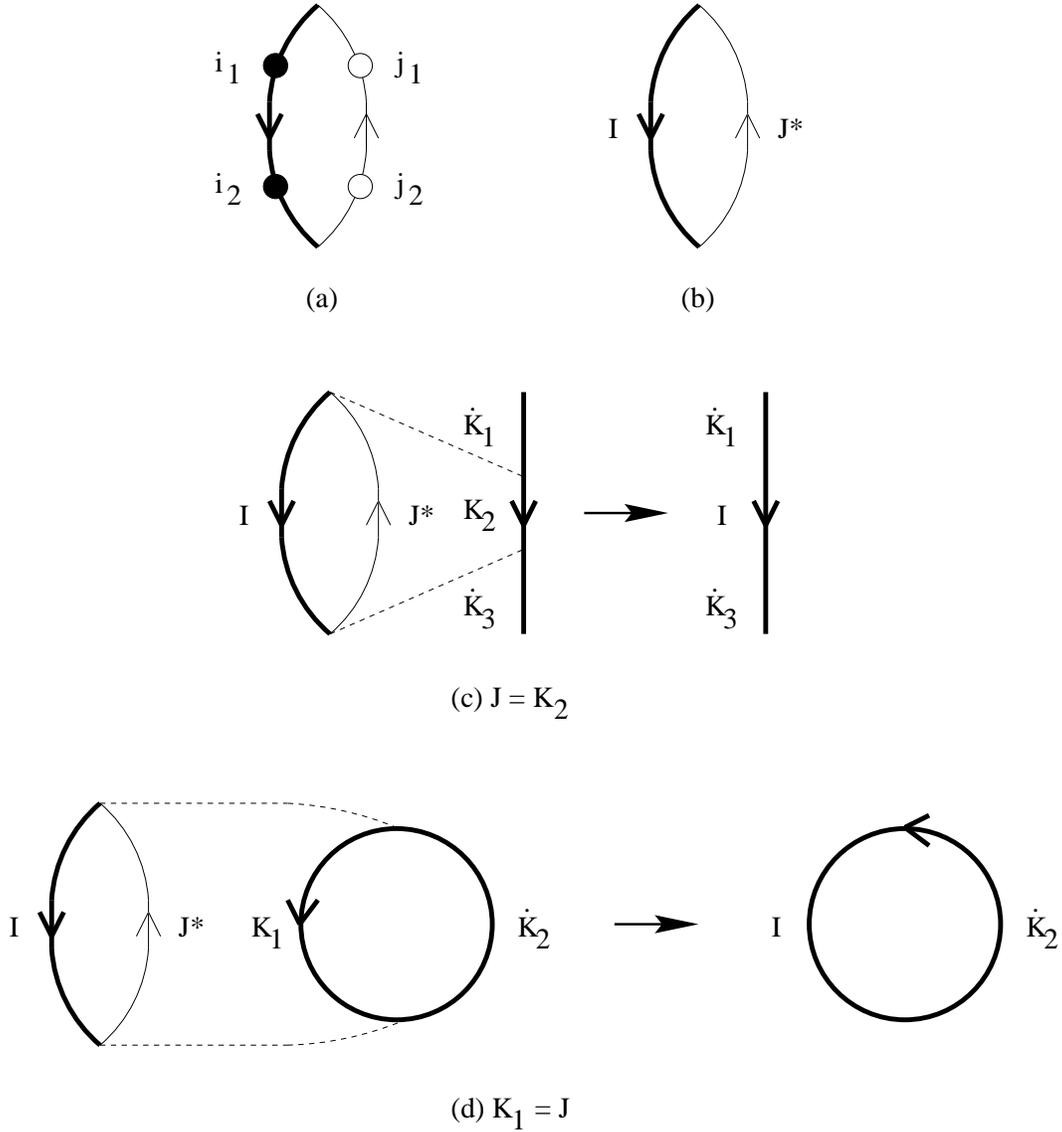}}
\caption{\em  (a) A typical operator of the fourth kind with two creation and two annihilation operators in detail. 
The conjugate and fundamental partons in an open singlet state are unaffected by the action of this operator.  (b) 
An operator of the fourth kind in brief.  (c) The action of an operator of the fourth kind on an open singlet state 
(Eq.(\ref{2.2.19})).  (d) The action of an operator of the fourth kind on a closed singlet state 
(Eq.(\ref{2.2.20})).}
\label{f2.5}
\end{figure}

{\em Operators of the fourth kind} are the most non-trivial among all physical operators.  They are finite linear 
combinations of operators of the form
\beq
   \g^I_J & \equiv & N^{-(a + b - 2)/2} a^{\da\mu_2}_{\mu_1}(i_1) a^{\da\mu_3}_{\mu_2}(i_2) \cd 
   a^{\da\n_b}_{\mu_a}(i_a) \nn \\
   & & \cdot a^{\n_{b-1}}_{\n_b}(j_b) a^{\n_{b-2}}_{\n_{b-1}}(j_{b-1}) \cd a^{\mu_1}_{\n_1}(j_1).
\la{2.2.18}
\eeq
We may write $\g^I_J$ as $\s^I_J$ in the ensuing paragraphs.  Unlike the operators of the first three kinds, {\em 
both $I$ and $J$ in an operator of the fourth kind must be non-empty sequences.}  Figs.~\ref{f2.5}(a) and (b) show 
the diagrammatic representations of such an operator.  In the planar large-$N$ limit, it replaces some sequences of 
adjoint partons on an open singlet state with some other sequences, producing a linear combination of this kind of 
states:
\beq
   \g^I_J \le( \bar{\ph}^{\l_1} \otimes s^{\dot{K}} \otimes \ph^{\l_2} \ri) =
   \bar{\ph}^{\l_1} \otimes \le( \sum_{\dot{K}_1 K_2 \dot{K}_3 = \dot{K}} 
   \d^{K_2}_J s^{\dot{K}_1 I \dot{K}_3} \ri) \otimes \ph^{\l_2}.
\la{2.2.19}
\eeq
This action is depicted in Fig.~\ref{f2.5}(c).  When it acts on a closed singlet state, it also replaces some 
sequences of adjoint partons with others:
\beq
   \g^I_J \Ps^K & = & \d^K_J \Ps^I + \sum_{K_1 K_2 = K} \d^{K_2 K_1}_J \Ps^I
   + \sum_{K_1 K_2 = K} \d^{K_1}_J \Ps^{I K_2} \nn \\
   & & + \sum_{K_1 K_2 K_3 = K} \d^{K_2}_J \Ps^{I K_3 K_1} 
   + \sum_{K_1 K_2 = K} (-1)^{\ep(K_1) \ep(K_2)} \d^{K_2}_J \Ps^{I K_1} \nn \\
   & & + \sum_{J_1 J_2 = J} \sum_{K_1 K_2 K_3 = K} \d^{K_3}_{J_1} \d^{K_1}_{J_2} \Ps^{I K_2};
\la{2.2.20}
\eeq
This action is depicted in Fig.~\ref{f2.5}(d).

\begin{figure}
\epsfxsize=4.4in
\centerline{\epsfbox{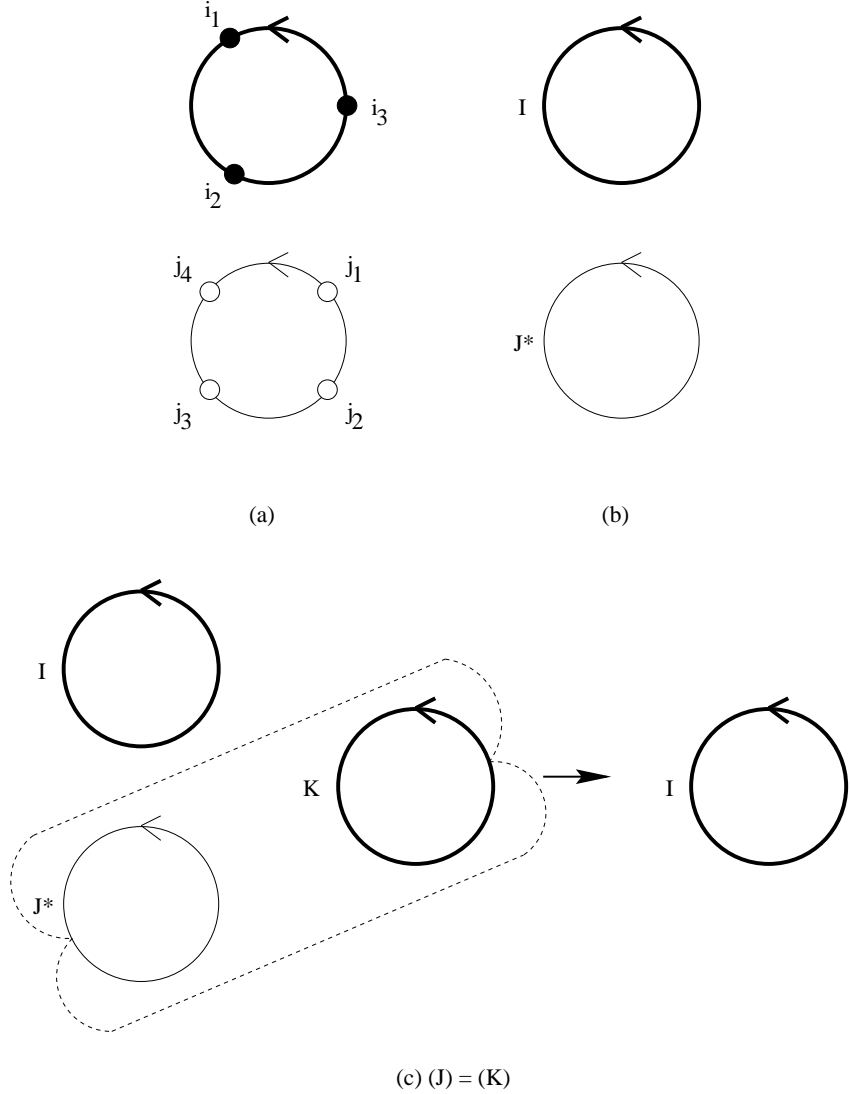}}
\caption{\em (a) A typical operator of the fifth kind with three creation and two annihilation operators in detail. 
The upper circle, labelled by $I$, represents creation operators, whereas the lower circle, labelled by $J$ 
represents annihilation operators.  Note that the sequence $J$ is put in the clockwise instead of the 
counterclockwise direction.  (b) An operator of the fifth kind in brief. $J^*$ is the reverse of $J$.  (c) The 
action of an operator of the fifth kind on a closed singlet state (Eq.(\ref{2.2.23})).}
\label{f2.6}
\end{figure}

{\em Operators of the fifth kind} form the last family of physical operators.  They are finite linear combinations 
of operators of the form
\beq
   \ti{f}^I_J & \equiv & N^{-(a+b)/2} a^{\da\mu_2}_{\mu_1}(i_1) a^{\da\mu_3}_{\mu_2}(i_2) \cd
   a^{\da\mu_1}_{\mu_a}(i_a) \cdot \nn \\
   & & a^{\n_b}_{\n_1}(j_b) a^{\n_{b-1}}_{\n_b}(j_{b-1}) \cd a^{\n_1}_{\n_2}(j_1).
\la{2.2.21}
\eeq
A typical operator of this kind is shown in Figs.~\ref{f2.6}(a) and (b).  This operator possesses the following
interesting cyclic property:
\beq
   \ti{f}^{I_2 I_1}_J & = & \ti{f}^{I_1 I_2}_J \mbox{; and} \nn \\
   \ti{f}^I_{J_2 J_1} & = & \ti{f}^I_{J_1 J_2}.
\la{2.2.21a}
\eeq
It annihilates an open singlet state:
\beq
   \ti{f}^I_J \le( \bar{\ph}^{\l_1} \otimes s^{\dot{K}} \otimes \ph^{\l_2} \ri) = 0.
\la{2.2.22}
\eeq
It may replace a closed singlet state with another one, though:
\beq
   \ti{f}^I_J \Ps^K & = & \d^K_J \Ps^I + \sum_{J_1 J_2 = J} \d^K_{J_2 J_1} \Ps^I.
\la{2.2.23}
\eeq
This action is depicted in Fig.~\ref{f2.6}(c).
 
We will say more about the mathematical properties of these physical operators in later sections.  In the remainder
of this section, we will see how these physical states and operators show up in actual physical systems, starting
with Yang--Mills theory.

\subsection{Examples: Yang--Mills Theory}
\la{s2.3}

The simplest example for this class of models is 2-dimensional Yang--Mills theory with adjoint matter fields.  For
the sake of pedagogy, we will give a detailed exposition below.  The following presentation is closely parallel to 
Ref.\cite{dakl}.  As will be shown below, there is no dynamics for the gauge bosons in 2 dimensions.  The quantum 
matrices in the above abstract formalism is realized as canonically quantized adjoint matter fields.  Different 
matrices carry different momenta.  The linear momentum and Hamiltonian of the model can be expressed as linear 
combinations of the five families of operators introduced above.

We need a number of preliminary definitions.  Let $g$ be the Yang--Mills coupling constant, $\a$ and $\b$ ordinary 
space-time indices, $A_{\a}$ a Yang--Mills potential and $\ph$ a scalar field in the adjoint representation of the 
gauge group $U(N)$, and $m$ the mass of this scalar field.  Both $\ph$ and $A_{\a}$ are $N \times N$ Hermitian 
matrix fields.  If we treat the Yang--Mills potential as the Lie-algebraically valued connection form and $\ph$ a 
fiber on the base manifold of space-time in the language of differential geometry\footnote{An authoritative text on 
differential geometry which explains the notion of a fiber bundle is Ref.\cite{kono}.  However, readers who are 
not interested in differential geometry may well take the covariant derivative to be given shortly for granted.}, 
then the covariant derivative is 
\[ D_{\a} \phi = \pa_{\a} \ph + {\rm i} \le[ A_{\a},\ph \ri]. \]

The Minkowski space action of this model is
\beq
   S = \int d^2 x {\rm Tr} \left[ \frac{1}{2} D_{\alpha}\phi D^{\alpha}\phi - \frac{1}{2} m^2 \phi^2 
   - \frac{1}{4g^2} F_{\alpha\beta} F^{\alpha\beta} \right].
\la{2.3.1}
\eeq
Introduce the light-cone coordinates
\[ x_+ = \frac{1}{\sqrt{2}} \le( x_0 + x_1 \ri) \]
and
\[ x_- = \frac{1}{\sqrt{2}} \le( x_0 - x_1 \ri). \]
Take $x^+$ as the time variable.  Choose the {\em light-cone gauge}
\[ A_- = 0. \]
The action is now simplified to
\beq
   S = \int dx^+ dx^- \Tr \le[ \pa_+ \ph \pa_- \ph - \frac{1}{2} m^2 \ph^2 + \frac{1}{2g^2} \le( \pa_- A_+ \ri)^2
       + A_+ J^+ \ri],
\la{2.3.2}
\eeq
where
\[ J^{+\m_1}_{\m_2} = \im \le[ \ph, \pa_- \ph \ri]^{\m_1}_{\m_2} \]
is the longitudinal momentum current.  Note that $A_+$ has no time dependence in Eq.(\ref{2.3.2}) and so the gauge 
field is not dynamical at all in the light-cone gauge.  Let us split $A_+$ into its zero mode $A_{+,0}$ (this is a
mode which is independent of $x_-$.) and non-zero mode $A_{+,n}$, i.e.,
\[ A_+ = A_{+,0} + A_{+,n}. \]
The Lagrange constraints for them are
\beq
   \int dx^- J^+ & = & 0 \mbox{, and}
\la{2.3.3} \\
   \pa_-^2 A_{+,n} - g^2 J^+ & = & 0,
\la{2.3.4}
\eeq
respectively.  Now we can use these two equations to eliminate $A_+$ in Eq.(\ref{2.3.2}) and get
\beq
   S = \int dx^+ dx^- \Tr \le( \pa_+ \ph \pa_- \ph - \frac{1}{2} m^2 \ph^2 + \frac{g^2}{2} J^+ \frac{1}{\pa_-^2} J^+
       \ri).
\la{2.3.5}
\eeq
The light-cone momentum and Hamiltonian are 
\[ P^{\pm} \equiv \int dx^- T^{+\pm}. \]
It follows from Eq.(\ref{2.3.5}) that their expressions in this model are 
\beq
   P^+ & = & \int dx^- \Tr \le( \pa_- \ph \ri)^2 \mbox{, and}
\la{2.3.6} \\
   P^- & = & \int dx^- \Tr \le( \frac{1}{2} m^2 \ph^2 - \frac{1}{2} g^2 J^+ \frac{1}{\pa_-^2} J^+ \ri).
\la{2.3.7}
\eeq

We now quantize the system.  Eq.(\ref{2.3.5}) implies that the canonical quantization condition is
\beq
   \le[ \ph^{\m_1}_{\m_2}(x^-), \pa_- \ph^{\m_3}_{\m_4}(\ti{x}^-) \ri] = 
   \frac{\im}{2} \d^{\m_1}_{\m_4} \d^{\m_3}_{\m_2} \d(x^- - \ti{x}^-). 
\la{2.3.8}
\eeq
Then a convenient field decomposition is
\beq
   \ph^{\m_1}_{\m_2}(x^+ = 0) = \frac{1}{\sqrt{2 \pi}} \int_0^{\ift} \frac{dk^+}{\sqrt{2k^+}} 
   \le[ a^{\m_1}_{\m_2}(k^+) \re^{- \im k^+ x^-} + a^{\da\m_1}_{\m_2}(k^+) \re^{\im k^+ x^-} \ri].
\la{2.3.9}
\eeq
If the annihilation and creation operators here satisfy the canonical commutation relation Eq.(\ref{2.2.1}), except 
that the Kroneckar delta function $\d_{k_1 k_2}$ in Eq.(\ref{2.2.1}) is replaced with the Dirac delta function 
$\d(k_1 - k_2)$ here, then the adjoint matter field will satisfy Eq.(\ref{2.3.8}).  A state can be built out of a 
series of creation operators acting on the trivial vacuum, and it takes the form
\[ a^{\da\m_1}_{\n_1}(k^+_1) a^{\da\m_2}_{\n_2}(k^+_2) \cd a^{\da\m_c}_{\n_c}(k^+_c) |0\rangle. \]
However, in general this state does not satisfy the Lagrangian constraint Eq.(\ref{2.3.3}).  Only those of the form 
given in Eq.(\ref{2.2.6}) satisfy this constraint.  We thus obtain a model whose physical states are the physical 
states introduced in the previous subsection.

Using Eqs.(\ref{2.3.6}) and (\ref{2.3.8}), we can quantize the light-cone momentum and get
\beq
   P^+ = \int_0^{\ift} dk k \g^k_k,
\la{2.3.10}
\eeq
where $\g^k_k$ is defined in Eq.(\ref{2.2.18}).  We have dropped the superscript $+$ in $k$ to simplify the 
notation.  This is a concrete example in which a physical observable is expressed as a linear combination of the 
operators introduced in the previous section.  (Here $k$ can take on an infinite number of values, and the 
regulator $\L \rar \ift$.  Nonetheless, taking the regulator to infinity has an influence on technical details 
only, since we are talking about a field theory without divergences in this limit.)  To obtain a similar formula 
for the light-cone Hamiltonian, we need to express, in momentum space, the longitudinal momentum current as a 
quantum operator first.  Define the longitudinal momentum current in momentum space to be
\beq
   \ti{J}^+ (k) & = & \frac{1}{\sqrt{2 \pi}} \int dx^- J^+ (x^-) \re^{- \im k^+ x^-}.
\la{2.3.11}
\eeq
Then, for $q > 0$,
\beq
   \ti{J}^{+\m_1}_{\m_2}(-q) & = & \frac{1}{2 \sqrt{2 \pi}} \int_0^{\ift} dp \frac{2p + q}{\sqrt{p(p + q)}}
   \le[ a^{\da\n}_{\m_2}(p) a^{\m_1}_{\n}(p + q) \ri. \nn \\
   & & \le. - a^{\da\m_1}_{\n}(p) a^{\n}_{\m_2}(p + q) \ri] \nn \\
   & & + \int_0^q dp \frac{q - 2p}{\sqrt{p(q - p)}} a^{\n}_{\m_2}(p) a^{\m_1}_{\n}(q - p),
\la{2.3.12}
\eeq
and
\beq
   \ti{J}^{+\m_1}_{\m_2}(q) = \le[ \ti{J}^{+\m_1}_{\m_2}(-q) \ri]^{\da}.
\la{2.3.13}
\eeq
We can now use Eqs.(\ref{2.3.7}), (\ref{2.3.8}), (\ref{2.3.11}), (\ref{2.3.12}) and (\ref{2.3.13}) to obtain
\begin{eqnarray}
   P^- & = & \frac{1}{2} m^2 \int_0^{\infty} \frac{dk}{k} g^k_k 
   + \frac{g^2 N}{4\pi} \int_0^{\infty} \frac{dk}{k} C g^k_k \nonumber \\
   & & + \frac{g^2 N}{8\pi} \int_0^{\infty} \frac{dk_1 dk_2 dk_3 dk_4}
   {\sqrt{k_1 k_2 k_3 k_4}}  
   \left\{ A \delta (k_1 + k_2 - k_3 - k_4) g^{k_3 k_4}_{k_1 k_2} \right. \nonumber \\  
   & & + \left. B \delta (k_1 - k_2 - k_3 - k_4) \le( g^{k_2 k_3 k_4}_{k_1} + g^{k_1}_{k_2 k_3 k_4} \ri) \ri\}, 
\label{2.3.14}
\end{eqnarray}     
where 
\begin{eqnarray*}
   A & = & \frac{(k_2 - k_1)(k_4 - k_3)}{(k_1 + k_2)^2} - \frac{(k_3 + k_1)(k_4 + k_2)}{(k_4 - k_2)^2}; \\
   B & = & \frac{(k_2 + k_1)(k_4 - k_3)}{(k_4 + k_3)^2} - \frac{(k_4 + k_1)(k_3 - k_2)}{(k_3 + k_2)^2} 
           \mbox{; and} \\
   C & = & \int_0^k dp \frac{(k+p)^2}{p(k-p)^2}.
\end{eqnarray*}
Eq.(\ref{2.3.14}) shows that the light-cone Hamiltonian is a linear combination of operators of the fourth kind.

The above formalism can be easily generalized to different cases.  Studying glueballs in quantum chromodynamics 
\cite{anda96a} requires the incorporation of more than one adjoint matter fields.  Take the large-$N$ limit as an 
approximation for pure Yang--Mills theory.  If we assume that the gluon field, i.e., the gauge field, is not 
dependent on the transverse dimensions (in other words, we take {\em dimensional reduction} as a further 
approximation), the gluon field will precisely be the adjoint matter fields.  The number of adjoint matter fields is
the same as the number of transverse dimensions in the system.  

To study mesons, we need to include fundamental and conjugate matter fields \cite{anda96b}.  They will serve as 
quark and antiquark fields, respectively.  The reader can find the expressions for the light-cone momentum and
Hamiltonian of a dimensionally reduced model of mesons in terms of the first four kinds of operators in 
Ref.\cite{opstal}.

\subsection{Examples: Quantum Gravity and String Theory}
\la{s2.4}

There are several ways to study quantum gravity and its most promising candidate, string theory, in terms of quantum
matrix models.  One idea is based on the crucial observation that the dual of a Feynman diagram can be taken as a 
triangulation of a plannar surface, which serves as a lattice approximation to a geometrical surface.  The 
partition function for two-dimensional quantum gravity can then be approximated by a matrix model in the large-$N$ 
limit \cite{david, kazakov85}.  Whether the matrix model is classical or quantum, and the exact form of its action
depends on what the conformal matter field coupled to quantum gravity is.  This, in turn, is equivalent to a string
theory with certain dimensions \cite{frgizi}.  This approach has the virtue that it reveals some non-perturbative 
behavior of string theory.  For example, a 3-dimensional non-critical string theory is equivalent to a model of 
two-dimensional quantum gravity coupled with conformal matter with the conformal charge $c = 2$.  Then this model 
of quantum gravity is further mapped to a one-matrix model in the large-$N$ limit with $\phi^3$ interaction, i.e.,
the action of this matrix model is
\beq
   S = \int d^2 x \Tr \le( \frac{1}{2} \pa_{\a} \Ph \pa^{\a} \Ph + \frac{1}{2} \m \Ph^2 - 
       \frac{1}{3 \sqrt{N}} \l \Ph^3 \ri),
\la{2.4.1}
\eeq
where $\Ph$ is an $N \times N$ matrix, and $\m$ and $\l$ are constants.

The second way is to consider the low-energy dynamics of string theory \cite{polchinski}.  For example, if we 
exclude all Feynman diagrams with more than one loop in a bosonic open string theory with $N$ Chan--Paton factors, 
we will retain only a tachyonic matrix field $\vp$ with three-tachyon coupling, and Yang--Mills gauge bosons 
minimally coupled to tachyons.  The action is
\beq
   S & = & \frac{1}{g^2} \int d^{26} x \le[ - \frac{1}{2} \Tr ( D_{\a} \vp D^{\a} \vp ) +
   \frac{1}{2 \a'} \Tr ( \vp^2 ) + \frac{1}{3} \sqrt{\frac{2}{\a'}} \Tr ( \vp^3 ) \ri. \nn \\
   & & \le. - \frac{1}{4} \Tr ( F_{\a\b} F^{\a\b} ) \ri],
\la{2.4.2}
\eeq
where $\a'$ is the Regge slope, and $g$ is the gauge boson coupling constant.

The third way is to approximate a string by a collection of string bits \cite{beth}.  A closed singlet state then 
represents a closed string, and an open singlet state represents an open string.

The fourth way is to consider the low-energy dynamics of D$p$-branes \cite{polchinski}.  Let $\x^1$, $\x^2$, \ld, 
and $\x^p$ be the coordinates inside the $p$-brane.  The action turns out to be another matrix model:
\beq
   S & = & - T_p \int d^{p+1} \x \re^{-\Ph} \le[ - {\rm det} ( G_{ab} + B_{ab} + 2 \pi \a' F_{ab} ) \ri]^{1/2},
\la{2.4.3}
\eeq
where $T_p$ is the brane tension, 
\beq
   G_{ab}(\x) = \frac{\pa X^{\a}}{\pa \x^a} \frac{\pa X^{\b}}{\pa \x^b} G_{\a\b}(X(\x))
\la{2.4.4}
\eeq
and
\beq
   B_{ab}(\x) = \frac{\pa X^{\a}}{\pa \x^a} \frac{\pa X^{\b}}{\pa \x^b} B_{\a\b}(X(\x))
\la{2.4.5}
\eeq
are the induced metric and antisymmetric tensor on the brane, and $F_{ab}$ is the background Yang--Mills field.  If
supersymmetry is incorporated into the theory, and if we further assume that space--time is essentially flat, the
action shown in Eq.(\ref{2.4.3}) will be turned into a supersymmetric Yang--Mills theory.

Closely related to the D-brane action is the M-theory conjecture and the matrix string corollary.  According to this
corollary, the action of type-IIA string theory in the infinite momentum frame is given by the matrix form of the
Green--Schwarz action \cite{diveve}.

The presence of a multitude of methods to transcribe string theory to matrix models shows their intimate 
relationship.

\subsection{Examples: Quantum Spin Chains}
\la{s2.7}

Quantum spin chain systems form another major class of models which can be expressed as quantum matrix models.  
There are various reasons to study quantum spin chain systems.  The essence of many condensed matter and high 
energy phenomena are captured by them.  For example, we can use the Ising model to study ferromagnetism, lattice 
gas, order-disorder phase transition \cite{domb}, lattice quantum gravity and even string theory \cite{amdujo}.
In addition, the integrability of these models provides rich insight in pure mathematics like quantum groups and 
knot theory \cite{chpr}.  The Bethe ansatz \cite{bethe} and the closely related Yang--Baxter equations \cite{yang, 
baxter} are well known to be powerful tools in studying and exactly solving many of these spin chain systems.  
These tools provide us a way to determine the integrability of and solve the associated quantum matrix models.  A 
better knowledge in quantum matrix models should thus help us understand exactly integrable models and condensed 
matter systems better, and vice versa.

There is a one-to-one correspondence between spin chain systems whose collective spin chain states are cyclically
symmetric and satisfy the periodic boundary condition, spin chain systems satisfying open boundary conditions, and 
quantum matrix systems in the large-$N$ limit.  (A connection between spin systems and matrix models was observed 
previously in Ref.\cite{klsu}.)  We will illustrate the idea using the simplest spin chain model, the 
one-dimensional quantum Ising model satisfying the periodic boundary condition \cite{onsager, frsu, kogut}.  A 
typical collective state of the whole spin chain can be characterized by a sequence of $c$ 2-dimensional column 
vectors, where $c$ is the number of sites in the spin chain.  Spin-up and spin-down states at the $p$-th site are 
characterized by the $p$-th column 
vectors 
\[ \left( \begin{array}{c} 1 \\ 0 \end{array} \right) \; \mbox{and} \;
   \left( \begin{array}{c} 0 \\ 1 \end{array} \right), \]
respectively.  A cyclically symmetric spin chain state can be constructed by summing over all cyclic permutations
of this sequence of $c$ column vectors.  The Hamiltonian $H^{\rm spin}_{\rm Ising}$ of this spin chain model is
\beq
   H^{\rm spin}_{\rm Ising}(\tau, \lambda) = \sum_{p=1}^c \tau_p^z + \lambda \sum_{p=1}^c \tau_p^x \tau_{p+1}^x.
\label{2.7.1}
\eeq
In this equation, $\tau_p^x$, $\tau_p^y$ and $\tau_p^z$ are Pauli matrices at the $p$-th site.  They act on the 
$p$-th column vector.  $\l$ is a real constant.  Two Pauli matrices at different sites (i.e., with different 
subscripts) commute with each other.  Moreover, they satisfy the periodic boundary condition 
$\t^{x,y,z}_{c+1} = \t^{x,y,z}_1$.

We can paraphrase the model in terms of the states and operators of quantum matrix models as follows.  A closed 
singlet state corresponds to a cyclically symmmetric spin chain state.  We allow the existence of only one 
Each adjoint parton corresponds to a spin.  There are 2 possible quantum states (i.e., $\L = 2$) other than color 
for an adjoint parton.  $a^{\da}(1)$ (color indices are suppressed) corresponds to a spin-up state, whereas 
$a^{\da}(2)$ corresponds to a spin-down state.  Therefore a typical collective state of a cyclically symmetric spin 
chain can be denoted by $\Ps^K$, where $K$ is an integer sequence of $c$ numbers, each of which is either 1 or 2.

Let us turn to the Hamiltonian in Eq.(\ref{2.7.1}).  Consider the operator $\tau_p^x \tau_{p+1}^x$.  It turns
\begin{eqnarray*}
   \left( \begin{array}{c} 1 \\ 0 \end{array} \right)_p
   \left( \begin{array}{c} 1 \\ 0 \end{array} \right)_{p+1} & , &
   \left( \begin{array}{c} 1 \\ 0 \end{array} \right)_p
   \left( \begin{array}{c} 0 \\ 1 \end{array} \right)_{p+1}, \\
   \left( \begin{array}{c} 0 \\ 1 \end{array} \right)_p
   \left( \begin{array}{c} 1 \\ 0 \end{array} \right)_{p+1} & \mbox{and} &
   \left( \begin{array}{c} 0 \\ 1 \end{array} \right)_p
   \left( \begin{array}{c} 0 \\ 1 \end{array} \right)_{p+1},
\end{eqnarray*}
where the subscripts $p$ and $p+1$ tell us that they are the $p$-th and $(p+1)$-th column vectors, respectively, 
into
\begin{eqnarray*}
   \left( \begin{array}{c} 0 \\ 1 \end{array} \right)_p
   \left( \begin{array}{c} 0 \\ 1 \end{array} \right)_{p+1} & , &
   \left( \begin{array}{c} 0 \\ 1 \end{array} \right)_p
   \left( \begin{array}{c} 1 \\ 0 \end{array} \right)_{p+1}, \\
   \left( \begin{array}{c} 1 \\ 0 \end{array} \right)_p
   \left( \begin{array}{c} 0 \\ 1 \end{array} \right)_{p+1} & \mbox{and} &
   \left( \begin{array}{c} 1 \\ 0 \end{array} \right)_p
   \left( \begin{array}{c} 1 \\ 0 \end{array} \right)_{p+1},
\end{eqnarray*}
respectively.  Let us rewrite this in the language of matrix model.  let $k_p$ and $k_{p+1}$ be the $p$-th and 
$(p+1)$-th numbers in $K$.  If ($k_p$, $k_{p+1}$) = (1, 1), (1, 2), (2, 1) or (2, 2), then they will be replaced 
with (2, 2), (2, 1), (1, 2) or (1, 1), respectively.  This implies that we can identify $\sum_{p=1}^c \tau_p^x
\tau_{p+1}^x$ in the spin chain model to be $\g_{11}^{22} + \g_{12}^{21} + \g_{21}^{12} + \g_{22}^{11}$ in the 
matrix model.  Likewise, $\sum_{p=1}^c \t_p^z$ can be identified with $\g^1_1 - \g^2_2$.  As a result, the 
Hamiltonian $H^{\rm matrix}_{\rm Ising}$ of this integrable matrix model in the large-$N$ limit is
\beq
   H^{\rm matrix}_{\rm Ising} = H_0 + \lambda V
\eeq
where
\beq
   H_0 & = & \g^1_1 - \g^2_2 \mbox{; and} \nonumber \\
     V & = & \left[ \g^{22}_{11} + \g^{21}_{12} + \g^{12}_{21} + \g^{11}_{22} \right].
\label{2.7.2}
\eeq
We can further rewrite this formula in terms of the creation and annihilation operators $a^{\dagger}$ and $a$:
\beq
   H^{\rm matrix}_{\rm Ising} & = & \Tr \le[ a^{\da}(1) a(1) - a^{\da}(2) a(2) \ri]
   + \frac{\l}{N} \Tr \le[ a^{\da}(2) a^{\da}(2) a(1) a(1) \right. \nonumber \\
   & & + a^{\da}(2) a^{\da}(1) a(2) a(1) + a^{\da}(1) a^{\da}(2) a(1) a(2) \nonumber \\
   & & \left. + a^{\da}(1) a^{\da}(1) a(2) a(2) \right] .
\label{2.7.3}
\eeq
Let us underscore that the matrix model defined by the Hamiltonian $H^{\rm matrix}_{\rm Ising}$ in 
Eqs.(\ref{2.7.2}) or (\ref{2.7.3}) is an integrable matrix model in the large-$N$ limit.

Notice that the Hamiltonian of the Ising spin chain model possesses translationally invariant terms which describe
nearest neighbour interactions only.  In general, any spin chain model which possesses the same kind of terms can
be transcribed to a quantum matrix model in the large-$N$ limit by the equation  
\beq
   \g^I_J = \sum_{p=1}^c X^{i_1 j_1}_p X^{i_2 j_2}_{p+1} \cd X^{i_a j_a}_{p+a-1}
\la{2.7.4}
\eeq
where $X^{ij}_p$ is the {\em Hubbard operator} defined by its action: if the spin chain state at the $p$-th state is
$j$, then it is changed to $i$; otherwise, the operator annihilates the spin chain state and we get 0.  The reader 
can take a look at Refs.\cite{prl} or \cite{clstal} for some more examples of exactly integrable quantum matrix 
models obtained from exactly integrable spin chain models satisfying the periodic boundary condition.

There is also a one-to-one correspondence between open spin chains and quantum matrix models in the large-$N$ 
limit.  The relation between a $\g$ and Hubbard operators is almost the same as that in Eq.(\ref{2.7.4}), except 
that the summation index $p$ runs from 1 to $c - a + 1$ this time.  It can be easily seen that the corresponding 
Hubbard operators for an $l^I_J$ and an $r^I_J$ are
\beq
   l^I_J = X^{i_1 j_1}_1 X^{i_2 j_2}_2 \cd X^{i_a j_a}_a
\la{2.7.5}
\eeq
and
\beq
   r^I_J = X^{i_1 j_1}_{n-a+1} X^{i_2 j_2}_{n-a+2} \cd X^{i_a j_a}_n,
\la{2.7.6}
\eeq
respectively.

Let us consider an integrable spin-$\frac{1}{2}$ XXZ model \cite{albaba} to see how the transcription is put into 
practice.  The Hamiltonian $H^{\rm spin}_{\rm XXZ}$ of this spin chain model is
\begin{eqnarray}
   H^{\rm spin}_{\rm XXZ} & = & \frac{1}{2 \sin \gamma} \left[ \sum_{p=1}^{c-1} 
   (\tau_p^x \tau_{p+1}^x + \tau_p^y \tau_{p+1}^y + \cos \gamma \tau_j^z \tau_{p+1}^z)
   \right. \nonumber \\
   & & + \left. i \sin \gamma ( \coth \xi_- \tau_1^z + \coth \xi_+ \tau_c^z )
   \right],
\label{2.7.7}
\end{eqnarray}
where $\gamma \in (0, \pi)$ and both $\xi_-$ and $\xi_+$ are arbitrary constants.  The Hamiltonian 
$H^{\ma}_{\rm XXZ}$ of the associated integrable matrix model is
\begin{eqnarray}
   H^{\ma}_{\rm XXZ} & = & \frac{1}{2 \sin \gamma} \left\{ 2 ( \s_{12}^{21} + \s_{21}^{12} 
   + \cos \gamma (\s_{11}^{11} - \s_{12}^{12} - \s_{21}^{21} + \s_{22}^{22} ) \right. \nonumber \\
   & & + \im \left. \sin \gamma \lbrack \coth \xi_- (l_1^1 - l_2^2) + \coth \xi_+ (r_1^1 - r_2^2) \rbrack \right\}.
\label{2.7.8}
\end{eqnarray}

Other examples of exactly integrable spin chain models satisfying open boundary conditions and the corresponding
quantum matrix models in the large-$N$ limit can be found in Ref.\cite{opstal}.

We also have a transcription rule between fermionic spin chain models and integrable supersymmetric quantum matrix 
models \cite{plb}.

\section{A Lie Algebra for Open Strings}
\la{c4}

As we have just been seen in the previous chapter, quantum matrix models in the large-$N$ limit are of very wide 
applicability in physics.  This leads us naturally to the following question: can we do something more in addition
to writing down the physical observables of these models as linear combinations of the operators introduced in 
Subsection~\ref{s2.2}?  One major milestone in understanding the physics of these models will be to obtain the 
spectra of some of these observables, or at least to obtain their key qualitative properties.

A number of systematic analytic approaches have been developed to solve classical matrix models in the large-$N$ 
limit, another class of models widely used in physics \cite{britpazu, mehta, douglas}.  On the other hand, the 
method employed for studying the spectra of the Hamiltonians of various versions of Yang--Mills theory in recent 
years is numerical in nature (e.g., Ref.\cite{dakl}).  A lot of effort has been spent to develop more accurate and 
sophisticated numerical analysis, and we have a fairly large body of knowledge in the numerical solutions of these 
quantum matrix models.  It should be nice to see how much this approach can tell us about the various interesting 
physical systems.  Nonetheless, it should also be nice if we have several other approaches to the same problem; 
some results which may be difficult to obtain in one approach may be pretty trivial in another.  A combination of 
the knowledge gained from different approaches can lead us to much deeper understandings of the physics.

In Section~\ref{c1}, we indicated that understanding the symmetry of a physical system can lead to the  
elucidation of important features of it.  This provides us another approach: what is the symmetry of quantum matrix 
models?  A continuous symmetry in physics is usually expressed as a Lie group, or its infinitesimal form, a Lie 
algebra.  Is there a Lie algebra for quantum matrix models?  Or more specifically speaking, are the physical 
observables of quantum matrix models elements of a Lie algebra?  

Yes, such a Lie algebra exists.  

Indeed, we introduced two Lie algebras, the grand (open string) algebra and the algebra $\cyclix$, for open and 
closed bosonic strings in Refs.\cite{opstal} and \cite{clstal}, respectively.  The fact that we can give a unifying 
account of all quantum observables for open and closed singlet states suggests that there may be a way to treat 
these two Lie algebras on an equal footing.  In fact, they are quotient algebras of a bigger algebra which in turn 
is a subalgebra of a precursor algebra.  Some familiarity of the grand open string algebra and $\cyclix$, however, 
is necessary to see how this works.  We will therefore review a hierachy of subalgebras and quotient algebras, from 
the simplest to the most non-trivial, of a simplified version of the grand open string algebra in which there are 
only one degree of freedom for the fermion in the fundamental representation, and one degree of freedom for the 
anti-fermion in the conjugate representation.  We will also derive some results about the structure of these 
algebras not published elsewhere.

Another point worthy mentioning here is that these algebras are related intimately with the Cuntz algebra 
\cite{cuntz, evans} and the Witt algebra \cite{zassenhaus, chang, seligman}, the central extension of which is the 
Virasoro algebra \cite{virasoro}.  Further understandings of these well-known algebras can help us a lot in 
explaining the physics of quantum matrix models.

After acquiring some expertise in this kind of Lie algebraic arguments, we will study the full algebra, the 
{\em grand string algebra}, for all five kinds of operators, and some more subalgebras of it in subsequent 
sections.

\subsection{Operators of the First Three Kinds}
\la{s4.2}

We will review what we have learnt so far about the Lie algebras for the operators of the first three kinds in 
Ref.\cite{opstal} which will be useful in constructing the grand string algebra later, and state some other 
properties of them which have not been published yet.

The adjoint matter portion of the operators of the first kind form a Lie algebra with the following Lie bracket
\beq
   \left[ f^{\dot{I}}_{\dot{J}}, f^{\dot{K}}_{\dot{L}} \right] = 
   \delta^{\dot{K}}_{\dot{J}} f^{\dot{I}}_{\dot{L}} - \delta^{\dot{I}}_{\dot{L}} f^{\dot{K}}_{\dot{J}}.
\la{4.2.2}
\eeq
This Lie algebra $\salt$ can be derived from the commutator of the associative algebra of these operators.  $\salt$ 
is isomorphic to $\gl$, the Lie algebra, with the usual bracket, of all complex matrices $(a_{ij})_{i, j} \in Z_+$ 
such that the number of nonzero $a_{ij}$ is finite.  The isomorphism is given by a one-one correspondence between 
the multi-indices $I$ and the set of natural numbers $Z_+r$,  for example by a {\em lexicographic 
ordering}\footnote{This is a total ordering $>$ among integer sequences such that $\dot{I} >\dot{J}$ for the 
sequences $\dot{I}$ and $\dot{J}$ if either
\begin{enumerate}
\item $\#(\dot{I}) > \#(\dot{J})$; or
\item $\#(\dot{I}) = \#(\dot{J}) = a\neq 0$, and there exists an integer $r \leq a$ 
such that
      $i_1 = j_1$, $i_2 = j_2$, \ldots, $i_{r-1} = j_{r-1}$ and $i_r > j_r$.
\end{enumerate}
Explicitly, the total ordering can be presented as
\begin{eqnarray*}
   & & \emptyset < 1 < 2 < \cdots < \Lambda < 11 < 12 < \cdots < 1\Lambda \\
   & & < 21 < \cdots < \Lambda1 < \cdots < \Lambda\Lambda < 111 \cdots.
\end{eqnarray*}
This also gives a one-one correspondence (counting rule) between the set of values of the indices and the set of 
natural numbers: $\emptyset \to 1, 1 \to 2, \ldots, \Lambda \to \Lambda + 1, 1\Lambda \to \Lambda+2,$ etc..}. 

Operators of the second kind form another Lie algebra.  This is again the commutator of an associative algebra with
the multiplication law
\beq
   l^{\dot{I}}_{\dot{J}} l^{\dot{K}}_{\dot{L}} = \delta^{\dot{K}}_{\dot{J}} l^{\dot{I}}_{\dot{L}} +
   \sum_{\dot{J}_1 J_2 = \dot{J}} \delta^{\dot{K}}_{\dot{J}_1} l^{\dot{I}}_{\dot{L} J_2} + 
   \sum_{\dot{K}_1 K_2 = \dot{K}} \delta^{\dot{K}_1}_{\dot{J}} l^{\dot{I} K_2}_{\dot{L}}.
\la{4.3.2}
\eeq
The associativity is demonstrated in Appendix~\ref{sa4.1}.  The resulting Lie bracket is
\beq
   \lbrack l^{\dot{I}}_{\dot{J}}, l^{\dot{K}}_{\dot{L}} \rbrack & = & 
   \delta^{\dot{K}}_{\dot{J}} l^{\dot{I}}_{\dot{L}} +
   \sum_{\dot{J}_1 J_2 = \dot{J}} \delta^{\dot{K}}_{\dot{J}_1} l^{\dot{I}}_{\dot{L} J_2} + 
   \sum_{\dot{K}_1 K_2 = \dot{K}} \delta^{\dot{K}_1}_{\dot{J}} l^{\dot{I} K_2}_{\dot{L}} \nonumber \\
   & & - \delta^{\dot{I}}_{\dot{L}} l^{\dot{K}}_{\dot{J}} -
   \sum_{\dot{L}_1 L_2 = \dot{L}} \delta^{\dot{I}}_{\dot{L}_1} l^{\dot{K}}_{\dot{J} L_2} - 
   \sum_{\dot{I}_1 I_2 = \dot{I}} \delta^{\dot{I}_1}_{\dot{L}} l^{\dot{K} I_2}_{\dot{J}}.
\la{4.3.3}
\eeq
The first three terms on the R.H.S. of Eq.(\ref{4.3.3}) are diagrammatically represented in Fig.\ref{f4.1}.  We will
call the Lie algebra defined by Eq.(\ref{4.3.3}) the {\em leftix algebra} or $\hatleftix$. (We justify this name as 
an abbreviation of `left multi-matrix algebra'.) 

\begin{figure}[ht]
\epsfxsize=4in
\centerline{\epsfbox{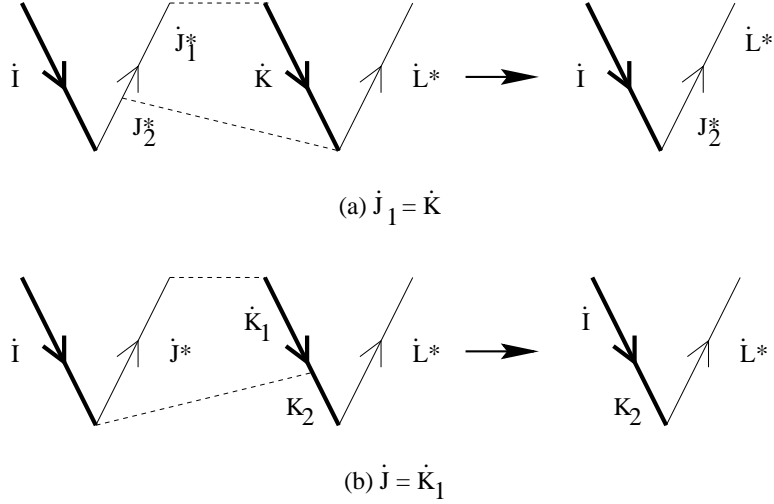}}
\caption{\em Diagrammatic representations of the defining Lie bracket of the leftix algebra.  (a) The second term of
the R.H.S. of Eq.(\ref{4.3.3}).  (b) The third term of the R.H.S. of Eq.(\ref{4.3.3}).}
\label{f4.1}
\end{figure}

To understand the properties of this Lie algebra better, let us determine its Cartan subalgebra and the root spaces 
with respect to this Cartan subalgebra.  We will follow the definition of the Cartan subalgebra given by Humphreys 
\cite{humphreys}.  To understand this definition, we need a number of preliminary notions.  Let ${\cal L}$ be a Lie 
algebra.  Define the {\em descending central series} by ${\cal L}^0 = {\cal L}$, ${\cal L}^1 = \lb {\cal L}, 
{\cal L} \rb$, ${\cal L}^2 = \lb {\cal L}, {\cal L}^1 \rb$, \ld, ${\cal L}^i = \lb {\cal L}, {\cal L}^{i-1} \rb$.  
${\cal L}$ is called {\em nilpotent} if ${\cal L}^n = 0$ for some $n$.  The {\em normalizer} of a subalgebra 
${\cal K}$ of ${\cal L}$ is defined by $N_{\cal L} ({\cal K}) = \{ x \in {\cal L} | \lb x, {\cal K} \rb \subset 
{\cal K} \}$.  A {\em Cartan subalgebra} of a Lie algebra ${\cal L}$ is a nilpotent subalgebra which is equal to 
its normalizer in ${\cal L}$.  A Cartan subalgebra of a Lie algebra is a nilpotent subalgebra of it such that the 
normalizer of the nilpotent subalgebra is the nilpotent subalgebra itself.  It turns out that all vectors of the 
form $l^{\dot{I}}_{\dot{I}}$, where $\dot{I}$ is either empty or an arbitrary finite integer sequence of integers 
between $1$ and $\L$ inclusive, span a Cartan subalgebra of the algebra $\hatleftix$.  The proof of this 
proposition can be found in Appendix~\ref{sa4.2}. 

As we have learnt in Ref.\cite{opstal}, the action of the linear combination
\beq
   f^{\dot{K}}_{\dot{L}} = l^{\dot{K}}_{\dot{L}} - \sum_{j=1}^{\L} l^{\dot{K}j}_{\dot{L}j}
\la{4.3.5}
\eeq
is
\beq
   f^{\dot{K}}_{\dot{L}} s^{\dot{M}} = \d^{\dot{M}}_{\dot{L}} s^{\dot{K}},
\la{4.2.1}
\eeq
so this linear combination satisfies
\beq
   \le[ l^{\dot{I}}_{\dot{J}}, f^{\dot{K}}_{\dot{L}} \ri] = 
   \sum_{\dot{K}_1 \dot{K}_2 = \dot{K}} \delta^{\dot{K}_1}_{\dot{J}} f^{\dot{I}\dot{K}_2}_{\dot{L}}
   - \sum_{\dot{L}_1 \dot{L}_2 = \dot{L}} \delta^{\dot{I}}_{\dot{L}_1} f^{\dot{K}}_{\dot{J} \dot{L}_2}.
\la{4.3.7}
\eeq
This formula is depicted in Fig.\ref{f4.2}.  It follows from this equation that
\beq
   \left[ l^{\dot{I}}_{\dot{I}}, f^{\dot{K}}_{\dot{L}} \right] = \left(  
   \sum_{\dot{K}_1 \dot{K}_2 = \dot{K}} \delta^{\dot{K}_1}_{\dot{I}} 
   - \sum_{\dot{L}_1 \dot{L}_2 = \dot{L}} \delta^{\dot{I}}_{\dot{L}_1} \right) f^{\dot{K}}_{\dot{L}}.
\la{4.3.15}
\eeq
As a result, every $f^{\dot{K}}_{\dot{L}}$ is a root vector of $\leftix$.  Moreover, these are the only root 
vectors of the algebra $\hatleftix$. A proof that there are no root vectors other than the $f^{\dot{K}}_{\dot{L}}$ 
will be provided in Appendix~\ref{sa4.3}.  We therefore conclude that every root space is one-dimensional.
   
\begin{figure}[ht]
\epsfxsize=4in
\centerline{\epsfbox{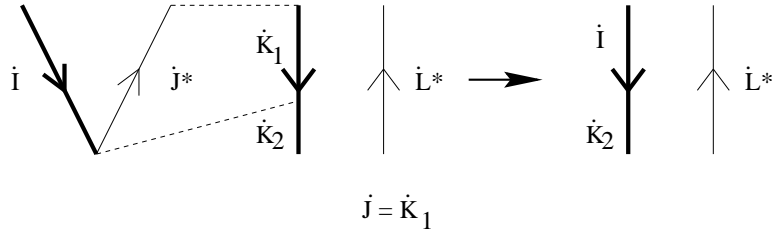}}
\caption{\em The Lie bracket between $l^{\dot{I}}_{\dot{J}}$ and $r^{\dot{K}}_{\dot{L}}$.  Only the first term on 
the R.H.S. of Eq.(\ref{4.3.7}) is shown.}
\la{f4.2}
\end{figure}

We have also learnt from Ref.\cite{opstal} that $\salt$ is a proper ideal of the algebra $\hatleftix$.  There 
exists the exact sequence of Lie algebras
\[ 0 \to \salt \to \hatleftix \to \leftix \to 0. \]
Moreover, $\hatleftix$ and $\leftix$ are the extended Cuntz--Lie algebra and the Cuntz--Lie algebra, respectively.

We can get further insight of $\hatleftix$ and $\leftix$ by considering the special case $\L = 1$.  Then all 
integer sequences are repetitions of the number 1 a number of times.  We can then simplify the notations and write 
$s^{\dot{K}}$ as $s^{\#(\dot{K})}$ and $l^{\dot{I}}_{\dot{J}}$ as $l^{\#(\dot{I})}_{\#(\dot{J})}$.  The action of 
$l^{\dot{a}}_{\dot{b}}$, where $\dot{a}$ and $\dot{b}$ are the numbers of integers in the various sequences, on 
$s^{\dot{c}}$, where $\dot{c}$ is also a non-negative integer, is given by
\begin{equation}
   l^{\dot{a}}_{\dot{b}} s^{\dot{c}} = \theta (\dot{b} \leq \dot{c}) s^{\dot{a} + \dot{c} - \dot{b}},
\la{4.3.16}
\end{equation}
where $\theta( \mbox{condition} )$ is 1 if the condition holds, and 0 otherwise.  The Lie bracket Eq.(\ref{4.3.3}) 
is simplified to
\begin{equation}
   \left[ l^{\dot{a}}_{\dot{b}}, l^{\dot{c}}_{\dot{d}} \right] = 
   \theta (\dot{c} \leq \dot{b}) l^{\dot{a}}_{\dot{b} + \dot{d} - \dot{c}} +
   \theta (\dot{b} < \dot{c}) l^{\dot{a} + \dot{c} - \dot{b}}_{\dot{d}} - 
   \theta (\dot{a} \leq \dot{d}) l^{\dot{c}}_{\dot{b} + \dot{d} - \dot{a}} -
   \theta (\dot{d} < \dot{a}) l^{\dot{a} + \dot{c} - \dot{d}}_{\dot{b}}.
\la{4.3.17}
\end{equation}
The set of all vectors of the form $l^{\dot{a}}_{\dot{a}}$ span a Cartan subalgebra.  The root vectors are given by 
$f^{\dot{c}}_{\dot{d}} = l^{\dot{c}}_{\dot{d}} - l^{\dot{c} + 1}_{\dot{d} + 1}$.  This can be deduced from 
Eq.(\ref{4.3.5}).  The action of $f^{\dot{a}}_{\dot{b}}$ on $s^{\dot{c}}$, which can be derived from 
Eq.(\ref{4.2.1}), is
\begin{equation}
   f^{\dot{a}}_{\dot{b}} s^{\dot{c}} = \delta^{\dot{c}}_{\dot{b}} s^{\dot{a}}.
\label{4.3.18}
\end{equation}
The corresponding eigenequation, which can be deduced from Eq.(\ref{4.3.15}), is
\begin{equation}
   \left[ l^{\dot{a}}_{\dot{a}}, f^{\dot{c}}_{\dot{d}} \right] = 
   \left( \theta (\dot{a} \leq \dot{c}) - \theta (\dot{a} \leq \dot{d}) \right) f^{\dot{c}}_{\dot{d}}.
\la{4.3.19}
\end{equation}

As in an earlier discussion, the subspace ${{\mathit F}_1}$ spanned by all the vectors of the form 
$f^{\dot{c}}_{\dot{d}}$ form a proper ideal of this $\Lambda=1$ algebra $\hat{\mathit L}_1$.  We can deduce from 
Eq.(\ref{4.3.7}) that
\begin{equation}
   \left[ l^{\dot{a}}_{\dot{b}}, f^{\dot{c}}_{\dot{d}} \right] = 
   \theta (\dot{b} \leq \dot{c}) f^{\dot{a} + \dot{c} - \dot{b}}_{\dot{d}} - 
   \theta (\dot{a} \leq \dot{d}) f^{\dot{c}}_{\dot{b} + \dot{d} - \dot{a}}. 
\la{4.3.20}
\end{equation}
We can now form the quotient algebra of cosets of the form $v + \salt$, where $v$ is an arbitrary vector of the 
algebra $\hat{\mathit l}_1$.  This quotient algebra is spanned by the cosets $l^{\dot{a}}_0 + {\salt}$ and 
$l^0_{\dot{b}} + {{\mathit F}_1}$, where $a$ and $b$ run over all $0, 1, \cdots, \infty$.  It is straightforward to 
show that the following Lie brackets hold:
\begin{eqnarray}
   \left[ l^{\dot{a}}_0 + {{\mathit F}_1}, l^{\dot{c}}_0 + {{\mathit F}_1} \right] & = & {{\mathit F}_1} ;  
   \nonumber \\
   \left[ l^{\dot{a}}_0 + {{\mathit F}_1}, l^0_{\dot{d}} + {{\mathit F}_1} \right] & = & {{\mathit F}_1} 
   \mbox{; and} \nonumber \\
   \left[ l^0_{\dot{b}} + {{\mathit F}_1}, l^0_{\dot{d}} + {{\mathit F}_1} \right] & = & {{\mathit F}_1}.
\label{4.3.21}
\end{eqnarray}
Therefore, this quotient algebra is an Abelian algebra.  

The relationship between $\hatleftix$ and the Cuntz algebra acquires a new meaning in this special case.  Let
$a^{\da} \equiv l^1_0$ and $a \equiv l^0_1$.  $a^{\da}$ and $a$ are the building blocks of $\hat{\mathit L}_1$ 
because 
\beq
   l^{\dot{a}}_{\dot{b}} = \le( a^{\da} \ri)^{\dot{a}}  (a)^{\dot{b}}.
\eeq
$a^{\da}$ and $a$ satisfy
\beq
   a a^{\dagger} = 1.
\la{4.3.22}
\eeq
This is the defining relation for the {\em Toeplitz algebra} \cite{murphy}.  Note that the commutator between $a$ 
and $a^{\dagger}$ is a finite-rank operator, the projection operator to the vacuum state.  Thus if we quotient the 
Toeplitz algebra by ${\mathit F}_1$, we will get an Abelian algebra generated by the operators satisfying
\beq
   a a^{\da} = 1, \quad \; \hbox{\rm and} \; a^{\da} a = 1.
\la{4.3.23}
\eeq
This is just the algebra of functions on a circle, and is consistent with the fact that the quotient algebra 
characterized by Eq.(\ref{4.3.21}) is Abelian.  Thus we can regard the Cuntz algebra as a non-commutative
multi-dimensional generalization of the algebra of functions on the circle.

The Lie algebra $\hatrightix$ spanned by the operators of the third kind is similar to $\hatleftix$.  Its Lie
bracket is
\begin{eqnarray}
   \lbrack r^{\dot{I}}_{\dot{J}}, r^{\dot{K}}_{\dot{L}} \rbrack & = & 
   \delta^{\dot{K}}_{\dot{J}} r^{\dot{I}}_{\dot{L}} +
   \sum_{J_1 \dot{J}_2 = \dot{J}} \delta^{\dot{K}}_{\dot{J}_2} l^{\dot{I}}_{J_1 \dot{L}} + 
   \sum_{K_1 \dot{K}_2 = \dot{K}} \delta^{\dot{K}_2}_{\dot{J}} l^{K_1 \dot{I}}_{\dot{L}} \nonumber \\
   & & - \delta^{\dot{I}}_{\dot{L}} l^{\dot{K}}_{\dot{J}} -
   \sum_{L_1 \dot{L}_2 = \dot{L}} \delta^{\dot{I}}_{\dot{L}_2} l^{\dot{K}}_{L_1 \dot{J}} - 
   \sum_{I_1 \dot{I}_2 = \dot{I}} \delta^{\dot{I}_2}_{\dot{L}} l^{I_1 \dot{K}}_{\dot{J}}.
\label{4.4.2}
\end{eqnarray}
A vector in $\salt$ can be expressed as an element in $\hatrightix$ as well:
\begin{equation}
   f^{\dot{K}}_{\dot{L}} = r^{\dot{K}}_{\dot{L}} - \sum_{i=1}^{\Lambda} r^{i\dot{K}}_{i\dot{L}}.
\label{4.4.3}
\end{equation}
Furthermore, $\salt$ is a proper ideal of $\hatrightix$, as is clear from the Lie bracket.
\begin{equation}
   \left[ r^{\dot{I}}_{\dot{J}}, f^{\dot{K}}_{\dot{L}} \right] = 
   \sum_{\dot{K}_1 \dot{K}_2 = \dot{K}} \delta^{\dot{K}_2}_{\dot{J}} f^{\dot{K}_1 \dot{I}}_{\dot{L}}
   - \sum_{\dot{L}_1 \dot{L}_2 = \dot{L}} \delta^{\dot{I}}_{\dot{L}_2} f^{\dot{K}}_{\dot{L}_1 \dot{J}}.
\label{4.4.4}
\end{equation}

All the operators of the first three kinds together form a bigger Lie algebra $\multix$.  This Lie algebra embraces 
all the ones discussed so far as subalgebras.  The additional Lie bracket needed is one between an operator of the 
second kind and that of the third kind, as shown:
\begin{equation}
   \left[ l^{\dot{I}}_{\dot{J}}, r^{\dot{K}}_{\dot{L}} \right] = 
   \sum_{\begin{array}{l}
   	    \dot{J}_1 \dot{J}_2 = \dot{J} \\
   	    \dot{K}_1 \dot{K}_2 = \dot{K}
   	 \end{array}} 
   \delta^{\dot{K}_1}_{\dot{J}_2} f^{\dot{I} \dot{K}_2}_{\dot{J}_1 \dot{L}} -
   \sum_{\begin{array}{l}
   	    \dot{I}_1 \dot{I}_2 = \dot{I} \\
   	    \dot{L}_1 \dot{L}_2 = \dot{L}
   	 \end{array}}
   \delta^{\dot{I}_2}_{\dot{L}_1} f^{\dot{I}_1 \dot{K}}_{\dot{J} \dot{L}_2}.
\label{4.4.5}
\end{equation}
Verifying that the action of the R.H.S. on an arbitrary open singlet state $s^{\dot{M}}$ gives 
$l^{\dot{I}}_{\dot{J}} (r^{\dot{K}}_{\dot{L}} s^{\dot{M}}) - r^{\dot{K}}_{\dot{L}} (l^{\dot{I}}_{\dot{J}} 
s^{\dot{M}})$ is not enough to validate this binary operation between these two operators as a Lie bracket because 
Eqs.(\ref{4.3.5}) and (\ref{4.4.3}) together imply that the set of all finite linear combinations of the operators 
in $\hatleftix$ and $\hatrightix$ is not linearly independent, which in turn implies that the Jacoby identity may 
not be satisfied.  We will properly justify Eq.(\ref{4.4.5}) in the Section~\ref{c4}, where we will see that we can 
treat $\multix$ as a subalgebra of yet another larger Lie algebra.  A typical term in the above equation is 
depicted in Fig.\ref{f4.3}.

\begin{figure}[ht]
\epsfxsize=4in
\centerline{\epsfbox{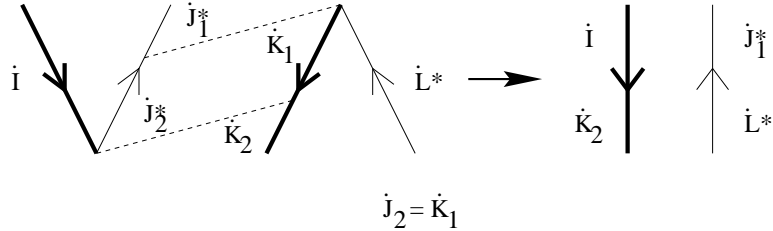}}
\caption{\em A diagrammatic representation of a Lie bracket of $\multix$, Eq.(\ref{4.4.5}).  Only the first term 
on the R.H.S. of this equation is shown.}
\label{f4.3}
\end{figure}

\begin{figure}
\epsfxsize=5in
\centerline{\epsfbox{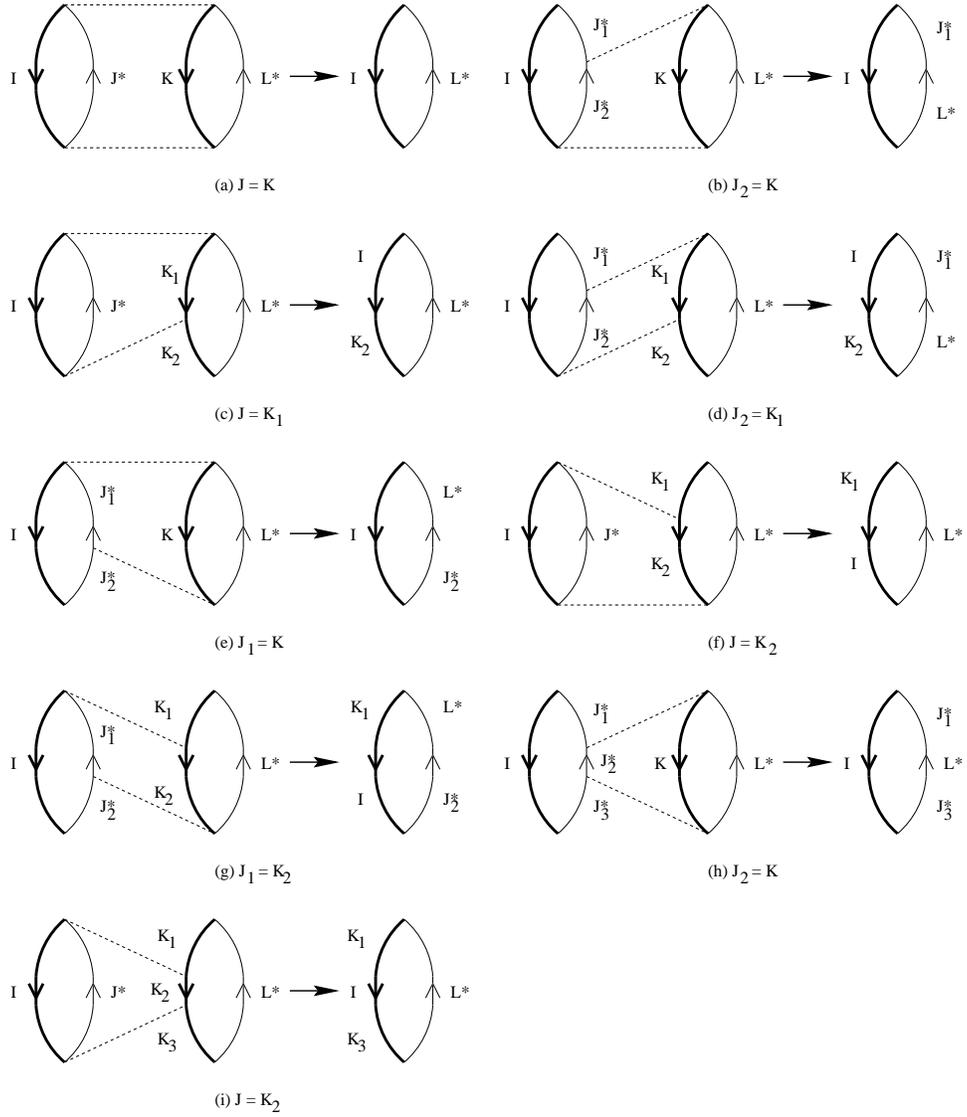}}
\caption{\em  Diagrammatic representations of the first nine terms on the R.H.S. of Eq.(\ref{4.5.4}).}
\label{f4.4}
\end{figure}

\subsection{A Lie Algebra for the Operators of the Fourth Kind}
\la{s4.5}

We now review the most non-trivial of all subalgebras of the grand open string algebra, the centrix algebra 
$\hatcentrix$.  If we retain only the adjoint matter of the action of an operator of the fourth kind on an open 
singlet state in Eq.(\ref{2.2.19}), we will get
\begin{equation}
   \s^I_J s^{\dot{K}} = \sum_{\dot{A}, \dot{B}} \delta^{\dot{K}}_{\dot{A} J \dot{B}} s^{\dot{A} I \dot{B}} 
\label{4.5.2}   
\end{equation}
(c.f. Fig.\ref{f2.5}(c)).  It is impossible to find a finite linear combination of this kind of operators whose 
action on an arbitrary singlet state is exactly the same as the composite operator $\s^I_J \s^K_L$.  (We will put 
off proving this assertion because we are going to prove a similar but more general statement in the next section 
in Appendix~\ref{sa3.2}.)  Nevertheless, a remarkable thing here is that the Lie bracket between two operators of 
the fourth kind can still be equated with a finite linear combination of this kind of operators by the requirement
\begin{equation}
   \left( \lbrack \s^I_J, \s^K_L\rbrack \right) s^{\dot{P}} \equiv \s^I_J \left( \s^K_L s^{\dot{P}} \right) -
   \s^K_L \left( \s^I_J s^{\dot{P}} \right)
\label{4.5.3}
\end{equation}
for any arbitrary sequence $\dot{P}$.  Then 
\begin{eqnarray}
   \lefteqn{ \left[ \s^I_J, \s^K_L \right] =
   \delta^K_J \s^I_L + \sum_{J_1 J_2 = J} \delta^K_{J_2} 
   \s^I_{J_1 L} + \sum_{K_1 K_2 = K} \delta^{K_1}_J \s^{I K_2}_L } \nonumber \\
   & & + \sum_{\begin{array}{l}
		  J_1 J_2 = J \\
		  K_1 K_2 = K
	       \end{array}}
   \delta^{K_1}_{J_2} \s^{I K_2}_{J_1 L} 
   + \sum_{J_1 J_2 = J} \delta^K_{J_1} \s^I_{L J_2}
   + \sum_{K_1 K_2 = J} \delta^{K_2}_J \s^{K_1 I}_L \nonumber \\
   & & + \sum_{\begin{array}{l}
		  J_1 J_2 = J \\
		  K_1 K_2 = K
	       \end{array}}
   \delta^{K_2}_{J_1} \s^{K_1 I}_{L J_2}
   + \sum_{J_1 J_2 J_3 = J} \delta^K_{J_2} \s^I_{J_1 L J_3} 
   + \sum_{K_1 K_2 K_3 = K} \delta^{K_2}_J \s^{K_1 I K_3}_L \nonumber \\
   & & - (I \leftrightarrow K, J \leftrightarrow L). 
\label{4.5.4}
\end{eqnarray}
The proof of this equation is given in detail in Appendix~\ref{sa4.4} as a demonstration of how we carry out 
computations involving multi-indices.  Fig.\ref{f4.4} gives diagrammatic representations of the first nine terms.
We will call the Lie algebra defined by Eq.(\ref{4.5.4}) the {\em centrix algebra} $\hatcentrix$.

Let us explore the structure of $\hatcentrix$.  All vectors of the form $\s^I_I$, where $I$ is an arbitrary finite 
integer sequence of integers between $1$ and $\L$ inclusive, span a Cartan subalgebra $\hat{\Sigma}^0_{{\Lambda}}$ 
of the centrix algebra.  The proof is pretty much the same as the one shown in Appendix~\ref{sa4.2}.   

\begin{figure}[ht]
\epsfxsize=5in
\centerline{\epsfbox{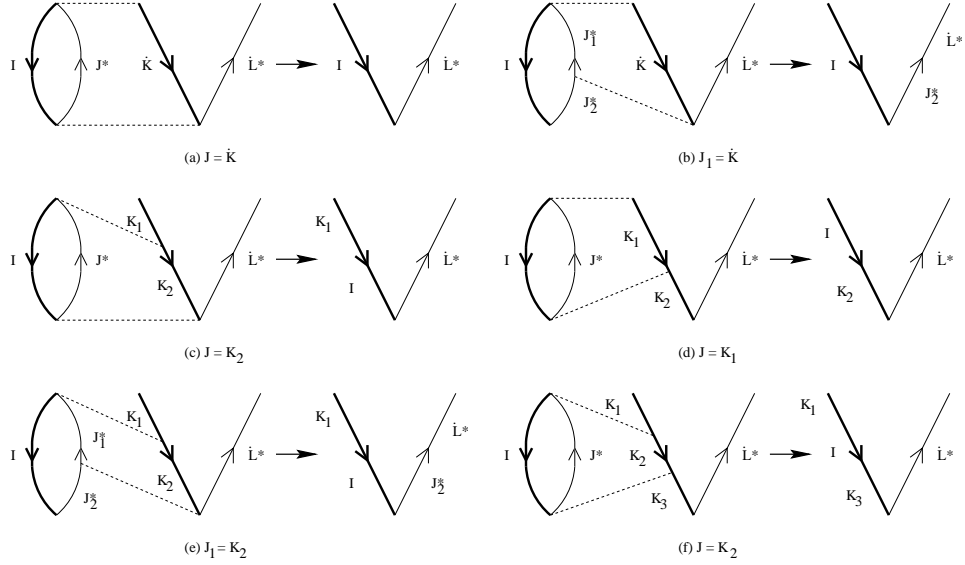}}
\caption{\em  Diagrammatic representations of Eq.(\ref{4.5.10}).  Only the first six terms on the R.H.S. of this 
equation are shown here.}
\label{f4.5}
\end{figure}

Next, the reader can easily verify the following actions, which we have mentioned in Ref.\cite{opstal}:
\beq
   \left( \s^I_J - \sum_{i=1}^{\Lambda} \s^{iI}_{iJ} \right) s^{\dot{K}} 
   = \sum_{K_1 \dot{K}_2 = \dot{K}} \delta^{K_1}_J s^{I\dot{K}_2}
\la{4.5.5}
\eeq
and
\beq
   \left( \s^I_J - \sum_{j=1}^{\Lambda} \s^{Ij}_{Jj} \right) s^{\dot{K}} 
   = \sum_{\dot{K}_1 K_2 = \dot{K}} \delta^{K_2}_J s^{\dot{K}_1 I}.
\la{4.5.6}
\eeq
These are exactly the action of the opeartors $l^I_J$ and $r^I_J$. Thus,
\begin{eqnarray}
   l^I_J & = & \s^I_J - \sum_{i=1}^{{\Lambda}} \s^{iI}_{iJ}\mbox{; and}
\label{4.5.7} \\
   r^I_J & = & \s^I_J - \sum_{j=1}^{{\Lambda}} \s^{Ij}_{Jj}.
\label{4.5.8}
\end{eqnarray}
We remind the reader that unlike Eqs.(\ref{4.3.5}) and (\ref{4.4.3}), here $I$ and $J$ must be {\em non-empty}.  
Combining Eqs.(\ref{4.5.7}) and (\ref{4.3.5}), or Eqs.(\ref{4.5.8}) and (\ref{4.4.3}), we obtain
\beq
  f^I_J = \sigma^I_J - \sum_{j=1}^{\Lambda} \sigma^{Ij}_{Jj} - \sum_{i=1}^{\Lambda} \sigma^{iI}_{iJ}
  + \sum_{i,j=1}^{\Lambda} \sigma^{iIj}_{iJj}.
\label{4.5.9}
\eeq
(We are going to use a generalized version of these relations in a later section.)  Let $\salt'$, $\hatleftix'$ and 
$\hatrightix'$ be vector spaces spanned by $f^I_J$'s, $l^I_J$'s and $r^I_J$'s, respectively, and let $\multix'$ be 
the sum of all these vector spaces.  All four vectors spaces are proper ideals of $\hatcentrix$.

\begin{figure}
\epsfxsize=5in
\centerline{\epsfbox{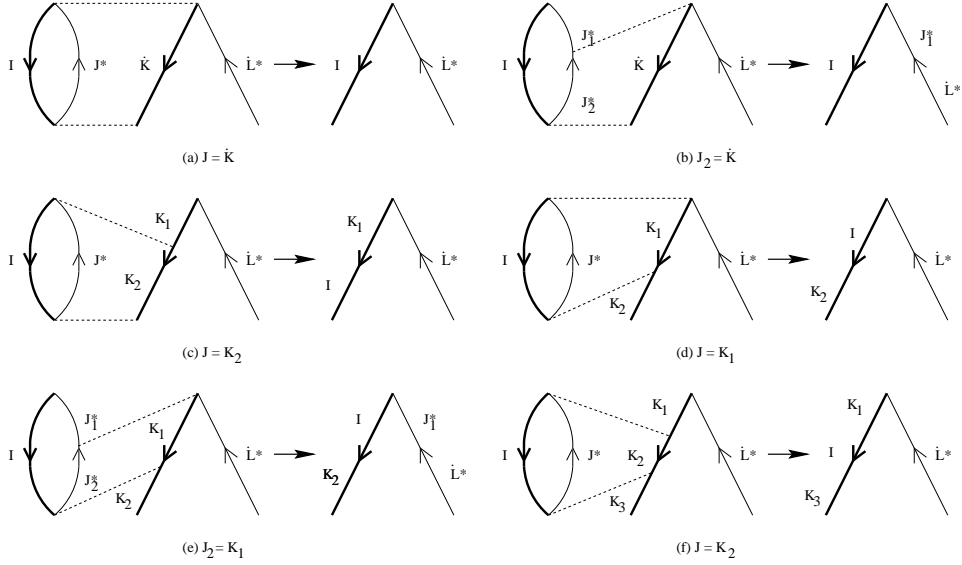}}
\caption{\em  Diagrammatic representations of Eq.(\ref{4.5.11}).  Only the first six terms on the R.H.S. of this 
equation are shown here.}
\label{f4.6}
\end{figure}

For the sake of completeness, we review the Lie bracket relations between an operator of the fourth kind and an
operator of another kind:
\begin{eqnarray}
   \lefteqn{\left[ \s^I_J, l^{\dot{K}}_{\dot{L}} \right] = \delta^{\dot{K}}_J l^I_{\dot{L}} + 
   \sum_{J_1 J_2 = J} \delta^{\dot{K}}_{J_1} l^I_{\dot{L} J_2} +
   \sum_{K_1 K_2 = \dot{K}} \delta^{K_2}_J l^{K_1 I}_{\dot{L}} } \nonumber \\
   & & + \sum_{K_1 K_2 = \dot{K}} \delta^{K_1}_J l^{I K_2}_{\dot{L}} +
   \sum_{\begin{array}{l}
	    J_1 J_2 = J \\
	    K_1 K_2 = \dot{K}
	 \end{array}}
   \delta^{K_2}_{J_1} l^{K_1 I}_{\dot{L} J_2}
   + \sum_{K_1 K_2 K_3 = \dot{K}} \delta^{K_2}_J l^{K_1 I K_3}_{\dot{L}} \nonumber \\
   & & - \delta^I_{\dot{L}} l^{\dot{K}}_J -
   \sum_{I_1 I_2 = I} \delta^{I_1}_{\dot{L}} l^{\dot{K} I_2}_J -
   \sum_{L_1 L_2 = \dot{L}} \delta^I_{L_2} l^{\dot{K}}_{L_1 J} \nonumber \\
   & & - \sum_{L_1 L_2 = \dot{L}} \delta^I_{L_1} l^{\dot{K}}_{J L_2} -
   \sum_{\begin{array}{l}
   	    L_1 L_2 = \dot{L} \\
   	    I_1 I_2 = I
   	 \end{array}}
   \delta^{I_1}_{L_2} l^{\dot{K} I_2}_{L_1 J} 
   - \sum_{L_1 L_2 L_3 = \dot{L}} \delta^I_{L_2} l^{\dot{K}}_{L_1 J L_3};
\label{4.5.10} \\
   \lefteqn{\left[ \s^I_J, r^{\dot{K}}_{\dot{L}} \right] = \delta^{\dot{K}}_J r^I_{\dot{L}} + 
   \sum_{J_1 J_2 = J} \delta^{\dot{K}}_{J_2} r^I_{J_1 \dot{L}} +
   \sum_{K_1 K_2 = \dot{K}} \delta^{K_2}_J r^{K_1 I}_{\dot{L}} } \nonumber \\
   & & + \sum_{K_1 K_2 = \dot{K}} \delta^{K_1}_J r^{I K_2}_{\dot{L}} +
   \sum_{\begin{array}{l}
	    J_1 J_2 = J \\
	    K_1 K_2 = \dot{K}
	 \end{array}}
   \delta^{K_1}_{J_2} r^{I K_2}_{J_1 \dot{L}}
   + \sum_{K_1 K_2 K_3 = \dot{K}} \delta^{K_2}_J r^{K_1 I K_3}_{\dot{L}} \nonumber \\
   & & - \delta^I_{\dot{L}} r^{\dot{K}}_J -
   \sum_{I_1 I_2 = I} \delta^{I_2}_{\dot{L}} r^{I_1 \dot{K}}_J -
   \sum_{L_1 L_2 = \dot{L}} \delta^I_{L_2} r^{\dot{K}}_{L_1 J} \nonumber \\
   & & - \sum_{L_1 L_2 = \dot{L}} \delta^I_{L_1} r^{\dot{K}}_{J L_2} -
   \sum_{\begin{array}{l}
   	    L_1 L_2 = \dot{L} \\
   	    I_1 I_2 = I
   	 \end{array}}
   \delta^{I_2}_{L_1} r^{I_1 \dot{K}}_{J L_2} 
   - \sum_{L_1 L_2 L_3 = \dot{L}} \delta^I_{L_2} r^{\dot{K}}_{L_1 J L_3} \mbox{; and}
\label{4.5.11} \\
   \lefteqn{\left[ \s^I_J, f^{\dot{K}}_{\dot{L}} \right] = 
   \sum_{\dot{K}_1 K_2 \dot{K}_3 = \dot{K}} \delta^{K_2}_J f^{\dot{K}_1 I \dot{K}_3}_{\dot{L}}
   - \sum_{\dot{L}_1 L_2 \dot{L}_3 = \dot{L}} \delta^I_{L_2} f^{\dot{K}}_{\dot{L}_1 J \dot{L}_3}. }
\label{4.5.12}
\end{eqnarray}
Eqs.(\ref{4.5.10}), (\ref{4.5.11}) and (\ref{4.5.12}) are illustrated in Figs.\ref{f4.5}, \ref{f4.6} and 
\ref{f4.7}, respectively.  It follows from these three equations that the quotient algebra $\vectrix \equiv 
\hatcentrix / \hatmultix'$ is the Lie algebra of a set of outer derivations of the Cuntz algebra.  As we have shown
in Ref.\cite{opstal}, $\vectrix$ simplifies precisely to the Witt algebra in the special case $\L = 1$.  Thus, the 
Lie algebra of this set of outer derivations of the Cuntz algebra is a generalization of the algebra of vector 
fields on the unit circle.

\begin{figure}
\epsfxsize=2.5in
\centerline{\epsfbox{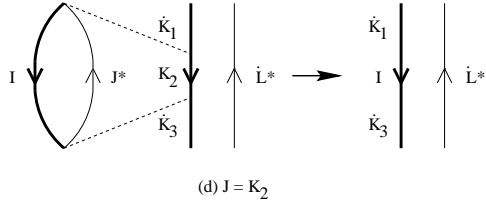}}
\caption{\em  A diagrammatic representation of Eq.(\ref{4.5.12}).  Only the first term on the R.H.S. of this 
equation is shown here.}
\label{f4.7}
\end{figure}

Let us make some remarks on root vectors.  We obtain from Eq.(\ref{4.5.12}) that 
\begin{equation}
   \left[ \s^I_I, f^K_L \right] = \left( \sum_{\dot{K}_1 K_2 \dot{K}_3 = K} \delta^{K_2}_I 
   - \sum_{\dot{L}_1 L_2 \dot{L}_3 = L} \delta^I_{L_2} \right) f^K_L	 
\label{4.5.13}
\end{equation}
As a result, every $f^K_L$ is a root vector.  Moreover, there are no root vectors other than $f^K_L$'s, and a
proof of this statement will be given in Appendix~\ref{sa4.5}.  Hence every root space is one-dimensional.   

There is a way of splitting $\hatcentrix$ into `raising' operators, `diagonal' operators and `lowering' operators.
Let $\hat{\Sigma}_{{\Lambda}}^+$ be the vector space spanned by all elements of the form $\s^I_J$ such that $I > J$ 
(See the footnote in the previous subsection).  Then it can be proved by checking term by term on the R.H.S. of 
Eq.(\ref{4.5.4}) that $\hat{\Sigma}_{{\Lambda}}^+$ is indeed a subalgebra of the centrix algebra\footnote{Take the 
fourth term as an example.  If both $\s^I_J$ and $\s^K_L \in \hat{\Sigma}_{{\Lambda}}^+$, then $IK_2 > JK_2 = 
J_1 J_2 K_2 = J_1 K_1 K_2 = J_1 K > J_1 L$.  Hence $\s^{I K_2}_{J_1 L} \in \hat{\Sigma}_{{\Lambda}}^+$ also.}.  
Likewise, let $\hat{\Sigma}_{{\Lambda}}^-$ be the vector space spanned by all elements of the form $\s^I_J$ such 
that $I < J$.  Then it also follows from Eq.(\ref{4.5.4}) that $\hat{\Sigma}_{{\Lambda}}^-$ is also a subalgebra.  
Moreover, we have
\[ \hat{\Sigma}_{{\Lambda}} = \hat{\Sigma}^+_{{\Lambda}} \oplus 
\hat{\Sigma}^0_{{\Lambda}} \oplus \hat{\Sigma}^-_{{\Lambda}}. \] 

\section{Open String and Closed String Algebras}
\la{c3}

In the previous section, we studied a Lie algebra for bosonic open strings, and gained some expertise in how to
study its structure.  We will use this skill to fulfill our promise in the preceding section, which is to study the 
full Lie algebra for open and closed strings, or in other words, the Lie algebra for the five kinds of operators 
discussed in Section~\ref{s2.2}.  

To achieve this purpose, we need to define a precursor Lie algebra which we will call the {\em heterix algebra} 
among some physical operators.  These physical operators are nothing but the ones defined in Subsection~\ref{s2.2}, 
except that the ranges of values of the quantum states other than color (see the second paragraph in 
Subection~\ref{s2.2}) are changed.  We will derive this algebra in Subsection~\ref{s3.2}.  Then the grand string 
algebra will be seen as a subalgebra of the heterix algebra.  Readers who are only interested in the definition of 
this algebra may skip this section. occasionally returning to it to look up the relevant definitions.  In 
Subsection~\ref{s5.2}, we will give the definition of a algebra of operators acting on open and closed singlet 
states.  We will call this the `grand string algebra'.  The difference between this Lie algebra and the precursor 
algebra is that there are more than one degree of freedom at the ends of the open singlet states.  At first glance, 
this algebra is somewhat `larger' than the precursor algebra.  Ironically, we will derive this grand string algebra 
as a {\em subalgebra} of the precursor algebra by manipulating the numbers of degrees of freedom in an accompanying 
appendix.

In Subection~\ref{s5.3}, we will derive a Lie algebra just for open strings.  This will be a quotient algebra of 
the grand string algebra.  In Section~\ref{s5.4}, we will derive a Lie algebra for closed strings as another 
quotient algebra of the grand string algebra.  We will see that the corresponding Lie algebra, which we will call 
the `cyclix algebra', for bosonic closed strings has a closed relationship with the Witt algebra, too.

To have a glimpse of how to use these algebras in the study of physical systems, we will turn to the Ising model
again in Section~\ref{s5.5}.  It is well known that there are many ways of solving the quantum Ising model in one
dimension.  One method which is close to the spirit of the original way Onsager himself solved the model 
\cite{onsager} is via the `Onsager algebra' \cite{davies90}.  We will see that actually this Onsager algebra is
a subalgebra of the cyclix algebra, and we can use the cyclix algebra directly to obtain some conserved quantities
of the Ising matrix model.  This example may give us a clue of how to use these Lie algebras more effectively in the
future.

\subsection{Derivation of a Precursor Algebra}
\la{s3.2}

We are going to derive a precursor algebra.  The grand string algebra will be identified as a subalgebra of this 
algebra in the next section.

Let $\a^{\mu}_{\n}(k)$ be an annihilation operator for a boson in the adjoint representation for $1 \leq k \leq 
\L + 2 \L_F$, and let $\bar{\c}_{\mu}$ and $\c^{\mu}$ be annihilation operators for an antifermion in the conjugate 
representation and a fermion in the fundamental representation, respectively.  Moreover, let $\a^{\da\mu}_{\n}(k)$, 
$\bar{\c}^{\da\mu}$ and $\c^{\da}_{\mu}$ be the corresponding creation operators.  The annihilation and creation 
operators satisfy the usual canonical (anti)-commutation relations, the non-vanishing ones being
\beq
   \le[ \a^{\mu_1}_{\mu_2}(k_1), \a^{\da\mu_3}_{\mu_4}(k_2) \ri] & = &  
   \d_{k_1 k_2} \d^{\mu_3}_{\mu_2} \d^{\mu_1}_{\mu_4}, 
\la{3.2.1}
\eeq
\beq
   \le[ \bar{\c}_{\mu_1}, \bar{\c}^{\da\mu_2} \ri]_+ = \d^{\mu_2}_{\mu_1}, 
\la{3.2.3}
\eeq
and
\beq
   \le[ \c^{\mu_1}, \c^{\da}_{\mu_2} \ri]_+ = \d^{\mu_1}_{\mu_2}.
\la{3.2.4}
\eeq

Again we introduce two families of color-invariant singlet states.  A typical open singlet state is a linear 
combination of the states of the form
\beq
   s'^K \equiv N^{-(c+1)/2} \bar{\c}^{\da\u_1} \a^{\da\u_2}_{\u_1}(k_1) \a^{\da\u_3}_{\u_2}(k_2) \cd 
   \a^{\da\u_{c+1}}_{\u_c}(k_c) \c^{\da}_{\u_{c+1}} |0 \rangle.
\la{3.2.5}
\eeq
We denote by ${\cal T}'_o$ the Hilbert space of all these open singlet states.  A typical closed singlet state is a 
linear combination of the states of the form
\beq
   \Ps'^K \equiv N^{-c/2} \a^{\da\u_2}_{\u_1}(k_1) \a^{\da\u_3}_{\u_2}(k_2) \cd \a^{\da\u_1}_{\u_c}(k_c) 
   |0 \rangle.
\la{3.2.6}
\eeq
The Hilbert space of all closed singlet states will be denoted by ${\cal T}'_c$.

We need only two families of color-invariant operators acting on ${\cal T}'_o$ and ${\cal T}'_c$ to establish our
main results.  One family consists of operators of the form 
\beq 
   \ti{f}'^I_J & \equiv & N^{-(a+b)/2} \a^{\da\mu_2}_{\mu_1} (i_1) \a^{\da\mu_3}_{\mu_2} (i_2) \cd 
   \a^{\da\mu_1}_{\mu_a} (i_a, \ep(i_a)) \nn \\
   & & \a^{\n_b}_{\n_1} (j_b) \a^{\n_{b-1}}_{\n_b} (j_{b-1}) \cd \a^{\n_1}_{\n_2} (j_1).
\la{3.2.7}
\eeq
In the large-$N$ limit, the actions of this operator on singlet states read (c.f. Ref.\cite{thorn79} or 
Appendix~\ref{sa1.3} of this article)
\beq
   \ti{f}'^I_J s'^K & = & 0 \mbox{; and}
\la{3.2.8} \\
   \ti{f}'^I_J \Ps'^K & = & \sum_{\dot{J}_1 J_2 = J} \d^K_{J_2 \dot{J}_1} \Ps'^I,
\la{3.2.9}
\eeq
It is clear that the sum on the right hand side of Eq.(\ref{3.2.9}) is a finite one.  The other family consists of 
operators of the form
\beq
   \g'^I_J & \equiv & N^{-(a+b-2)/2} \a^{\da\mu_2}_{\mu_1} (i_1) \a^{\da\mu_3}_{\mu_2} (i_2) \cd 
   \a^{\da\n_b}_{\mu_a} (i_a) \nn \\
   & & \a^{\n_{b-1}}_{\n_b} (j_b) \a^{\n_{b-2}}_{\n_{b-1}} (j_{b-1}) \cd \a^{\mu_1}_{\n_1} (j_1).
\la{3.2.10}
\eeq
In the large-$N$ limit, this operator propagates singlet states in the following manner:
\beq
   \lefteqn{\g'^I_J s'^K \equiv \sum_{\dot{K}_1 K_2 \dot{K}_3 = K} \d^{K_2}_J s'^{\dot{K_1} I \dot{K_3}} 
   \mbox{; and} } 
\la{3.2.11} \\
   \lefteqn{\g'^I_J \Ps'^K \equiv \d^K_J \Ps'^I + \sum_{K_1 K_2 = K} \d^{K_2 K_1}_J \Ps'^I
   + \sum_{K_1 K_2 = K} \d^{K_1}_J \Ps'^{I K_2} } \nn \\
   & & \sum_{K_1 K_2 K_3 = K} \d^{K_2}_J \Ps'^{I K_3 K_1} + \sum_{K_1 K_2 = K} \d^{K_2}_J \Ps'^{I K_1} \nn \\
   & & \sum_{J_1 J_2 = J} \sum_{K_1 K_2 K_3 = K} \d^{K_3}_{J_1} \d^{K_1}_{J_2} \Ps'^{I K_2}.
\la{3.2.12}
\eeq
(c.f. Eqs.(2) and (3) in Ref.\cite{clstal}).  The set of all $\ti{f}'^I_J$'s and $\g'^I_J$'s is linearly 
independent.  This was proved in Appendix~A of Ref.\cite{clstal}.  (Notice that the open singlet states are needed 
for the linear independence.  If we simply considered their actions on closed singlet states only, the set would not
be linear independent.)

\begin{figure}
\epsfxsize=5in
\centerline{\epsfbox{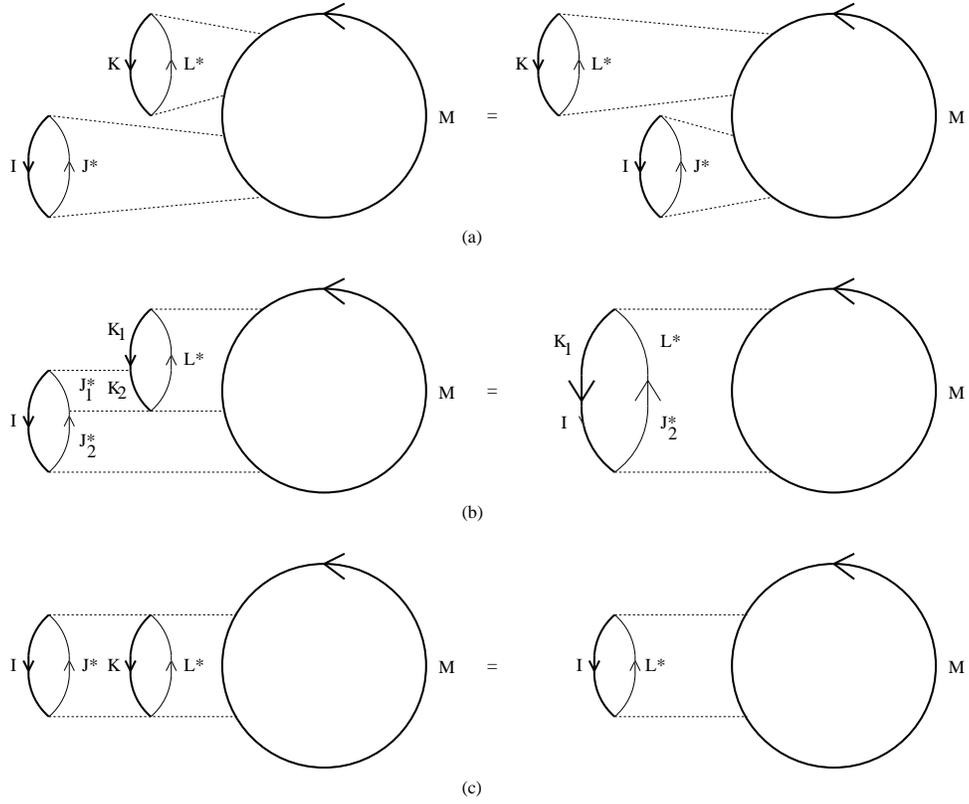}}
\caption{\em Possible terms in $\g'^I_J (\g'^K_L \Ps'^M)$.  
(a) This is the case when $\g'^I_J$ and $\g'^K_L$ act on disjoint segments in $\Ps'^M$.  This term is cancelled in 
the Lie superbracket.  (b) Partial overlap between $J$ and $K$.  This is the case when $J_1 = K_2$.  This term is 
equivalent to the corresponding one in $\g'^{K_1 I}_{L J_2} \Ps'^M$.  (c) Complete overlap between $J$ and $K$.  
This time $J = K$.  This term is equivalent to a term in $\g'^I_L \Ps'^M$.}
\la{f3.2.1}
\end{figure}

\begin{figure}[ht]
\epsfxsize=5in
\centerline{\epsfbox{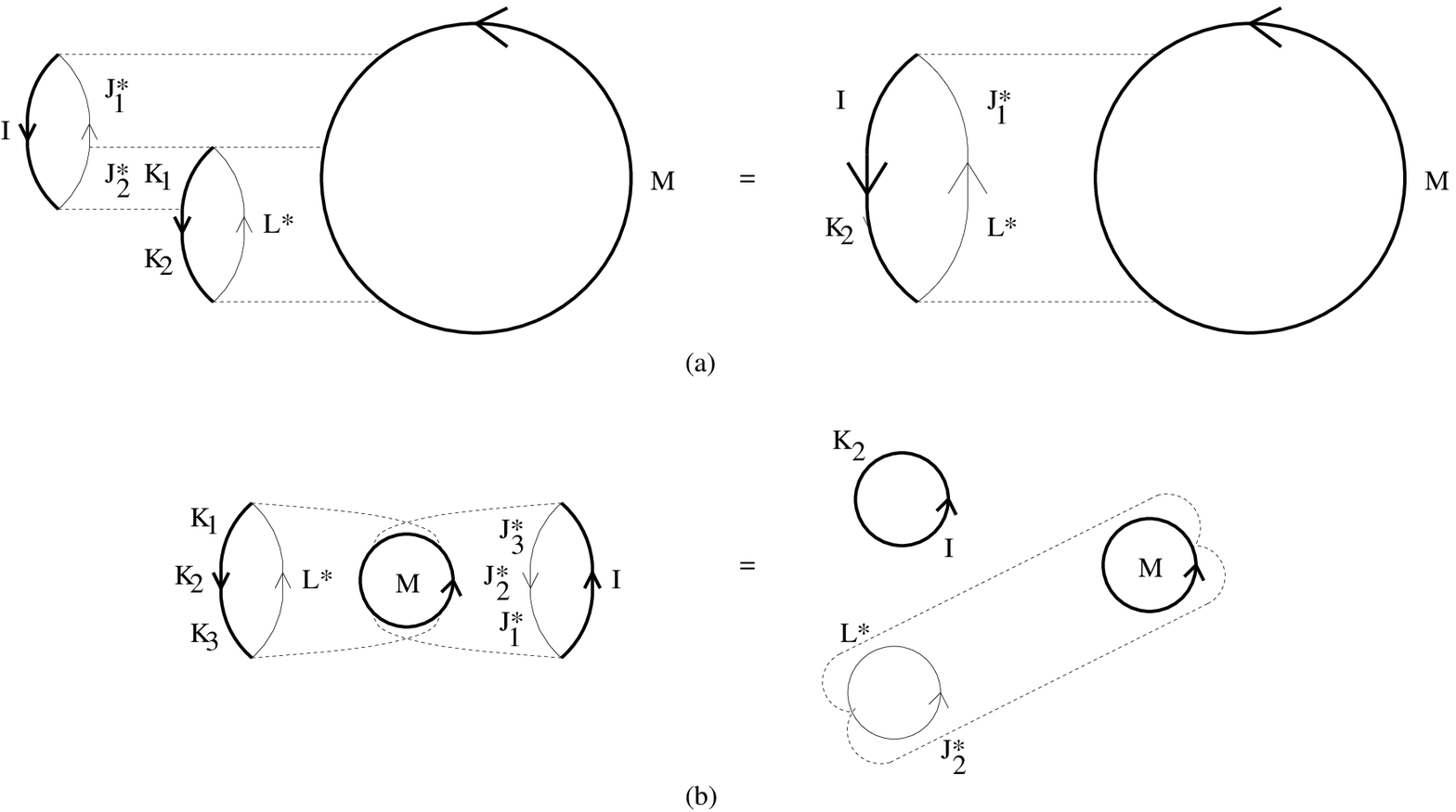}}
\caption{\em Other possible terms in $\g'^I_J (\g'^K_L \Ps'^M)$  (a) Another partial overlap between $J$ and $K$.
This time $J_2 = K_1$.  This term is equivalent to a term in $\g'^{I K_2}_{J_1 L} \Ps'^M$.  (b) This partial overlap
corresponds to the case when $J_1 = K_3$ and $J_3 = K_1$.  Note that we have turned $\g'^I_J$ by 180 degrees.  This 
term is equivalent to $\ti{f}'^{I K_2}_{J_2 L} \Ps'^M$.}
\la{f3.2.2}
\end{figure}

The next thing we are going to do is to construct a precursor Lie algebra out of the two families of 
color-invariant operators.  If the product of any two of these operators were well defined, i.e., if we could write
down the product as a finite linear combination of color-invariant operators, the easiest way to obtain a Lie
algebra would certainly be to define the Lie bracket of two operators as a sum or difference of their products in 
different orders of the operators.  However, such a product is actually not well defined; this is shown in 
Appendix~\ref{sa3.2}.  Now consider the commutator of two operators.  {\em It can be shown that the actions of 
these commutators on singlet states are identical to the actions of some observables}.  If we define the commutator 
to be the color-invariant operator whose action on ${\cal T}'_o \oplus {\cal T}'_c$ is identical to it, then we have
\begin{eqnarray}
   \lefteqn{ \left[ \g'^I_J, \g'^K_L \right] = 
   \delta^K_J \g'^I_L + \sum_{J_1 J_2 = J} \delta^K_{J_2} \g'^I_{J_1 L} 
   + \sum_{K_1 K_2 = K} \delta^{K_1}_J \g'^{I K_2}_L } \nonumber \\
   & & + \sum_{\begin{array}{l}
		  J_1 J_2 = J \\
		  K_1 K_2 = K
	       \end{array}}
   \delta^{K_1}_{J_2} \g'^{I K_2}_{J_1 L} 
   + \sum_{J_1 J_2 = J} \delta^K_{J_1} \g'^I_{L J_2}
   + \sum_{K_1 K_2 = J} \delta^{K_2}_J \g'^{K_1 I}_L \nonumber \\
   & & + \sum_{\begin{array}{l}
		  J_1 J_2 = J \\
		  K_1 K_2 = K
	       \end{array}}
   \delta^{K_2}_{J_1} \g'^{K_1 I}_{L J_2}
   + \sum_{J_1 J_2 J_3 = J} \delta^K_{J_2} \g'^I_{J_1 L J_3} 
   + \sum_{K_1 K_2 K_3 = K} \delta^{K_2}_J \g'^{K_1 I K_3}_L \nonumber \\
   & & + \sum_{\begin{array}{l}
   		  J_1 J_2 = J \\
   		  K_1 K_2 = K
   	       \end{array}}
   \delta^{K_1}_{J_2} \delta^{K_2}_{J_1} \ti{f}'^I_L
   + \sum_{\begin{array}{l}
   	      J_1 J_2 J_3 = J \\
   	      K_1 K_2 = K
   	   \end{array}}
   \delta^{K_1}_{J_3} \delta^{K_2}_{J_1} \ti{f}'^I_{J_2 L}
   \nonumber \\
   & & + \sum_{\begin{array}{l}
   		  J_1 J_2 = J \\
   		  K_1 K_2 K_3 = K
   	       \end{array}}
   \delta^{K_1}_{J_2} \delta^{K_3}_{J_1} \ti{f}'^{I K_2}_L \nonumber \\
   & & + \sum_{\begin{array}{l}
		  J_1 J_2 J_3 = J \\
		  K_1 K_2 K_3 = K
	       \end{array}}
   \delta^{K_1}_{J_3} \delta^{K_3}_{J_1} \ti{f}'^{I K_2}_{J_2 L} 
    - (I \leftrightarrow K, J \leftrightarrow L), 
\label{3.2.13}
\end{eqnarray}
\begin{equation}
   \left[ \g'^I_J, \ti{f}'^K_L \right] = 
   \delta^K_{(J)} \ti{f}'^I_L + \sum_{K_1 K_2 = (K)} \delta^{K_1}_J \ti{f}'^{I K_2}_L - \delta^I_{(L)} \ti{f}'^K_J -
   \sum_{L_1 L_2 = (L)} \delta^I_{L_2} \ti{f}^K_{L_1 J}  
\label{3.2.14}
\end{equation}
and
\begin{equation} 
   \left[ \ti{f}'^I_J, \ti{f}'^K_L \right] = 
   \delta^K_{(J)} \ti{f}'^I_L - \delta^I_{(L)} \ti{f}'^K_J
\label{3.2.15}
\end{equation}

That these equations hold true was demonstrated in Appendix~D of Ref.\cite{clstal}.  Unfortunately, the rigorous 
proof is not illuminating.  To enlighten the reader, we would like to discuss the following special cases in 
Eq.(\ref{3.2.13}) to see why it makes sense for the right hand sides of these equations to appear in the above 
manner.

\begin{figure}
\epsfxsize=3.3in
\epsfysize=6.7in
\centerline{\epsfbox{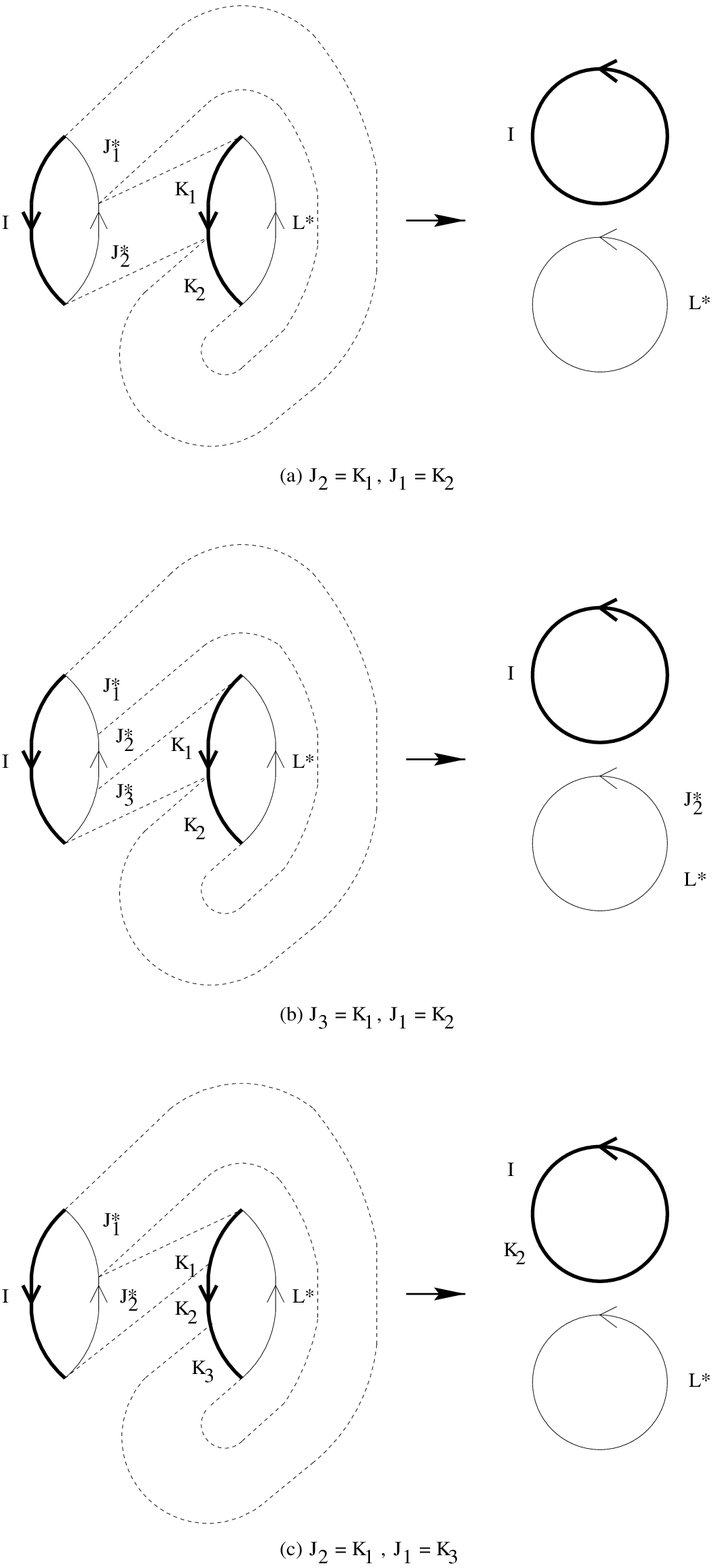}}
\caption{\em Diagrammatic representations of the tenth to twelfth terms on the R.H.S. of Eq.(\ref{3.2.13}).}
\label{f3.3.1}
\end{figure}

\begin{figure}
\epsfxsize=3.5in
\centerline{\epsfbox{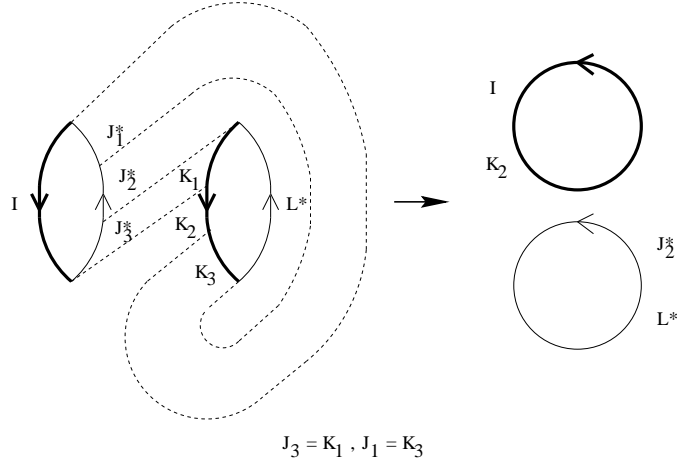}}
\caption{\em Diagrammatic representations of the thirteenth term on the R.H.S. of Eq.(\ref{3.2.13}).}
\label{f3.3.2}
\end{figure}

\begin{figure}
\epsfxsize=4in
\epsfysize=6.5in
\centerline{\epsfbox{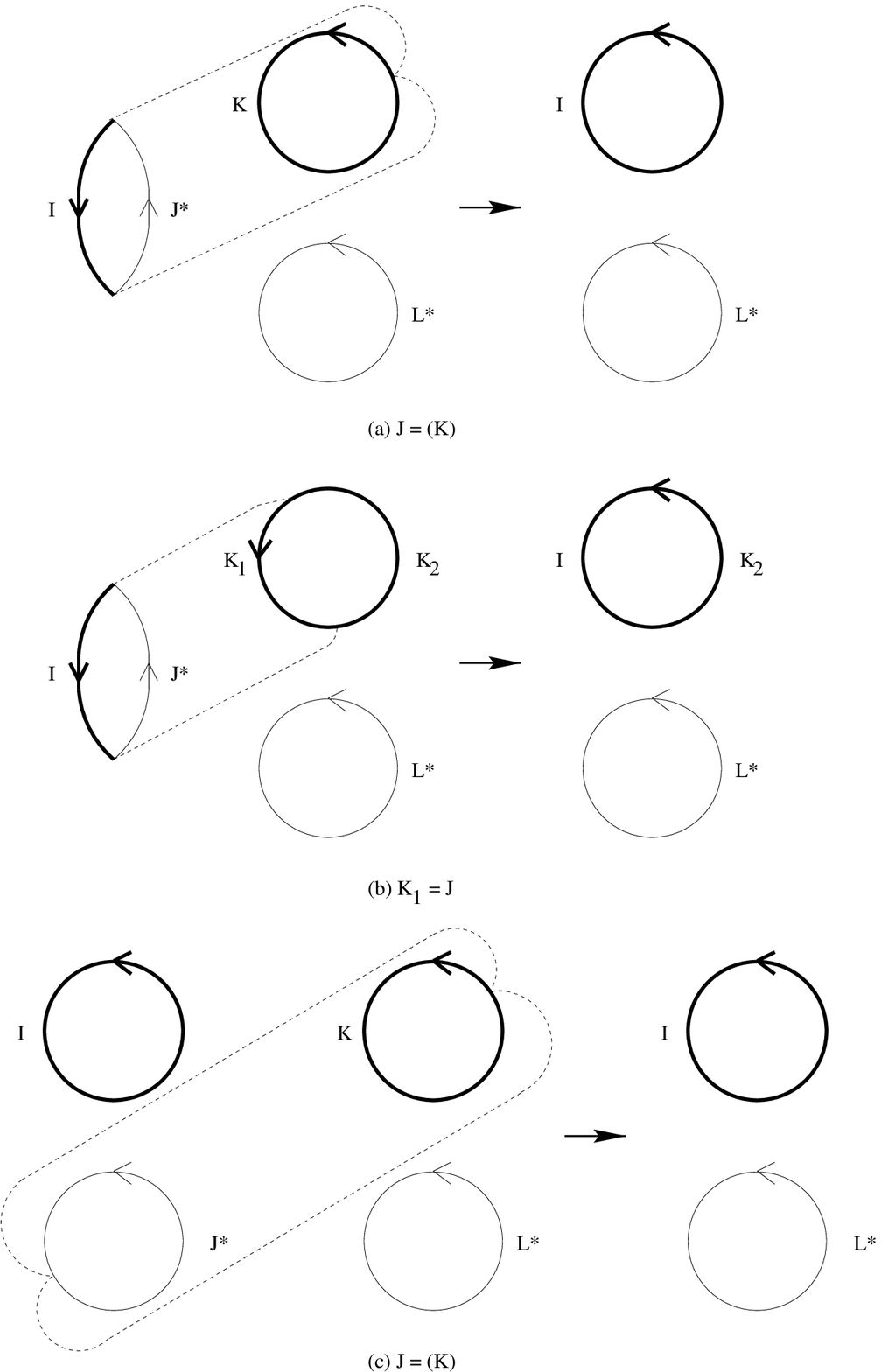}}
\caption{\em Diagrammatic representations of the first two terms on the R.H.S. of Eq.(\ref{3.2.14}), and the first
term on the R.H.S. of Eq.(\ref{3.2.15}).}
\label{f3.3.3}
\end{figure}

Consider $\g'^I_J (\g'^K_L \Ps'^M)$, and consider a term $t_1$ in $\g'^I_J (\g'^K_L \Ps'^M)$ produced by the 
operations of $\g'^I_J$ and $\g'^K_L$ on disjoint sequences of creation operators in $\Ps'^M$ 
(Fig.~\ref{f3.2.1}(a)), i.e., no annihilation operator in $\g'^I_J$ acts on any creation operator in $\g'^K_L$.  
This term is identical to the term $t_2$ in $\g'^K_L (\g'^I_J \Ps'^M)$ produced by the operations of $\g'^I_J$ and 
$\g'^K_L$ on the same disjoint sequences in $\Ps'^M$.  Since during the process of producing $t_1$ and $t_2$, there 
is no contraction at all between any operator in $\g'^I_J$ and any in $\g'^K_L$ (any contraction between $\g'^I_J$ 
and $\g'^K_L$ will produce terms which we are not considering right now), $t_2$ can be obtained from 
$\g'^I_J (\g'^K_L \Ps'^M)$ as well by interchanging $\g'^I_J$ and $\g'^K_L$ first before these two color-invariant 
operators act on $\Ps'^M$.  As a result, $t_1$ is cancelled by $t_2$ in the commutator.

Therefore, during the process of producing terms that are not killed by the commutator, there must be
contraction(s) among some operators in $\g'^I_J$ and some in $\g'^K_L$.  If we perform such contractions first
before contracting the annihilation operators in $\g'^I_J$ and $\g'^K_L$ with the creation operators in $\Ps'^M$,
in general we will obtain a color-invariant operator of the form ${\rm Tr} ( \a^{\da} \cd \a^{\da} \a \cd \a 
\a^{\da} \cd \a^{\da} \a \cd \a \cd) \cd {\rm Tr} ( \a^{\da} \cd \a^{\da} \a \cd \a \a^{\da} \cd)$ multiplied by a 
factor of $N$ raised to some power.  However, Figs.~\ref{f3.2.1}(b), (c) and Fig.\ref{f3.2.2} clearly reveals that 
only color-invariant operators of the forms ${\rm Tr} (\a^{\da} \cd \a^{\da} \a \cd \a)$ and ${\rm Tr} (\a^{\da} 
\cd \a^{\da}) {\rm Tr} (\a \cd \a)$ survive the large-$N$ limit.  All possibilities of producing color-invariant 
operators in these two forms are in Eqs.(\ref{3.2.13}) to (\ref{3.2.15}).  

The {\em heterix algebra} $\hatheterix$ is the Lie algebra defined by the Lie brackets in Eqs.(\ref{3.2.13}) to 
(\ref{3.2.15}).  Figs.\ref{f3.3.1}, \ref{f3.3.2} and \ref{f3.3.3} shows the diagrammatic representations of these
three equations. 

The reader can take a look at Ref.\cite{clstal} to learn the structure of this Lie algebra.

\subsection{Definition of the Grand String Algebra}
\la{s5.2}

We are going to give a precise definition of a Lie algebra of the operators defined in Section~\ref{s2.2} acting on 
closed or open singlet states.  Unlike the open singlet states in Section~\ref{s3.2}, the numbers of degrees of 
freedom of the fundamental and conjugate fields are arbitrary.  The proof that this is really a Lie algebra will be 
given in Appendix~\ref{sa5.1}.

Consider a vector space of all finite linear combinations of all five families of operators listed in 
Table~\ref{t5.1}.

\begin{table}[ht]
\begin{center}
\begin{tabular}{||c|c||}
\hline
operator of which kind & expression \\ \hline \hline
first & $\bar{\X}^{\l_1}_{\l_2} \otimes f^{\dot{I}}_{\dot{J}} \otimes \X^{\l_3}_{\l_4}$ \\
second & $\bar{\X}^{\l_1}_{\l_2} \otimes l^{\dot{I}}_{\dot{J}}$ \\
third & $r^{\dot{I}}_{\dot{J}} \otimes \X^{\l_3}_{\l_4}$ \\
fourth & $\g^I_J$ \\
fifth & $\ti{f}^I_J$ \\
\hline
\end{tabular}
\caption{\em Definitions of physical observables.}
\la{t5.1}
\end{center}
\end{table}

In this table, $1 \leq \l_1, \l_2, \l_3, \; \mbox{and} \; \l_4 \leq \L_F$, $I$ and $J$ are non-empty sequences of 
integers between 1 and $\L$ inclusive, and $\dot{I}$ and $\dot{J}$ are empty or non-empty sequences of integers also
between 1 and $\L$ inclusive.  The reader can regard the five types of vectors as operators defined in 
Eq.(\ref{a5.1.1}), or as axiomatic entities satisfying a set of Lie brackets to be described immediately.  We call 
a vector or an operator an {\em operator of the first, second, third, fourth or fifth kind} if it is a finite 
linear combination of operators, all of which are of the first, second, third, fourth or fifth form enlisted above, 
respectively.

Let us describe the Lie brackets among different kinds of operators.  The Lie bracket between an operator of the 
fifth kind, and an operator of the first, second or third kind is trivial:
\beq
   \le[ \bar{\X}^{\l_1}_{\l_2} \otimes f^{\dot{I}}_{\dot{J}} \otimes \X^{\l_3}_{\l_4}, 
   \ti{f}^K_L \ri] & = & 0; \nn \\
   \le[ \bar{\X}^{\l_1}_{\l_2} \otimes l^{\dot{I}}_{\dot{J}}, \ti{f}^K_L \ri] & = & 0 \mbox{; and} \nn \\
   \le[ r^{\dot{I}}_{\dot{J}} \otimes \X^{\l_1}_{\l_2}, \ti{f}^K_L \ri] & = & 0.
\la{5.2.1}
\eeq
The operators of the first kind form a proper ideal of the Lie algebra:
\beq
   \lefteqn{\le[ \bar{\X}^{\l_1}_{\l_2} \otimes f^{\dot{I}}_{\dot{J}} \otimes \X^{\l_3}_{\l_4},
   \bar{\X}^{\l_5}_{\l_6} \otimes f^{\dot{K}}_{\dot{L}} \otimes \X^{\l_7}_{\l_8} \ri] = } \nn \\
   & & \d^{\l_5}_{\l_2} \d^{\dot{K}}_{\dot{J}} \d^{\l_7}_{\l_4} 
   \bar{\X}^{\l_1}_{\l_6} \otimes f^{\dot{I}}_{\dot{L}} \otimes \X^{\l_3}_{\l_8} -  
   \d^{\l_1}_{\l_6} \d^{\dot{I}}_{\dot{L}} \d^{\l_3}_{\l_8} 
   \bar{\X}^{\l_5}_{\l_2} \otimes f^{\dot{K}}_{\dot{J}} \otimes \X^{\l_7}_{\l_4};
\nn \\
   \lefteqn{\le[ \bar{\X}^{\l_1}_{\l_2} \otimes f^{\dot{I}}_{\dot{J}} \otimes \X^{\l_3}_{\l_4},
   \bar{\X}^{\l_5}_{\l_6} \otimes l^{\dot{K}}_{\dot{L}} \ri] = } \nn \\
   & & \d^{\l_5}_{\l_2} \bar{\X}^{\l_1}_{\l_6} \otimes 
   \sum_{\dot{J}_1 \dot{J}_2 = \dot{J}} \d^{\dot{K}}_{\dot{J}_1} f^{\dot{I}}_{\dot{L} \dot{J}_2} \otimes
   \X^{\l_3}_{\l_4} -
   \d^{\l_1}_{\l_6} \bar{\X}^{\l_5}_{\l_2} \otimes
   \sum_{\dot{I}_1 \dot{I}_2 = \dot{I}} \d^{\dot{I}_1}_{\dot{L}} f^{\dot{K} \dot{I}_2}_{\dot{J}} \otimes
   \X^{\l_3}_{\l_4};
\nn \\
   \lefteqn{\le[ \bar{\X}^{\l_1}_{\l_2} \otimes f^{\dot{I}}_{\dot{J}} \otimes \X^{\l_3}_{\l_4},
   r^{\dot{K}}_{\dot{L}} \otimes \X^{\l_5}_{\l_6} \ri] = } \nn \\
   & & \d^{\l_5}_{\l_4} \bar{\X}^{\l_1}_{\l_2} \otimes \sum_{\dot{J}_1 \dot{J}_2 = \dot{J}} 
   \d^{\dot{K}}_{\dot{J}_2} f^{\dot{I}}_{\dot{J}_1 \dot{L}} \otimes \X^{\l_3}_{\l_6}
   - \d^{\l_3}_{\l_6} \bar{\X}^{\l_1}_{\l_2} \otimes \sum_{\dot{I}_1 \dot{I}_2 = \dot{I}}
   \d^{\dot{I}_2}_{\dot{L}} f^{\dot{I}_1 \dot{K}}_{\dot{J}} \otimes \X^{\l_5}_{\l_4} \mbox{; and}
\nn \\ 
   \lefteqn{\le[ \bar{\X}^{\l_1}_{\l_2} \otimes f^{\dot{I}}_{\dot{J}} \otimes \X^{\l_3}_{\l_4}, 
   \g^K_L \ri] = } \nn \\
   & & \bar{\X}^{\l_1}_{\l_2} \otimes \le( \sum_{\dot{J}_1 J_2 \dot{J}_3 = \dot{J}} 
   \d^K_{J_2} f^{\dot{I}}_{\dot{J}_1 L \dot{J}_3} 
   - \sum_{\dot{I}_1 I_2 \dot{I}_3 = \dot{I}} 
   \d^{I_2}_L f^{\dot{I}_1 K \dot{I}_3}_{\dot{J}} \ri) \otimes \X^{\l_3}_{\l_4}.
\la{5.2.2}
\eeq
The operators of the second kind form a subalgebra.  So are the operators of the third kind:
\beq
   \lefteqn{\le[ \bar{\X}^{\l_1}_{\l_2} \otimes l^{\dot{I}}_{\dot{J}},
   \bar{\X}^{\l_3}_{\l_4} \otimes l^{\dot{K}}_{\dot{L}} \ri] = } \nn \\
   & & \d^{\l_3}_{\l_2} \bar{\X}^{\l_1}_{\l_4} \otimes \le( \d^{\dot{K}}_{\dot{J}} l^{\dot{I}}_{\dot{L}}
   + \sum_{\dot{J}_1 J_2 = \dot{J}} \d^{\dot{K}}_{\dot{J}_1} l^{\dot{I}}_{\dot{L} J_2} +
   \sum_{\dot{K}_1 K_2 = \dot{K}} \d^{\dot{K}_1}_{\dot{J}} l^{\dot{I} K_2}_{\dot{L}} \ri) \nn \\
   & & - \d^{\l_1}_{\l_4} \bar{\X}^{\l_3}_{\l_2} \otimes \le( \d^{\dot{I}}_{\dot{L}} l^{\dot{K}}_{\dot{J}}
   + \sum_{\dot{L}_1 L_2 = \dot{L}} \d^{\dot{I}}_{\dot{L}_1} l^{\dot{K}}_{\dot{J} L_2} +
   \sum_{\dot{I}_1 I-2 = \dot{I}} \d^{\dot{I}_1}_{\dot{L}} l^{\dot{K} I_2}_{\dot{J}} \ri) \; \mbox{and}
\nn \\
   \lefteqn{\le[ r^{\dot{I}}_{\dot{J}} \otimes \X^{\l_1}_{\l_2},
   r^{\dot{K}}_{\dot{L}} \otimes \X^{\l_3}_{\l_4} \ri] = } \nn \\
   & & \d^{\l_3}_{\l_2} \le( \d^{\dot{K}}_{\dot{J}} r^{\dot{I}}_{\dot{L}} + \sum_{J_1 \dot{J}_2 = \dot{J}} 
   \d^{\dot{K}}_{\dot{J}_2} r^{\dot{I}}_{J_1 \dot{L}} 
   + \sum_{K_1 \dot{K}_2 = \dot{K}} 
   \d^{\dot{K}_2}_{\dot{J}} r^{K_1 \dot{I}}_{\dot{L}} \ri) \otimes \X^{\l_1}_{\l_4} \nn \\
   & & - \d^{\l_1}_{\l_4} \le( \d^{\dot{I}}_{\dot{L}} r^{\dot{K}}_{\dot{J}} + \sum_{L_1 \dot{L}_2 = \dot{L}} 
   \d^{\dot{I}}_{\dot{L}_2} r^{\dot{K}}_{L_1 \dot{J}} 
   + \sum_{I_1 \dot{I}_2 = \dot{I}}  
   \d^{\dot{I}_2}_{\dot{L}} r^{I_1 \dot{K}}_{\dot{J}} \ri) \otimes \X^{\l_3}_{\l_2}.
\la{5.2.3}
\eeq
Eqs.(\ref{5.2.1}) to (\ref{5.2.3}) together with the following relations show that operators of the first three 
kinds as a whole form another ideal:
\beq   
   \lefteqn{\le[ \bar{\X}^{\l_1}_{\l_2} \otimes l^{\dot{I}}_{\dot{J}},
   r^{\dot{K}}_{\dot{L}} \otimes \X^{\l_3}_{\l_4} \ri] = } \nn \\
   & & \bar{\X}^{\l_1}_{\l_2} \otimes \le( \sum_{\ba{l} \dot{J}_1 \dot{J}_2 = \dot{J} \\ 
   \dot{K}_1 \dot{K}_2 = \dot{K} \ea} 
   \d^{\dot{K}_1}_{\dot{J}_2} f^{\dot{I} \dot{K}_2}_{\dot{J}_1 \dot{L}} 
   - \sum_{\ba{l} \dot{I}_1 \dot{I}_2 = \dot{I} \\ \dot{L}_1 \dot{L}_2 = \dot{L} \ea}
   \d^{\dot{I}_2}_{\dot{L}_1} f^{\dot{I}_1 \dot{K}}_{\dot{J} \dot{L}_2} \ri) \otimes \X^{\l_3}_{\l_4};
\nn \\
   \lefteqn{\le[ \bar{\X}^{\l_1}_{\l_2} \otimes l^{\dot{I}}_{\dot{J}}, \g^K_L \ri] = } \nn \\
   & & \bar{\X}^{\l_1}_{\l_2} \otimes \le(
   \d^K_{\dot{J}} l^{\dot{I}}_L + \sum_{K_1 K_2 = K} \d^{K_1}_{\dot{J}} l^{\dot{I} K_2}_L
   + \sum_{J_1 J_2 = \dot{J}} \d^K_{J_2} l^{\dot{I}}_{J_1 L} \ri. \nn \\ 
   & & + \sum_{J_1 J_2 = \dot{J}} \d^K_{J_1} l^{\dot{I}}_{L J_2} 
   + \sum_{\ba{l} J_1 J_2 = \dot{J} \\ K_1 K_2 = K \ea} \d^{K_1}_{J_2} l^{\dot{I} K_2}_{J_1 L} 
   + \sum_{J_1 J_2 J_3 = \dot{J}} \d^K_{J_2} l^{\dot{I}}_{J_1 L J_3} \nn \\
   & & - \d^{\dot{I}}_L l^K_{\dot{J}} - \sum_{L_1 L_2 = L} \d^{\dot{I}}_{L_1} l^K_{\dot{J} L_2} 
   - \sum_{I_1 I_2 = \dot{I}} \d^{I_2}_L l^{I_1 K}_{\dot{J}} \nn \\
   & & - \le. \sum_{I_1 I_2 = \dot{I}} \d^{I_1}_L l^{K I_2}_{\dot{J}} 
   - \sum_{\ba{l} L_1 L_2 = L \\ I_1 I_2 = \dot{I} \ea} \d^{I_2}_{L_1} l^{I_1 K}_{\dot{J} L_2}  
   - \sum_{I_1 I_2 I_3 = \dot{I}} \d^{I_2}_L l^{I_1 K I_3}_{\dot{J}} \ri);
\nn \\
   \lefteqn{\le[ r^{\dot{I}}_{\dot{J}} \otimes \X^{\l_1}_{\l_2}, \g^K_L \ri] = } \nn \\
   & & \le( \d^K_{\dot{J}} r^{\dot{I}}_L + \sum_{K_1 K_2 = K} \d^{K_2}_{\dot{J}} r^{K_1 \dot{I}}_L 
   + \sum_{J_1 J_2 = \dot{J}} \d^K_{J_2} r^{\dot{I}}_{J_1 L} \ri. \nn \\
   & & + \sum_{J_1 J_2 = \dot{J}} \d^K_{J_1} r^{\dot{I}}_{L J_2} 
   + \sum_{\ba{l} J_1 J_2 = \dot{J} \\ K_1 K_2 = K \ea} \d^{K_2}_{J_1} r^{K_1 \dot{I}}_{L J_2} 
   + \sum_{J_1 J_2 J_3 = \dot{J}} \d^K_{J_2} r^{\dot{I}}_{J_1 L J_3} \nn \\
   & & - \d^{\dot{I}}_L r^K_{\dot{J}} - \sum_{L_1 L_2 = L} \d^{\dot{I}}_{L_2} r^K_{L_1 \dot{J}} 
   - \sum_{I_1 I_2 = \dot{I}} \d^{I_2}_L r^{I_1 K}_{\dot{J}} \nn \\
   & & - \le. \sum_{I_1 I_2 = \dot{I}} \d^{I_1}_L r^{K I_2}_{\dot{J}} 
   - \sum_{\ba{l} L_1 L_2 = L \\ I_1 I_2 = \dot{I} \ea} \d^{I_1}_{L_2} r^{K I_2}_{L_1 \dot{J}} 
   - \sum_{I_1 I_2 I_3 = \dot{I}}  
   \d^{I_2}_L r^{I_1 K I_3}_{\dot{J}} \ri) \otimes \X^{\l_1}_{\l_2}.
\la{5.2.4}
\eeq
These equations also reveal that operators of the first kind form a proper ideal of the algebra spanned by 
operators of the first three kinds.

The operators of the fifth kind form yet another proper ideal of this algebra:
\beq
   \le[ \ti{f}^I_J, \ti{f}^K_L \ri] = \d^K_J \ti{f}^I_L + 
   \sum_{K_1 K_2 = K} \d^{K_2 K_1}_J \ti{f}^I_L - (I \lrar K, J \lrar L)
\la{5.2.5}
\eeq
and
\beq
   \lefteqn{ \le[ \g^I_J, \ti{f}^K_L \ri] = \le\{ \d^K_J \ti{f}^I_L +
   \sum_{K_1 K_2 = K} \d^{K_2 K_2}_J \ti{f}^I_L +
   \sum_{K_1 K_2 = K} \d^{K_1}_J \ti{f}^{I K_2}_L \ri. } \nn \\
   & & + \le. \sum_{K_1 K_2 K_3 = K} \d^{K_2}_J \ti{f}^{K_1 I K_3}_L  
   + \sum_{K_1 K_2 = K} \d^{K_2}_J \ti{f}^{K_1 I}_L 
   + \sum_{K_1 K_2 K_3 = K} \d^{K_3 K_1}_J \ti{f}^{I K_2}_L \ri\} \nn \\
   & & - \le\{ \d^I_L \ti{f}^K_J - \sum_{L_1 L_2 = L} \d^I_{L_2 L_1} \ti{f}^K_J 
   - \sum_{L_1 L_2 = L} \d^I_{L_2} \ti{f}^K_{L_1 J} \ri. \nn \\
   & & - \le. \sum_{L_1 L_2 L_3 = L} \d^I_{L_2} \ti{f}^K_{L_1 J L_3}
   - \sum_{L_1 L_2 = L} \d^I_{L_1} \ti{f}^K_{J L_2} 
   - \sum_{L_1 L_2 L_3 = L} \d^I_{L_3 L_1} \ti{f}^K_{J L_2} \ri\}.
\la{5.2.6}
\eeq
Finally, the Lie bracket between two operators of the fourth kind is a linear combination of operators of the
fourth and fifth kinds:
\beq
   \lefteqn{ \left[ \g^I_J, \g^K_L \right] = 
   \delta^K_J \g^I_L + \sum_{J_1 J_2 = J} \delta^K_{J_2} \g^I_{J_1 L} 
   + \sum_{K_1 K_2 = K} \delta^{K_1}_J \g^{I K_2}_L } \nonumber \\
   & & + \sum_{\begin{array}{l}
		  J_1 J_2 = J \\
		  K_1 K_2 = K
	       \end{array}}
   \delta^{K_1}_{J_2} \g^{I K_2}_{J_1 L} 
   + \sum_{J_1 J_2 = J} \delta^K_{J_1} \g^I_{L J_2}
   + \sum_{K_1 K_2 = J} \delta^{K_2}_J \g^{K_1 I}_L \nonumber \\
   & & + \sum_{\begin{array}{l}
		  J_1 J_2 = J \\
		  K_1 K_2 = K
	       \end{array}}
   \delta^{K_2}_{J_1} \g^{K_1 I}_{L J_2}
   + \sum_{J_1 J_2 J_3 = J} \delta^K_{J_2} \g^I_{J_1 L J_3} 
   + \sum_{K_1 K_2 K_3 = K} \delta^{K_2}_J \g^{K_1 I K_3}_L \nonumber \\
   & & + \sum_{\begin{array}{l}
   		  J_1 J_2 = J \\
   		  K_1 K_2 = K
   	       \end{array}}
   \delta^{K_1}_{J_2} \delta^{K_2}_{J_1} \ti{f}^I_L
   + \sum_{\begin{array}{l}
   	      J_1 J_2 J_3 = J \\
   	      K_1 K_2 = K
   	   \end{array}}
   \delta^{K_1}_{J_3} \delta^{K_2}_{J_1} \ti{f}^I_{J_2 L}
   \nonumber \\
   & & + \sum_{\begin{array}{l}
   		  J_1 J_2 = J \\
   		  K_1 K_2 K_3 = K
   	       \end{array}}
   \delta^{K_1}_{J_2} \delta^{K_3}_{J_1} \ti{f}^{I K_2}_L \nonumber \\
   & & + \sum_{\begin{array}{l}
		  J_1 J_2 J_3 = J \\
		  K_1 K_2 K_3 = K
	       \end{array}}
   \delta^{K_1}_{J_3} \delta^{K_3}_{J_1} \ti{f}^{I K_2}_{J_2 L} 
    - (I \leftrightarrow K, J \leftrightarrow L), 
\la{5.2.7}
\eeq

We call the Lie algebra defined by the Lie brackets from Eqs.(\ref{5.2.1}) to (\ref{5.2.7}) the {\em grand string 
algebra}. 

Let ${\cal T}_o$ be the vector space of all finite linear combinations of singlet states of the form 
$\bar{\ph}^{\r_1} \otimes s^{\dot{K}} \otimes \ph^{\r_2}$ where $1 \leq \r_1, \r_2 \leq \L_F$ and all integers are 
between 1 and $\L$ inclusive in $\dot{K}$, which may be empty.  Also let ${\cal T}_c$ be the vector space of all 
finite linear combinations of singlet states of the form $\Ps^K$ such that all integers are again between 1 and 
$\L$ inclusive in $K$, which has to be non-empty, and that $\Ps^K$ satisfies Eq.(\ref{2.2.8}).  If we treat the 
operators ${\cal T}_o \oplus {\cal T}_c$ as the ones defined in the Subsection~\ref{s3.2}, we will find that the 
actions of an operator of the first kind are precisely given by Eqs.(\ref{2.2.10}) and (\ref{2.2.11}); those of the 
second kind by Eqs.(\ref{2.2.13}) and (\ref{2.2.14}); those of the third kind by Eqs.(\ref{2.2.16}) and 
(\ref{2.2.17}); those of the fourth kind by Eqs.(\ref{2.2.19}) and (\ref{2.2.20}); and those of the fifth kind by 
Eqs.(\ref{2.2.22}) and (\ref{2.2.23}).  Alternatively, we can {\em define} the actions of the operators of the five 
kinds on ${\cal T}_o \oplus {\cal T}_c$ by these ten equations, and show that ${\cal T}_o \oplus {\cal T}_c$ 
provides a representation for this Lie algebra.

\subsection{Open String Algebra}
\la{s5.3}

Now that we have the grand string algebra at hand, we are going to derive the Lie algebra of operators acting on 
open singlet states only as a quotient algebra of it.  We will call this the open string algebra, and identify 
some subalgebras of it.  The Lie algebra of operators acting on closed singlet states only will be considered 
in the next section.

Since every element of the grand string algebra maps a state in ${\cal T}_o$ to a state in ${\cal T}_o$, and a 
state in ${\cal T}_c$ to a state in ${\cal T}_c$, ${\cal T}_o$ and ${\cal T}_c$ furnish two representation spaces 
to the algebra.  The representations of the operators in the algebra, of course, form a Lie algebra for each of 
these two representation spaces.  However, since none of these two representation spaces provide faithful 
representations to the grand string algebra, some operators vanish.

Consider the Lie algebra generated by the representation ${\cal T}_o$ first.  We are going to call this algebra the 
{\em open string algebra}.  Since all operators of the fifth kind sends any state in ${\cal T}_o$ to zero, there 
are only 4 kinds of operators in this algebra.  The Lie brackets of these operators are the same as those in the 
previous section, except that we set all $\ti{f}^{I}_{J}$'s to be zero, i.e., only Eq.(\ref{5.2.7}) needs to be 
modified.  Let us write $\g$ as $\s$ in this open string algebra.  Note that the operators in this algebra are not 
linearly independent; there are many relations among them.  For example,
\beq
   \sum_{\l=1}^{\L_F} \bar{\X}^{\l}_{\l} \otimes l^I_J & = & \s^I_J - \sum_{i=1}^{\L} \s^{iI}_{iJ}; \nn \\
   \sum_{\l=1}^{\L_F} r^I_J \otimes \X^{\l}_{\l} & = & \s^I_J - \sum_{j=1}^{\L} \s^{Ij}_{Jj}; \nn \\
   \sum_{\l_3=1}^{\L_F} \bar{\X}^{\l_1}_{\l_2} \otimes f^{\dot{I}}_{\dot{J}} \otimes \X^{\l_3}_{\l_3}
      & = & \bar{\X}^{\l_1}_{\l_2} \otimes l^{\dot{I}}_{\dot{J}} -
      \bar{\X}^{\l_1}_{\l_2} \otimes \sum_{j=1}^{\L} l^{\dot{I}j}_{\dot{J}j} \mbox{; and} \nn \\
   \sum_{\l_1=1}^{\L_F} \bar{\X}^{\l_1}_{\l_1} \otimes f^{\dot{I}}_{\dot{J}} \otimes \X^{\l_3}_{\l_4}
      & = & r^{\dot{I}}_{\dot{J}} \otimes \X^{\l_3}_{\l_4} -
      \sum_{i=1}^{\L} r^{i \dot{I}}_{i \dot{J}} \otimes \X^{\l_3}_{\l_4}.
\la{5.3.1}
\eeq

There are many subalgebras and ideals in the open string algebra.  They were described in Table~3 in 
Ref.\cite{opstal}.

\subsection{Closed String Algebra}
\la{s5.4}

Let us turn our attention to the Lie algebra of operators acting on closed singlet states only.  It is generated by 
the representation space ${\cal T}_c$.  We will call this the {\em closed string algebra} or the {\em cyclix 
algebra} $\cyclix$.  Since acting any operator of the first three kinds on ${\cal T}_c$ yields zero, this Lie 
algebra is obtained by considering the Lie bracket among operators of the fourth and fifth kinds only.  Thus the 
closed string algebra is characterized by Eqs.(\ref{5.2.5}), (\ref{5.2.6}) and (\ref{5.2.7}).  We will write $\g$ 
as $g$ in the closed string algebra.  Again the operators are not linearly independent; in fact, any operator of 
the fifth kind can be written as a linear combination of operators of the fourth kind as
\beq
   \ti{f}^I_J = g^I_J - \sum_{k=1}^{\L} g^{I k}_{J k}
\la{5.4.1}
\eeq
or
\beq
   \ti{f}^I_J = g^I_J - \sum_{k=1}^{\L} g^{k I}_{k J}. 
\la{5.4.2}
\eeq
These two relations together with
\beq
   \ti{f}^{I_1 I_2}_J = \ti{f}^{I_2 I_1}_J
\la{5.4.3}
\eeq
and
\beq
   \ti{f}^I_{J_1 J_2} = \ti{f}^I_{J_1 J_2}
\la{5.4.4}
\eeq
can generate many other relations.  From Eq.(\ref{5.2.6}), we see that the set of all $\ti{f}^I_J$'s span a proper 
ideal $\ti{F}'_{\L}$ for the cyclix algebra.  We conjecture that $\cyclix$ is precisely the quotient of the Lie 
algebra given by Eqs.(\ref{5.2.5}), (\ref{5.2.6}) and (\ref{5.2.7}) by the kernel of Eqs.(\ref{5.4.1}) and 
(\ref{5.4.2}).  We showed in Ref.\cite{clstal} that this conjecture is true for $\L = 1$.  Shown in the same 
reference was the fact that if $\L = 1$, then the quotient of the closed string algebra by $\ti{F}'_1$ and 
$\hat{M}'_1$, viewed as another proper ideal of the closed string algebra, yields the Witt algebra.  This fact 
should be much more transparent in the unified account of both the open string algebra and the closed string 
algebra in this article: set $\L = \L_F = 1$ in the grand string algebra, then quotient out $\hat{M}_1$ (which is 
spanned by all operators of the first three kinds and some linear combinations of operators of the fourth kind 
shown in Eq.(\ref{a5.1.1}) with the primes removed) and $\ti{F}'_1$ (which is spanned by all operators of the 
fifth kind and some linear combinations of operators of the fourth kind shown in Eqs.(\ref{5.4.1}) and 
(\ref{5.4.2})), and we will get the Witt algebra.  The two apparently different ways of obtaining the Witt algebra
in Refs.\cite{opstal} and \cite{clstal} are just different orders of quotienting out the same set of operators.

A Cartan subalgebra and the associated root vectors of the $\L = 1$ closed string algebra is discussed in 
Ref.\cite{clstal}.

\subsection{The Ising Model Revisited}
\la{s5.5}

In this and the last section, we have mostly been studying issues in mathematics.  Let us see how to use these
mathematical results in physics.  A good starting point should be a model which is simple enough that we know a
great deal of it so that we can see how the algebraic point of view we have been discussing fit in.  In 
Section~\ref{s2.7}, we introduced a number of exactly solvable matrix models.  The Ising model is the simplest, and
let us look into this model to see what can be learnt.

There are many different ways of solving the Ising model.  One which is closer to the original spirit of Onsager's
approach is via the so-called {\em Onsager algebra} \cite{onsager, davies90}.  Consider a system whose Hamiltonian 
has the
form
\beq
	H = H_0 + V
\la{5.5.1}
\eeq
with the two terms in the hamiltonian satisfying the {\em Dolan--Grady conditions} \cite{dogr, davies91}:
\beq
   \lbrack H_0, \lbrack H_0, \lbrack H_0, V \rbrack \rbrack \rbrack = 16 \lbrack H_0, V \rbrack,
\label{5.5.2}
\eeq
and 
\beq
   \lbrack V, \lbrack V, \lbrack V, H_0 \rbrack \rbrack \rbrack = 16 \lbrack V, H_0 \rbrack.
\label{5.5.3}
\eeq
Then we can construct operators satisfying an infinite-dimensional Lie algebra
\beq  
   \lbrack A_m, A_n \rbrack = 4 G_{m-n}, \; \lbrack G_m, A_n \rbrack = 2 A_{n+m} - 2 A_{n-m},
   \; \mbox{and} \; \lbrack G_m, G_n \rbrack = 0
\la{5.5.4}
\eeq
by the following recursion relations:
\beq
   & A_0 = H_0, \; A_1 = V, \; A_{n+1} - A_{n-1} = \frac{1}{2} \lbrack G_1, A_n \rbrack, & \nonumber \\
   & G_1 = \frac{1}{4} \lbrack A_1, A_0 \rbrack, \; \mbox{and} \; G_n = \frac{1}{4} \lbrack A_n, A_0 \rbrack. &
\la{5.5.5}
\eeq
This Lie algebra is called the Onsager Lie algebra. It is known to be isomorphic to a fixed-point subalgebra of the 
$sl(2)$ loop algebra ${\cal L}(sl(2))$ with respect to the action of a certain involution \cite{roan}.  We can now
use the Onsager algebra to generate an infinite number of conserved quantities \cite{honecker}:
\beq
   Q_m = -\frac{1}{2} \left( A_m + A_{-m} + \lambda A_{m+1} + \lambda A_{-m+1} \right).
\la{5.5.6}
\eeq
Thus any such system should be integrable. 

In the case of the Ising model we choose $H_0$ and $V$ as above.  The Lie brackets of the cyclix algebra then 
allows us to verify the Dolan--Grady conditions.  Thus we can construct the Onsager algebra as a proper subalgebra
of $\cyclix$.  Because of the relationship between the Onsager algebra and the loop algebras, this suggests a deep
relationship between $\cyclix$ and the loop algebras.

Certainly, if we understand some other features of the algebras we have described in this article better, we will 
be in a better position to understand Yang--Mills theory, the low energy behavior of string theory, and many 
integrable systems.

\vskip 1pc
\noindent \Large{\bf \hskip .2pc Acknowledgments}
\vskip 1pc
\noindent 

\normalsize
We thank V. John, S. Okubo, T. D. Palev and O.T. Turgut for discussions.  We were supported in part by funds 
provided by the U.S. Department of Energy under grant DE-FG02-91ER40685.

\vskip 1pc
\noindent \Large{\bf \hskip .2pc Appendix}

\normalsize
\appendix

\section{Actions of Operators on Physical States in the Large-$N$ Limit}
\label{sa1.3}

We are going to illustrate (though not rigorously prove) why in the planar large-$N$ limit, an operator 
representing a term in a dynamical variable sends singlet states to singlet states.  We are going to confine 
ourselves to the action of operators of the second kind defined in Section~\ref{s2.2} only.  The reasoning is 
similar for operators of other kinds.

Assume that the operator is of the form Eq.(\ref{2.2.12}) and the open singlet state is of the form 
Eq.(\ref{2.2.5}).  Let $\dot{a} = \#(\dot{I})$, $\dot{b} = \#(\dot{J})$ and $\dot{c} = \#(\dot{K})$.  So there are 
$\dot{a}$ creation operators and $\dot{b}$ annihilation operators for adjoint partons in the operator of the second 
kind, and $\dot{c}$ creation operators for adjoint partons in the open singlet state.  There is a factor of 
$N^{-(\dot{a}+\dot{b})/2}$ in the operator of the second kind and a factor of $N^{-(\dot{c}+1)/2}$ in the open 
singlet state, so initially there is a total factor of $N^{-(\dot{a} + \dot{b} + \dot{c} + 1) / 2}$.  A 
term in the final state is either an open singlet state or a product of an open singlet state and a number of 
closed singlet states.  No matter how many closed singlet states there are in this term, $N$ should be raised to 
the power of $-(\dot{a} - \dot{b} + \dot{c} + 1) / 2$ in order that the term survives the large-$N$ limit.  It is 
therefore that only the operations which produce a factor of $N^{\dot{b}}$ survive the large-$N$ limit.

To clarify the argument, we need to refine the diagrammatic representations of an operator of any kind and a 
singlet state more carefully.  Note that in Figs.~\ref{f2.1}(a) and (c), each square carries one color index 
whereas each circle carries two color indices.  We can put these indices at the ends of the thick or thin lines 
attaching to them, as is shown in Figs.~\ref{fa2.1}(a) and~(b).

\begin{figure}[ht]
\epsfxsize=5in
\centerline{\epsfbox{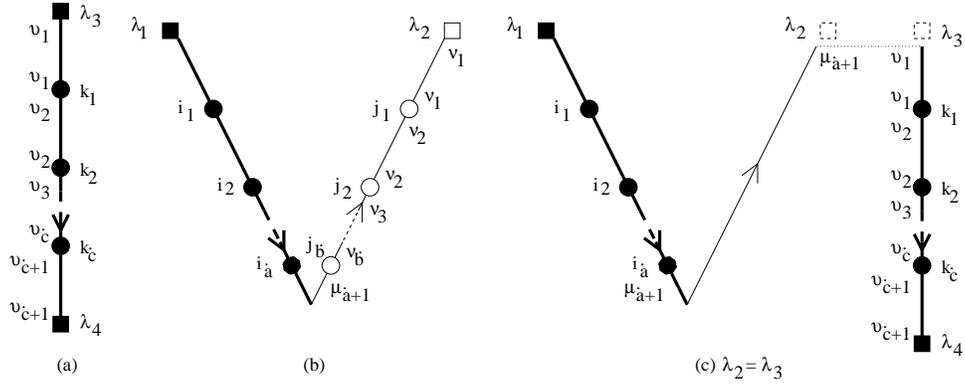}}
\caption{\em (a)  A typical open singlet state given by Eq.(\ref{2.2.5}).  The color indices are explicitly shown.  
Note that each square carries 1 color index, whereas each circle carries 2 color indices.  Moreover, the color 
indices at the two ends of a connecting solid line are the same.  (b)  An operator of the second kind given by 
Eq.(\ref{2.2.12}).  Only the color indices of the annihilation operators are shown.  (c)  The action of an operator 
of the second kind with no annihilation operator of an adjoint parton on an open singlet state.  To get a 
non-vanishing final state, we need $\lambda_2 = \rho_1$.  The creation and annihilation operators removed in the 
final state are depicted as dotted squares or dotted circles.  Algebraically the dotted line joining the operator 
of the second kind and the open singlet state is $\delta^{\upsilon_1}_{\mu_{a+1}}$.  Clearly the final state is an 
open singlet state.}
\label{fa2.1}
\end{figure}

Consider the case when there are no annihilation operators for adjoint partons in the operator of the second kind, 
i.e., $\dot{b}=0$.  Fig.~\ref{fa2.1}(c) shows the action of such an operator on an open singlet state.  It destroys 
the solid square in Fig.~\ref{fa2.1}(a) and the hollow square in Fig.~\ref{fa2.1}(b).  The ends of the lines 
originally attached to the squares are now joined together by a dotted line with an appearance different from other 
dotted lines in the diagrams.   Algebraically this dotted line is a Kronecker delta function.  The factor involving 
$N$ in the final state is $N^{-(\dot{a} + \dot{c} + 1) / 2}$.  This is precisely the factor for an open singlet 
state with $\dot{a} + \dot{c}$ partons in the adjoint representation.  

\begin{figure}[ht]
\epsfxsize=5in
\centerline{\epsfbox{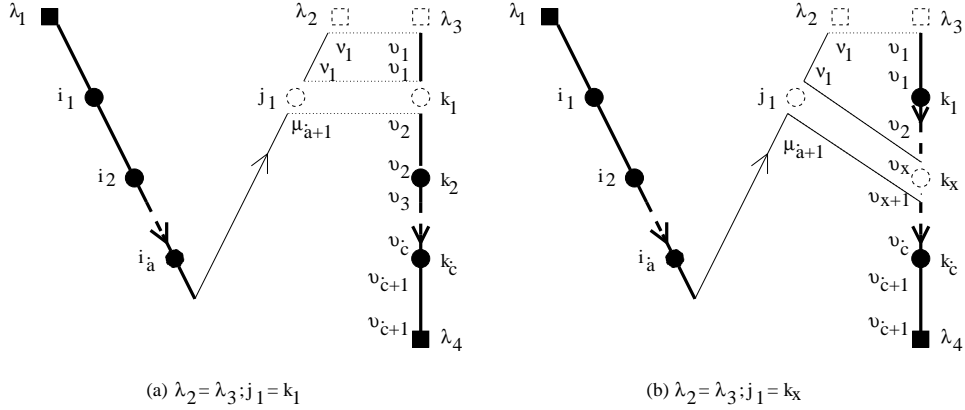}}
\caption{\em The action of an operator of the second kind on an open singlet state.  (a) The annihilation operator 
of an adjoint parton in the operator of the second kind contracts with the first creation operator of an adjoint 
parton in the sequence of adjoint partons in the initial open singlet state.  The final state is also an open 
singlet state.  This action survives the large-$N$ limit.  (b) Here this annihilation operator contracts with the 
creation operator of a later adjoint parton in the sequence.  The final state is a product of an 
open singlet and a closed singlet.  However, this term is negligable in the large-$N$ limit.}
\label{fa2.2}
\end{figure}

Now consider the case when there is one annihilation operator of an adjoint parton in the operator of the second
kind, i.e., $\dot{b} = 1$.  The action of this operator needs to produce a factor of $N$ in order for the final 
state to survive in the large-$N$ limit.  Consider the following two cases illustrated separately in 
Figs.~\ref{fa2.2}(a) and~(b).  In the former diagram, the annihilation operator of an adjoint parton in the 
operator of the second kind contracts with the creation operator of the first adjoint parton in the initial open
singlet state.  This results in a close loop with no squares or circles but two dotted lines only.  Algebraically 
this loop is the factor $\delta^{\upsilon_1}_{\nu_1} \delta^{\nu_1}_{\upsilon_1}$, which in turn is equal to $N$.  
The remaining parts of this diagram form an open singlet state.  Thus the singlet state survives the large-$N$ 
limit.  In the latter diagram, the annihilation operator of the adjoint parton in the operator of the second kind 
contracts with the creation operator of a later adjoint parton in the adjoint parton sequence of the initial open 
singlet state.  This time the final state is a product of an open singlet together with a closed singlet.  However, 
there is no closed loop with solid and dotted lines only.  This implies that no extra factor of $N$ is produced and 
so this term can be neglected in the large-$N$ limit. 

\begin{figure}[ht]
\epsfxsize=5in
\centerline{\epsfbox{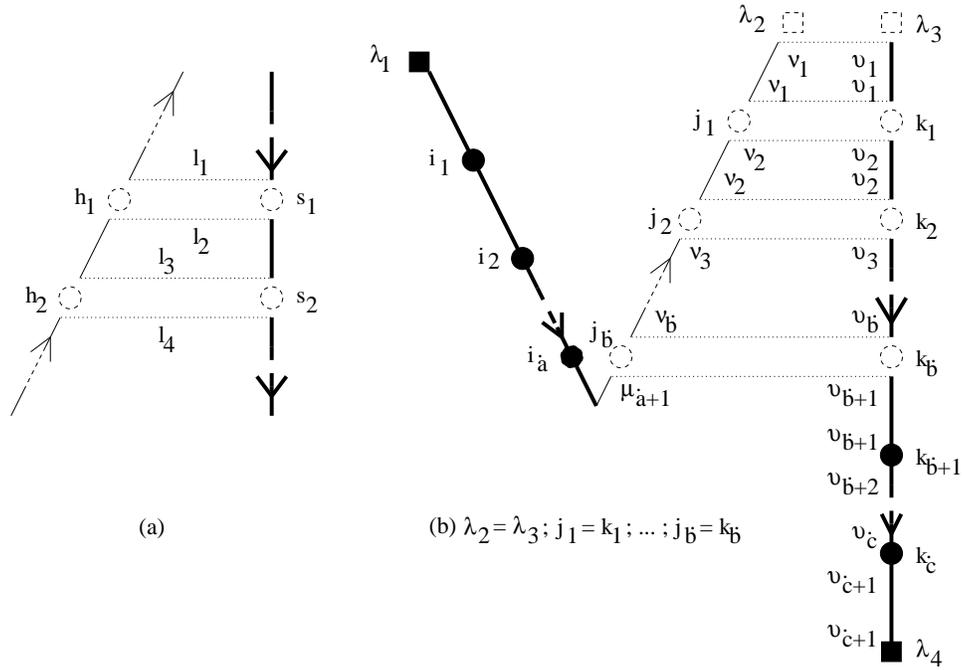}}
\caption{\em (a)  Contraction of circles.  See the text for details.  (b)  The surviving term in the large-$N$ 
limit of the action of an operator of the second kind with $b$ annihilation operators of adjoint partons on an open 
singlet state.  The final state is also an open singlet state.}
\label{fa2.3}
\end{figure}

Let us now turn to the case when $\dot{b}$ is an arbitrary positive integer.  The annihilation operator of a
conjugate parton in the operator of the second kind will contract with the creation operator of a conjugate parton
in the initial open singlet state.  This will produce 1 dotted line.  The $\dot{b}$ annihilation operators of 
adjoint partons in the operator of the second kind will contract with $\dot{b}$ creation operators of adjoint 
partons in the initial meson state.  This will further produce $2\dot{b}$ dotted lines.  Hence there are 
$2\dot{b}+1$ dotted lines.  One of these dotted lines has to be recruited to join the creation and annihilation 
operators in the final state.  In order for the final state to survive in the large-$N$ limit, we need a factor of 
$N^{\dot{b}}$, as explained above.  Since the minimum number of dotted lines to produce a closed loop without 
squares or circles is 2, the factor $N^{\dot{b}}$ can be obtained only if there are ${\dot{b}}$ closed loops, every 
closed loop has 2 dotted lines only, and there are no circles or squares in any closed loop.  This can be done 
only if the 2 dotted lines join 2 adjacent pairs of creation and annihilation operators.  Moreover, only 1 dotted 
line can be uninvolved in any closed loops.  Consider Fig~\ref{fa2.3}(a), where we contract 2 circles $h_1$ and 
$s_1$, producing 2 dotted lines $l_1$ and $l_2$.  Since one of these dotted lines has to be lie within a closed 
loop, a pair of circles adjacent to $h_1$ and $s_1$ has to be contracted.  In Fig~\ref{fa2.3}(a), $h_2$ and $s_2$ 
are contracted and hence we obtain $l_3$ and $l_4$.  Then either $l_1$ or $l_4$ (or both) has to lie within a 
closed loop.  If we continue this argument, we will obtain Fig.~\ref{fa2.3}(b).  As is clear from the figure, the 
final state is an open singlet state.  Thus we conclude that in the large-$N$ limit, an operator of the second kind
propagates open singlet states to open singlet states.  The actions of operators of other kinds can be understood 
similarly.

\section{Associativity of the Algebra for the Operators of the Second Kind}
\label{sa4.1}

The reader can prove the associativity of this algebra by verifying the following identity:
\begin{eqnarray*}
   \lefteqn{l^{\dot{I}}_{\dot{J}} \left( l^{\dot{K}}_{\dot{L}} l^{\dot{M}}_{\dot{N}} \right) = 
   \left( l^{\dot{I}}_{\dot{J}} l^{\dot{K}}_{\dot{L}} \right) l^{\dot{M}}_{\dot{N}} = 
   \delta^{\dot{K}}_{\dot{J}} \delta^{\dot{M}}_{\dot{L}} l^{\dot{I}}_{\dot{N}} +
   \sum_{\dot{M_1} M_2 = \dot{M}} \delta^{\dot{K}}_{\dot{J}} \delta^{\dot{M_1}}_{\dot{L}} l^{\dot{I} M_2}_{\dot{N}}
   } \\
   & & + \sum_{\dot{L_1} L_2 = \dot{L}} \delta^{\dot{K}}_{\dot{J}} \delta^{\dot{M}}_{\dot{L_1}}
   l^{\dot{I}}_{\dot{N} L_2} +
   \sum_{\dot{K_1} K_2 = \dot{K}} \delta^{\dot{K_1}}_{\dot{J}} \delta^{\dot{M}}_{\dot{L}} l^{\dot{I} K_2}_{\dot{N}}
   + \sum_{\dot{J_1} J_2 = \dot{J}} \delta^{\dot{K}}_{\dot{J_1}} \delta^{\dot{M}}_{\dot{L}}
   l^{\dot{I}}_{\dot{N} J_2} \\
   & & + \sum_{\begin{array}{c}
   		  \dot{K_1} K_2 = \dot{K} \\
   		  \dot{M_1} M_2 = \dot{M}
   	       \end{array}}
   \delta^{\dot{K_1}}_{\dot{J}} \delta^{\dot{M_1}}_{\dot{L}} l^{\dot{I} K_2 M_2}_{\dot{N}} 
   + \sum_{\begin{array}{c}
   	      \dot{K_1} K_2 = \dot{K} \\
   	      \dot{L_1} L_2 = \dot{L}
	   \end{array}}
   \delta^{\dot{K_1}}_{\dot{J}} \delta^{\dot{M}}_{\dot{L_1}} l^{\dot{I} K_2}_{\dot{N} L_2} \\
   & & + \sum_{\begin{array}{c}
		  \dot{J_1} J_2 = \dot{J} \\
		  \dot{M_1} M_2 = \dot{M}
	       \end{array}}
   \delta^{\dot{K}}_{\dot{J_1}} \delta^{\dot{M_1}}_{\dot{L}} \delta^{M_2}_{L_2} l^{\dot{I}}_{\dot{N}}
   + \sum_{\begin{array}{c}
   	      \dot{J_1} J_2 = \dot{J} \\
   	      \dot{L_1} L_2 = \dot{L}
  	   \end{array}}
   \delta^{\dot{K}}_{\dot{J_1}} \delta^{\dot{M}}_{\dot{L_1}} l^{\dot{I}}_{\dot{N} L_2 J_2} \\
   & & + \sum_{\begin{array}{c}
   		  \dot{J_1} J_2 = \dot{J} \\
   		  \dot{M_1} M_2 M_3 = M
   	       \end{array}}
   \delta^{\dot{K}}_{\dot{J_1}} \delta^{\dot{M_1}}_{\dot{L}} \delta^{M_2}_{J_2} l^{\dot{I} M_3}_{\dot{N}} 
   + \sum_{\begin{array}{c}
   	      \dot{J_1} J_2 J_3 = J \\
   	      \dot{M_1} M_2 = M \\
	   \end{array}}
   \delta^{\dot{K}}_{\dot{J_1}} \delta^{\dot{M_1}}_{\dot{L}} \delta^{M_2}_{J_2} l^{\dot{I}}_{\dot{N} J_3} 	 
\end{eqnarray*}     	      
Q.E.D.

\section{Cartan Subalgebra of $\hatleftix$}
\label{sa4.2}

This can be seen by the following argument.  Let us call the subspace spanned by all the $l^{\dot{I}}_{\dot{I}}$'s 
${\cal M}$.  From Eq.(\ref{4.3.3}), $\lbrack l^{\dot{I}}_{\dot{I}}, l^{\dot{J}}_{\dot{J}} \rbrack = 0$ for any 
integer sequences $\dot{I}$ and $\dot{J}$.  Thus ${\cal M}$ is an Abelian subalgebra.  In particular, ${\cal M}$ is 
niltpotent.  To proceed on, we need the following two lemmas:
\begin{lemma}
Let 
\[ \lbrack l^{\dot{I}}_{\dot{I}}, l^{\dot{K}}_{\dot{L}} \rbrack = \sum_{k=1}^n \alpha^{\dot{N}_k}_{\dot{M}_k} 
   l^{\dot{M}_k}_{\dot{N}_k} \]
where $n$ is a finite positive integer, $\dot{M}_k$ and $\dot{N}_k$ are positive integer sequences such that 
$l^{\dot{M}_k}_{\dot{N}_k} \neq l^{\dot{M}_{k'}}_{\dot{N}_{k'}}$ for $k \neq k'$, and 
$\alpha^{\dot{N}_k}_{\dot{M}_k}$ are non-zero numerical coefficients.  Then
\[ \#(\dot{M}_k) - \#(\dot{N}_k) = \#(\dot{K}) - \#(\dot{L}) \]
for every $k = 1, 2, \ldots, n$.
\label{la4.2.1}
\end{lemma}
This lemma can be proved by using Eq.(\ref{4.3.3}) with $\dot{J} = \dot{I}$.  
\begin{lemma}
With the same assumptions as in the previous lemma, we have
\[ \#(\dot{M}_k) + \#(\dot{N}_k) \geq \#(\dot{K}) + \#(\dot{L}) \]
for every $k = 1, 2, \ldots, n$.
\label{la4.2.2}
\end{lemma}
This lemma can also be proved by using Eq.(\ref{4.3.3}) with $\dot{J} = \dot{I}$.  Let $m$ be a positive integer.  
Now we are ready to show for arbitrary non-zero complex numbers $\beta^{\dot{L}_i}_{\dot{K}_i}$, where $i = 1, 2, 
\ldots, m$ and arbitrary integer sequences $\dot{L}_i$'s and $\dot{K}_i$'s such that $l^{\dot{K}_i}_{\dot{L}_i} 
\neq l^{\dot{K}_{i'}}_{\dot{L}_{i'}}$ for $i \neq i'$, and $\dot{K}_i \neq \dot{L}_i$ for at least one $i$, that 
there exists a sequence $\dot{I}$ such that 
\[ \lbrack l^{\dot{I}}_{\dot{I}}, \sum_{i=1}^m \beta^{\dot{L}_i}_{\dot{K}_i} l^{\dot{K}_i}_{\dot{L}_i} \rbrack \]
does not belong to ${\cal M}$.  Indeed, let $j$ be an integer such that
\begin{enumerate}
   \item $\#(\dot{K}_j) - \#(\dot{L}_j) \geq \#(\dot{K}_i) - \#(\dot{L}_i)$ for all $i = 1, 2, \ldots$ and $m$; and
   \item $\#(\dot{K}_j) + \#(\dot{L}_j) \leq \#(\dot{K}_i) + \#(\dot{L}_i)$ for any $i = 1, 2, \ldots$ or $m$ such
         that $\#(\dot{K}_j) - \#(\dot{L}_j) = \#(\dot{K}_i) - \#(\dot{L}_i)$.
\end{enumerate}
If $\#(\dot{K}_j) \geq \#(\dot{L}_j)$, then consider
\begin{equation}
   \lbrack l^{\dot{K}_j}_{\dot{K}_j}, \sum_{i=1}^m \beta^{\dot{L}_i}_{\dot{K}_i}
   l^{\dot{K}_i}_{\dot{L}_i} \rbrack = \beta^{\dot{L}_j}_{\dot{K}_j} l^{\dot{K}_j}_{\dot{L}_j} -          
   \beta^{\dot{L}_j}_{\dot{K}_j} \sum_{\dot{K}_{j1} K_{j2} = \dot{K}_j} \delta^{\dot{K}_{j1}}_{\dot{L}_j} 
   l^{\dot{K}_j K_{j2}}_{\dot{K}_j} + \Gamma, 
\label{a4.2.1}
\end{equation}   
where
\begin{eqnarray}
   \Gamma & = & \lbrack l^{\dot{K}_j}_{\dot{K}_j}, \sum_{\begin{array}{c} i=1 \\ i \neq j \end{array}}^m 
   \beta^{\dot{L}_i}_{\dot{K}_i} l^{\dot{K}_i}_{\dot{L}_i} \rbrack 
   \nonumber \\
   & = & \sum_{\begin{array}{c} i=1 \\ i \neq j \end{array}}^m \sum_{k=1}^{n_i} 
   \beta^{\dot{L}_i}_{\dot{K}_i} \alpha^{\dot{N}_{ik}}_{\dot{M}_{ik}} l^{\dot{M}_{ik}}_{\dot{N}_{ik}}
\label{a4.2.2}   
\end{eqnarray}
where each $n_i$ for $i = 1, 2, \ldots, m$ but $i \neq j$ is dependent on $i$.  Let us assume that
\begin{equation}
   l^{\dot{M}_{ik}}_{\dot{N}_{ik}} = l^{\dot{K}_j}_{\dot{L}_j}
\label{a4.2.3}
\end{equation}
for some $i \in \{ 1, 2, \ldots, m \}$ but $i \neq j$ and $k \in \{ 1, 2, \ldots n_i \}$.  Then $\#(\dot{M}_{ik}) = 
\#(\dot{K}_j)$ and $\#(\dot{N}_{ik}) = \#(\dot{L}_j)$.  By Lemmas~\ref{la4.2.1} and~\ref{la4.2.2}, we get 
$\#(\dot{K}_i) = \#(\dot{K}_j)$ and $\#(\dot{L}_i) = \#(\dot{L}_j)$.  However, we also know that $\dot{K}_i \neq 
\dot{K}_j$ or $\dot{L}_i \neq \dot{L}_j$ and so there is no $k \in \{ 1, 2, \ldots, n_i \}$ such that 
Eq.(\ref{a4.2.3}) holds.  This leads to a contradiction and so we conclude that
\[ l^{\dot{M}_{ik}}_{\dot{N}_{ik}} \neq l^{\dot{K}_j}_{\dot{L}_j} \]
for {\em all} $i \in \{ 1, 2, \ldots, m \}$ but $i \neq j$ and $k \in \{ 1, 2, \ldots, n_i \}$.  From 
Eqs.(\ref{a4.2.1}) and (\ref{a4.2.2}), we deduce that $\lbrack l^{\dot{K}_j}_{\dot{K}_j}, \sum_{i=1}^m 
\beta^{\dot{L}_i}_{\dot{K}_i} l^{\dot{K}_i}_{\dot{L}_i} \rbrack$ does not belong to ${\cal M}$.  Similarly, if 
$\#(\dot{K}_j) \leq \#(\dot{L}_j)$, then $\lbrack l^{\dot{L}_j}_{\dot{L}_j}, \sum_{i=1}^m 
\beta^{\dot{L}_i}_{\dot{K}_i} l^{\dot{K}_i}_{\dot{L}_i} \rbrack$ does not belong to ${\cal M}$.  Hence, the 
normalizer of ${\cal M}$ is ${\cal M}$ itself.  We therefore conclude that ${\cal M}$ is a Cartan subalgebra of the 
algebra $\hatleftix$.  Q.E.D. 

\section{Root Vectors of $\hatleftix$}
\label{sa4.3}

All we need to do is to show that any root vector has to be of the form given by Eq.(\ref{4.3.7}).  Let $f \equiv 
\sum_{\dot{P}, \dot{Q}} a^{\dot{Q}}_{\dot{P}} l^{\dot{P}}_{\dot{Q}}$, where only a finite number of the numerical 
coefficients $a^{\dot{Q}}_{\dot{P}} \neq 0$, be a root vector.  In addition, we can assume without loss of 
generality that $\dot{P} \neq \dot{Q}$ if $a^{\dot{Q}}_{\dot{P}} = 0$.  We can deduce from Eq.(\ref{2.2.13}) that
\[ l^{\dot{I}}_{\dot{J}} s^{\dot{K}} = 
   \sum_{\dot{K}_1 \dot{K}_2 = \dot{K}} \delta^{\dot{K}_1}_{\dot{J}} s^{\dot{I} \dot{K}_2}. \]
Hence,
\[ l^{\dot{P}}_{\dot{Q}} s^{\dot{K}} = \sum_{\dot{I}} \sum_{\dot{K}_1 \dot{K}_2 = \dot{K}} 
   \delta^{\dot{J}_1}_{\dot{Q}} \delta^{\dot{P} \dot{K}_2}_{\dot{I}} s^{\dot{I}}. \]
Therefore,
\begin{equation}
   \left[ l^{\dot{M}}_{\dot{M}}, f \right] s^{\dot{K}} = 
   \sum_{\dot{I}, \dot{P}, \dot{Q}} a^{\dot{Q}}_{\dot{P}} \sum_{\begin{array}{c} \dot{I}_1 \dot{I}_2 = \dot{I} \\
   \dot{K}_1 \dot{K}_2 = \dot{K} \end{array}}
   \left( \delta^{\dot{M}}_{\dot{I}_1} \delta^{\dot{K}_1}_{\dot{Q}} \delta^{\dot{P} \dot{K}_2}_{\dot{I}} -
   \delta^{\dot{P}}_{\dot{I}_1} \delta^{\dot{K}_1}_{\dot{M}} \delta^{\dot{K}}_{\dot{Q} \dot{I}_2} \right) 
   s^{\dot{I}}.			   
\label{a4.3.1}
\end{equation}
Since $f$ is a root vector, we have
\begin{equation}
   \left[ l^{\dot{M}}_{\dot{M}}, f \right] = 
   \lambda_{\dot{M}} \sum_{\dot{P}, \dot{Q}} a^{\dot{Q}}_{\dot{P}} l^{\dot{P}}_{\dot{Q}}
\label{a4.3.2}
\end{equation}   
where $\lambda_{\dot{M}}$ is a root.  As a result, we can combine Eqs.(\ref{a4.3.1}) and (\ref{a4.3.2}) to obtain
\begin{eqnarray}
   \sum_{\dot{P}, \dot{Q}} a^{\dot{Q}}_{\dot{P}} \sum_{\begin{array}{c} \dot{I}_1 \dot{I}_2 = \dot{I} \\
   \dot{K}_1 \dot{K}_2 = \dot{K} \end{array}}
   \left( \delta^{\dot{M}}_{\dot{I}_1} \delta^{\dot{K}_1}_{\dot{Q}} \delta^{\dot{P} \dot{K}_2}_{\dot{I}} -
   \delta^{\dot{P}}_{\dot{I}_1} \delta^{\dot{K}_1}_{\dot{M}} \delta^{\dot{K}}_{\dot{Q} \dot{I}_2} \right) 
   & & \nonumber \\ 
   - \sum_{\dot{P}, \dot{Q}} a^{\dot{Q}}_{\dot{P}} \lambda_{\dot{M}} \sum_{\dot{K}_1 \dot{K}_2 = \dot{K}} 
   \delta^{\dot{K}_1}_{\dot{Q}} \delta^{\dot{P} \dot{K}_2}_{\dot{I}} & = & 0
\label{a4.3.3}
\end{eqnarray}
for {\em any} integer sequences $\dot{I}$, $\dot{K}$ and $\dot{M}$.

Let us find an $a^{\dot{S}}_{\dot{R}}$ in $f$ such that $\dot{R} \neq \dot{S}$, $a^{\dot{S}}_{\dot{R}} \neq 0$, and 
$a^{S_1}_{R_1} = 0$ for all $R_1$'s and $S_1$'s such that $R_1 \dot{R}_2 = \dot{R}$ and $S_1 \dot{S}_2 = \dot{S}$ 
for some $\dot{R}_2$ and $\dot{S}_2$.  The reader can easily convince himself or herself that such an 
$a^{\dot{S}}_{\dot{R}}$ always exists.  Let us choose $\dot{I} = \dot{R}$ and $\dot{K} = \dot{S}$ in 
Eq.(\ref{a4.3.3}).  Then we obtain from this equation that
\begin{equation}
   \lambda_{\dot{M}} = \sum_{\dot{R}_1 \dot{R}_2 = \dot{R}} \delta^{\dot{R}_1}_{\dot{M}} 
   - \sum_{\dot{S}_1 \dot{S}_2 = \dot{S}} \delta^{\dot{M}}_{\dot{S}_1}.
\label{a4.3.4}
\end{equation}
Therefore,
\[ \lambda_{\dot{M}} - \sum_{i=1}^{\Lambda} \lambda_{\dot{M}i} = \delta^{\dot{R}}_{\dot{M}} - 
   \delta^{\dot{M}}_{\dot{S}}. \]
Thence
\begin{eqnarray*}
   \left[ l^{\dot{R}}_{\dot{R}} - \sum_{i=1}^{\Lambda} l^{\dot{R}i}_{\dot{R}i}, f \right] 
   & \neq & 0 \mbox{; and} \\
   \left[ l^{\dot{S}}_{\dot{S}} - \sum_{i=1}^{\Lambda} l^{\dot{S}i}_{\dot{S}i}, f \right] 
   & \neq & 0.
\end{eqnarray*}
However, both $l^{\dot{R}}_{\dot{R}} - \sum_{i=1}^{\Lambda} l^{\dot{R}i}_{\dot{R}i}$ and $l^{\dot{S}}_{\dot{S}} - 
\sum_{i=1}^{\Lambda} l^{\dot{S}i}_{\dot{S}i} \in \salt$.  Thus $f \in \salt$.  Now Eq.(\ref{4.3.15}) shows clearly 
that $f = f^{\dot{R}}_{\dot{S}}$.  The same equation also shows that each root vector space must be 
one-dimensional.  Q.E.D. 
         
\section{Lie Bracket of $\hatcentrix$}
\label{sa4.4}

We are going to show that the commutator between two operators of the fourth kind, Eq.(\ref{4.5.4}), defines a Lie 
bracket between them.  This can be done by showing that Eq.(\ref{4.5.4}) satisfies Eq.(\ref{4.5.3}).  This involves 
a tedious compuatation involving a delta function defined in Appendix~A of Ref.\cite{opstal}.  The properties of 
this delta function will be extensively used.  Let us consider the action of the commutator of two $\s$'s  on 
$s^{\dot{P}}$.  If $\dot{P}$ is empty, then Eq.(\ref{4.5.4}) certainly satisfies Eq.(\ref{4.5.3}).  Therefore we 
only need to consider the case when $\dot{P}$ is not empty.  In this case we can simply write $\dot{P}$ as $P$.  
Then the action of the commutator on $s^P$ is
\begin{eqnarray}
   \lefteqn{ \left[ \s^I_J, \s^K_L \right] s^P = \sum_Q \left(
   \delta^I_Q \delta^K_J \delta^P_L + \sum_C \delta^I_Q \delta^{C K}_J
     \delta^P_{C L} + \sum_D \delta^I_Q \delta^{K D}_J \delta^P_{L D} \right. }
     \nonumber \\
   & & + \sum_{C, D} \delta^I_Q \delta^{C K D}_J \delta^P_{C L D} +
   \sum_A \delta^{A I}_Q \delta^K_{A J} \delta^P_L +
   \sum_{A, C} \delta^{A I}_Q \delta^{C K}_{A J} \delta^P_{C L} \nonumber \\
   & & + \sum_{A, D} \delta^{A I}_Q \delta^{K D}_{A J} \delta^P_{L D} +
   \sum_{A, C, D} \delta^{A I}_Q \delta^{C K D}_{A J} \delta^P_{C L D}
   + \sum_B \delta^{I B}_Q \delta^K_{J B} \delta^P_L  \nonumber \\
   & & + \sum_{B, C} \delta^{I B}_Q \delta^{C K}_{J B} \delta^P_{C L} + 
   \sum_{B, D} \delta^{I B}_Q \delta^{K D}_{J B} \delta^P_{L D} +
   \sum_{B, C, D} \delta^{I B}_Q \delta^{C K D}_{J B} \delta^P_{C L D}
   \nonumber \\
   & & + \sum_{A, B} \delta^{A I B}_Q \delta^K_{A J B} \delta^P_L +
   \sum_{A, B, C} \delta^{A I B}_Q \delta^{C K}_{A J B} \delta^P_{C L} +
   \sum_{A, B, D} \delta^{A I B}_Q \delta^{K D}_{A J B} \delta^P_{L D}
   \nonumber \\
   & & + \left. \sum_{A, B, C, D} \delta^{A I B}_Q \delta^{C K D}_{A J B} 
   \delta^P_{C L D} \right) s^Q - (I \leftrightarrow K, J \leftrightarrow L).
\label{a4.4.1}
\end{eqnarray}
Each of the terms on the R.H.S. of the above equations can be rewritten as follows;
\begin{eqnarray}
   \lefteqn{ \delta^I_Q \delta^K_J \delta^P_L = \delta^K_J \delta^I_Q \delta^P_L; } 
\nonumber \\
   \lefteqn{ \sum_C \delta^I_Q \delta^{C K}_J \delta^P_{C L} =  
   \sum_{J_1 J_2 = J} \delta^K_{J_2} \delta^I_Q \delta^P_{J_1 L}; } 
\nonumber \\
   \lefteqn{ \sum_D \delta^I_Q \delta^{K D}_J \delta^P_{L D} =  
   \sum_{J_1 J_2 = J} \delta^K_{J_1} \delta^I_Q \delta^P_{L J_2}; } 
\nonumber \\
   \lefteqn{ \sum_{C, D} \delta^I_Q \delta^{C K D}_J \delta^P_{C L D} =
   \sum_{J_1 J_2 J_3 = J} \delta^K_{J_2} \delta^I_Q \delta^P_{J_1 L J_3}; }
\nonumber \\
   \lefteqn{ \sum_A \delta^{A I}_Q \delta^K_{A J} \delta^P_L =
   \sum_{K_1 K_2 = K} \delta^{K_2}_J \delta^{K_1 I}_Q \delta^P_L; } 
\nonumber \\
   \lefteqn{ \sum_{A, C} \delta^{A I}_Q \delta^{C K}_{A J} \delta^P_{C L} =
   \sum_{K_1 K_2 = K} \sum_E \delta^{K_2}_J \delta^{E K_2 I}_Q \delta^P_{E L} +
   \sum_E \delta^K_J \delta^{E I}_Q \delta^P_{E L} } \nonumber \\
   & & + \sum_{J_1 J_2 = J} \sum_E \delta^K_{J_2} \delta^{E I}_Q \delta^P_{E J_1 L}; 
\nonumber \\  
   \lefteqn{ \sum_{A, D} \delta^{A I}_Q \delta^{K D}_{A J} \delta^P_{L D} =
   \sum_F \delta^{K F I}_Q \delta^P_{L F J} + \delta^{K I}_Q \delta^P_{L J}
   + \sum_{\begin{array}{l}
   	      J_1 J_2 = J \\
   	      K_1 K_2 = K
  	   \end{array}} \delta^{K_2}_{J_1} \delta^{K_1}_Q \delta^P_{L J_2}; } 
\nonumber \\ 
   \lefteqn{ \sum_{A, C, D} \delta^{A I}_Q \delta^{C K D}_{A J} \delta^P_{C L D} = 
   \sum_{J_1 J_2 J_3 = J} \sum_E \delta^K_{J_2} \delta^{E I}_Q \delta^P_{E J_1 L J_3}
   + \sum_{J_1 J_2 = J} \sum_E \delta^K_{J_1} \delta^{E I}_Q \delta^P_{E L J_2}} \nonumber \\ 
   & & + \sum_{\begin{array}{l}
   	      J_1 J_2 = J \\
   	      K_1 K_2 = K
   	   \end{array}} \delta^{K_2}_{J_1} \delta^{E K_1 I}_Q \delta^P_{E L J_2}
   + \sum_E \delta^{E K I}_Q \delta^P_{E L J} + \sum_{E, F} \delta^{E K F I}_Q \delta^P_{E L F J};
\nonumber \\ 
   \lefteqn{ \sum_B \delta^{I B}_Q \delta^K_{J B} \delta^P_L =
   \sum_{K_1 K_2 = K} \delta^{K_1}_J \delta^{I K_2}_Q \delta^P_L; } 
\nonumber \\
   \lefteqn{ \sum_{B, C} \delta^{I B}_Q \delta^{C K}_{J B} \delta^P_{C L} =
   \sum_{\begin{array}{l}
   	    J_1 J_2 = J \\
   	    K_1 K_2 = K
   	 \end{array}} \delta^{K_1}_{J_2} \delta^{I K_2}_Q \delta^P_{J_1 L} +
   \delta^{I K}_Q \delta^P_{J L} + \sum_F \delta^{I F K}_Q \delta^P_{J F L}; } 
\nonumber \\
   \lefteqn{ \sum_{B, D} \delta^{I B}_Q \delta^{K D}_{J B} \delta_{L D} =
   \sum_{J_1 J_2 = J} \sum_F \delta^K_{J_1} \delta^{I F}_Q \delta^P_{L J_2 F} +
   \sum_F \delta^K_J \delta^{I F}_Q \delta^P_{L F} } \nonumber \\
   & & + \sum_{K_1 K_2 = K} \sum_F \delta^{K_1}_J \delta^{I K_2 F}_Q \delta^P_{L F}; 
\nonumber \\ 
   \lefteqn{ \sum_{B, C, D} \delta^{I B}_Q \delta^{C K D}_{J B} \delta^P_{C L D} =
   \sum_{F, G} \delta^{I F K G}_Q \delta^P_{J F L G} + 
   \sum_G \delta^{I K G}_Q \delta^P_{J L G} } \nonumber \\
   & & + \sum_{\begin{array}{l}
   		  J_1 J_2 = J \\
   		  K_1 K_2 = K
  	       \end{array}} \delta^{K_1}_{J_2} \sum_F \delta^{I K_2 F}_Q \delta^P_{J_1 L F}  
   + \sum_{J_1 J_2 = J} \delta^K_{J_2} \delta^{I F}_Q \delta^P_{J_1 L F} \nonumber \\
   & & + \sum_{J_1 J_2 J_3 = J} \sum_F \delta^K_{J_2} \delta^{I F}_Q \delta^P_{J_1 L J_3 F};
\nonumber \\
   \lefteqn{ \sum_{A, B} \delta^{A I B}_Q \delta^K_{A J B} \delta^P_L =
   \sum_{K_1 K_2 K_3 = K} \delta^{K_2}_J \delta^{K_1 I K_3}_Q \delta^P_L; }
\nonumber \\
   \lefteqn{ \sum_{A, B, C} \delta^{A I B}_Q \delta^{C K}_{A J B} \delta^P_{C L} =
   \sum_{K_1 K_2 K_3 = K} \sum_E \delta^{K_2}_J \delta^{E K_1 I K_3}_Q \delta^P_{E L} } \nonumber \\
   & & + \sum_{K_1 K_2 = K} \sum_E \delta^{K_1}_J \delta^{E I K_2}_Q \delta^P_{E L}
   + \sum_{\begin{array}{l}
   	      J_1 J_2 = J \\
   	      K_1 K_2 = K
   	   \end{array}} \sum_E \delta^{K_1}_{J_2} \delta^{E I K_2}_Q \delta^P_{E J_1 L} \nonumber \\
   & & + \sum_E \delta^{E I K}_Q \delta^P_{E J L}
   + \sum_{E, F} \delta^{E I F K}_Q \delta^P_{E J F L}; 
\nonumber \\
   \lefteqn{ \sum_{A, B, D} \delta^{A I B}_Q \delta^{K D}_{A J B} \delta^P_{L D} =
   \sum_{F, G} \delta^{K F I G}_Q \delta^P_{L F J G} + 
   \sum_G \delta^{K I G}_Q \delta^P_{L J G} } \nonumber \\
   & & + \sum_{\begin{array}{l}
   		  J_1 J_2 = J \\
   		  K_1 K_2 = K
  	       \end{array}} \sum_F \delta^{K_2}_{J_1} \delta^{K_1 I F}_Q \delta^P_{L J_2 F}
   + \sum_{K_1 K_2 = K} \sum_F \delta^{K_2}_J \delta^{K_1 I F}_Q \delta^P_{L F} \nonumber \\
   & & + \sum_{K_1 K_2 K_3 = K} \sum_F \delta^{K_2}_J \delta^{K_1 I K_3 F}_Q \delta^P_{L F} \mbox{; and}
\nonumber \\
   \lefteqn{ \sum_{A, B, C, D} \delta^{A I B}_Q \delta^{C K D}_{A J B} \delta^P_{C L D} =
   \sum_{E, F, G} \delta^{E K F I G}_Q \delta^P_{E L F J G} + 
   \sum_{E, G} \delta^{E K I G}_Q \delta^P_{E L J G} } \nonumber \\
   & & + \sum_{\begin{array}{l}
   		  J_1 J_2 = J \\
   		  K_1 K_2 = K
  	       \end{array}} \sum_{E, F} \delta^{K_2}_{J_1} \delta^{E K_1 I F}_Q \delta^Q_{E L J_2 F}
   + \sum_{K_1 K_2 = K} \sum_{E, F} \delta^{K_2}_J \delta^{E K_1 I F}_Q \delta^P_{E L F} \nonumber \\
   & & + \sum_{K_1 K_2 K_3 = K} \sum_{E, F} \delta^{K_2}_J \delta^{E K_1 I K_3 F}_Q \delta^P_{E L F}
   + \sum_{J_1 J_2 = J} \sum_{E, F} \delta^K_{J_1} \delta^{E I F}_Q \delta^P_{E L J_2 F} \nonumber \\
   & & + \sum_{E, F} \delta^K_J \delta^{E I F}_Q \delta^P_{E L F} +
   \sum_{K_1 K_2 = K} \sum_{E, F} \delta^{K_1}_J \delta^{E I K_2 F}_Q \delta^P_{E L F} \nonumber \\
   & & + \sum_{J_1 J_2 J_3 = J} \sum_{E, F} \delta^K_{J_2} \delta^{E I F}_Q \delta^P_{E J_1 L J_3 F}
   + \sum_{J_1 J_2 = J} \sum_{E, F} \delta^K_{J_2} \delta^{E I F}_Q \delta^P_{E J_1 L F} \nonumber \\
   & & + \sum_{\begin{array}{l}
   		  J_1 J_2 = J \\
   		  K_1 K_2 = K
   	       \end{array}} \sum_{E, F} \delta^{K_1}_{J_2} \delta^{E I K_2 F}_Q \delta^P_{E J_1 L F}
   + \sum_{E, G} \delta^{E I K G}_Q \delta^P_{E J L F} \nonumber \\
   & & + \sum_{E, F, G} \delta^{E I F K G}_Q \delta^P_{E J F L G}. 
\label{a4.4.2}
\end{eqnarray}
We can now substitute the expressions in Eq.(\ref{a4.4.2}) into Eq.(\ref{a4.4.1}) to get
\begin{eqnarray}
   \lefteqn{ \left[ \s^I_J, \s^K_L \right] s^P = } \nonumber \\   
   & & \sum_Q \left\{ \delta^K_J ( \delta^I_Q \delta^P_L 
   + \sum_E \delta^{E I}_Q \delta^P_{E L}
   + \sum_F \delta^{I F}_Q \delta^P_{L F} + 
   \sum_{E, F} \delta^{E I F}_Q \delta^P_{E L F} ) \right. \nonumber \\
   & & + \sum_{J_1 J_2 = J} \delta^K_{J_2} (\delta^I_Q \delta^P_{J_1 L} 
   + \sum_E \delta^{E I}_Q \delta^P_{E J_1 L} + 
   \sum_F \delta^{I F}_Q \delta^P_{J_1 L F} \nonumber \\
   & & + \sum_{E, F} \delta^{E I F}_Q \delta^P_{E J_1 L F} )
   + \sum_{K_1 K_2 = K} \delta^{K_1}_J (\delta^{I K_2}_Q \delta^P_L
   + \sum_E \delta^{E I K_2}_Q \delta^P_{E L} \nonumber \\
   & & + \sum_F \delta^{I K_2 F}_Q \delta^P_{L F} +
   \sum_{E, F} \delta^{E I K_2 F}_Q \delta^P_{E L F} )
   + \sum_{J_1 J_2 = J} \delta^K_{J_1} (\delta^I_Q \delta^P_{L J_2} \nonumber \\
   & & + \sum_E \delta^{E I}_Q \delta^P_{E L J_2} +
   \sum_F \delta^{I F}_Q \delta^P_{L J_2 F} +
   \sum_{E, F} \delta^{E I F}_Q \delta^P_{E L J_2 F} ) \nonumber \\
   & & + \sum_{K_1 K_2 = K} \delta^{K_2}_J (\delta^{K_1 I}_Q \delta^P_L
   + \sum_E \delta^{E K_1 I}_Q \delta^P_{E L} +
   \sum_F \delta^{K_1 I F}_Q \delta^P_{L F} \nonumber \\
   & & + \sum_{E, F} \delta^{E K_1 I F}_Q \delta^P_{E L F} )
   + \sum_{\begin{array}{l}
   	      J_1 J_2 = J \\
   	      K_1 K_2 = K
   	   \end{array}} \delta^{K_1}_{J_2}
   (\delta^{I K_2}_Q \delta^P_{J_1 L} + 
   \sum_E \delta^{E I K_2}_Q \delta^P_{E J_1 L} \nonumber \\
   & & + \sum_F \delta^{I K_2 F}_Q \delta^P_{J_1 L F} +
   \sum_{E, F} \delta^{E I K_2 F}_Q \delta^P_{E J_1 L F} )
   + \sum_{\begin{array}{l}
   	      J_1 J_2 = J \\
   	      K_1 K_2 = K
   	   \end{array}} \delta^{K_2}_{J_1}
   (\delta^{K_1 I}_Q \delta^P_{L J_2} \nonumber \\
   & & + \sum_E \delta^{E K_1 I}_Q \delta^P_{E L J_2} +
   \sum_F \delta^{K_1 I F}_Q \delta^P_{L J_2 F} +
   \sum_{E, F} \delta^{E K_1 I F}_Q \delta^P_{E L J_2 F} ) \nonumber \\
   & & + \sum_{J_1 J_2 J_3 = J} \delta^K_{J_2}
   (\delta^I_Q \delta^P_{J_1 L J_3} +
   \sum_E \delta^{E I}_Q \delta^P_{E J_1 L J_3} \nonumber \\
   & & + \sum_F \delta^{I F}_Q \delta^P_{J_1 L J_3 F} +
   \sum_{E, F} \delta^{E, I, F} \delta^P_{E J_1 L J_3 F} )
   + \sum_{K_1 K_2 K_3 = K} \delta^{K_2}_J
   (\delta^{K_1 I K_3}_Q \delta^P_L \nonumber \\ 
   & & + \sum_E \delta^{E K_1 I K_3}_Q \delta^P_{E L} +
   \sum_F \delta^{K_1 I K_3 F}_Q \delta^P_{L F} +
   \sum_{E, F} \delta^{E K_1 I K_3 F}_Q \delta^P_{E L F} ) \nonumber \\
   & & \left. + \Gamma \right\} s^Q - (I \leftrightarrow K, J \leftrightarrow L)
\label{a4.4.3}
\end{eqnarray}    
where
\begin{eqnarray}
   \Gamma & = & \delta^{I K}_Q \delta^P_{J L} + 
   \sum_E \delta^{E I K}_Q \delta^P_{E J L} +
   \sum_F \delta^{I F K}_Q \delta^P_{J F L} +
   \sum_G \delta^{I K G}_Q \delta^P_{J L G} \nonumber \\
   & & + \sum_{E, F} \delta^{E I F K}_Q \delta^P_{E J F L} +
   \sum_{F, G} \delta^{I F K G}_Q \delta^P_{J F L G} +
   \sum_{E, G} \delta^{E I K G}_Q \delta^P_{E J L G} \nonumber \\
   & & + \sum_{E, F, G} \delta^{E I F K G}_Q \delta^P_{E J F L G}
   + (I \leftrightarrow K, J \leftrightarrow L)  
\label{a4.4.4}
\end{eqnarray}
If we substitute Eq.(\ref{a4.4.3}) without $\Gamma$ into Eq.(\ref{a4.4.1}), we will obtain exactly the action of 
operators on the R.H.S. of Eq.(\ref{4.5.4}) on $s^P$.  $\Gamma$ is reproduced when $I$ and $K$ are interchanged 
with $J$ and $L$, respectively, in Eq.(\ref{a4.4.3}) and so it is cancelled.  Consequently, Eq.(\ref{4.5.4}) is 
true.  Q.E.D.
   
\section{Root Vectors of $\hatcentrix$}
\label{sa4.5}

We need to show that any root vector has to be of the form given by Eq.(\ref{4.5.13}).  Let $f \equiv \sum_{P, Q} 
a^Q_P \s^P_{\bar{Q}}$, where only a finite number of the numerical coeffiicients $a^Q_P \neq 0$, be a root vector.  
In addition, we can assume without loss of generality that $P \neq Q$ if $a^Q_P = 0$.  Recall from Eq.(\ref{4.5.2}) 
that
\[ \s^I_J s^K = \delta^K_J s^I + \sum_{K_1 K_2 = K} \delta^{K_1}_J 
   s^{I K_2} + \sum_{K_1 K_2 = K} \delta^{K_2}_J s^{K_1 I} +
   \sum_{K_1 K_2 K_3 = K} \delta^{K_2}_J s^{K_1 I K_3} \]
Hence,
\begin{eqnarray*}
   \s^P_Q s^J & = & \sum_I \left( \delta^J_Q \delta^P_I + 
   \sum_{J_1 J_2 = J} \delta^{J_2}_Q \delta^{J_1 P}_I +
   \sum_{J_1 J_2 = J} \delta^{J_1}_Q \delta^{P J_2}_I \right. \\
   & & \left. \sum_{J_1 J_2 J_3 = J} \delta^{J_2}_Q \delta^{J_1 P J_3}_I
   \right) s^I.
\end{eqnarray*}
Therefore,
\begin{eqnarray}
   \lefteqn{ \left[ \s^M_M, f \right] s^K = } \nonumber \\ 
   & & \sum_I \left( \delta^M_I \delta^K_Q \delta^P_M +
   \sum_{K_1 K_2 = K} \delta^M_I \delta^{K_2}_Q \delta^{K_1 P}_M + 
   \sum_{K_1 K_2 = K} \delta^M_I \delta^{K_1}_Q \delta^{P K_2}_M \right.
     \nonumber \\
   & & + \sum_{K_1 K_2 K_3 = K} \delta^M_I \delta^{K_2}_Q \delta^{K_1 P K_3}_M
   + \sum_{I_1 I_2 = I} \delta^M_{I_2} \delta^K_Q \delta^P_I +
   \sum_{\begin{array}{l}
   	    I_1 I_2 = I \\
   	    K_1 K_2 = K
   	 \end{array}}
   \delta^M_{I_2} \delta^{K_2}_Q \delta^{K_1 P}_I \nonumber \\
   & & + \sum_{\begin{array}{l}
   	    I_1 I_2 = I \\
   	    K_1 K_2 = K
   	 \end{array}}
   \delta^M_{I_2} \delta^{K_1}_Q \delta^{P K_2}_I +
   \sum_{\begin{array}{l}
   	    I_1 I_2 = I \\
   	    K_1 K_2 K_3 = K
   	 \end{array}}
   \delta^M_{I_2} \delta^{K_2}_Q \delta^{K_1 P K_3}_I \nonumber \\
   & & + \sum_{I_1 I_2 = I} \delta^M_{I_1} \delta^K_Q \delta^P_I +
   \sum_{\begin{array}{l}
   	    I_1 I_2 = I \\
   	    K_1 K_2 = K
   	 \end{array}}
   \delta^M_{I_1} \delta^{K_2}_Q \delta^{K_1 P}_I +
   \sum_{\begin{array}{l}
   	    I_1 I_2 = I \\
   	    K_1 K_2 = K
   	 \end{array}}
   \delta^M_{I_1} \delta^{K_1}_Q \delta^{P K_2}_I \nonumber \\
   & & + \sum_{\begin{array}{l}
   		  I_1 I_2 = I \\
   		  K_1 K_2 K_3 = K
  	       \end{array}}
   \delta^M_{I_1} \delta^{K_2}_Q \delta^{K_1 P K_3}_I +
   \sum_{I_1 I_2 I_3 = I} \delta^M_{I_2} \delta^K_Q \delta^P_I \nonumber \\
   & & + \sum_{\begin{array}{l}
   		  I_1 I_2 I_3 = I \\
   		  K_1 K_2 = K
  	       \end{array}}
   \delta^M_{I_2} \delta^{K_2}_Q \delta^{K_1 P}_I
   + \sum_{\begin{array}{l}
   	      I_1 I_2 I_3 = I \\
   	      K_1 K_2 = K
  	   \end{array}}
   \delta^M_{I_2} \delta^{K_1}_Q \delta^{P K_2}_I \nonumber \\
   & & \left. + \sum_{\begin{array}{l}
   		         I_1 I_2 I_3 = I \\
   		         K_1 K_2 K_3 = K
   	              \end{array}}
   \delta^M_{I_2} \delta^{K_2}_Q \delta^{K_1 P K_3}_I \right) a^Q_P s^I
   \nonumber \\
   & & - (P \leftrightarrow M \;\mbox{in the superscripts},
   M \leftrightarrow Q \;\mbox{in the subscripts})
\label{a4.5.1}
\end{eqnarray}
Since $f$ is a root vector, we have
\begin{equation}
   \left[ \s^M_M, f \right] =
   \lambda_M \sum_{P, Q} a^Q_P \s^P_Q.
\label{a4.5.2}
\end{equation}
where $\lambda_M$ is a root.  As a result, we can combine Eqs.(\ref{a4.5.1}) and (\ref{a4.5.2}) together to obtain 
an equation which is too long to be written down here for {\em any} integer sequences $I$, $K$ and $M$.
   
Let us find an $a^S_R$ in $f$ such that $R \neq S$, $a^S_R \neq 0$, $a^{S_1}_{R_1} = a^{S_2}_{R_2} = 0$ for all 
$R_1$'s, $S_1$'s, $R_2$'s and $S_2$'s such that $R_1 R_2 = R$ and $S_1 S_2 = S$, and $a^{S_2}_{R_2} = 0$ for all 
$R_2$'s and $S_2$'s such that $R_1 R_2 R_3 = R$ and $S_1 S_2 S_3 = S$ for some $R_1$, $R_3$, $S_1$ and $S_3$.  The 
reader can easily convince himself or herself that such an $a^S_R$ always exists.  Let us choose $I = R$ and $K = S$
in Eq.(\ref{a4.5.1}).  Then when we combine Eqs.(\ref{a4.5.1}) and (\ref{a4.5.2}), we get
\begin{eqnarray}
   \lambda_M & = & \delta^R_M + \sum_{R_1 R_2 = R} \delta^{R_1}_M 
   + \sum_{R_1 R_2 = R} \delta^{R_2}_M + \sum_{R_1 R_2 R_3 = R}
\delta^{R_2}_M
   \nonumber \\
   & & - \delta^M_S - \sum_{S_1 S_2 = S} \delta^M_{S_1}
   - \sum_{S_1 S_2 = S} \delta^M_{S_2} - \sum_{S_1 S_2 S_3 = S}
\delta^M_{S_2}.
\label{a4.5.3}
\end{eqnarray}
Therefore, we obtain after some manipulation that
\[ \lambda_M - \sum_{j=1}^{{\Lambda}} \lambda_{Mj} - \sum_{i=1}^{{\Lambda}}
\lambda_{iM}        
   - \sum_{i, j = 1}^{{\Lambda}} \lambda_{iMj} = \delta^R_M - \delta^M_S. \]
This means
\begin{eqnarray*}
   \left[ f^R_R , f \right] & \neq & 0 \mbox{; and} \\
   \left[ f^S_S , f \right] & \neq & 0.
\end{eqnarray*}
Since $\salt$ is a proper ideal, so $f \in \salt$.  Now Eq.(\ref{4.5.13}) shows clearly that $f = f^R_S$.  The same 
equation also shows that each root vector space must be one-dimensional.  Q.E.D.   

\section{Product of Two Color-Invariant Operators}
\label{sa3.2}

We will show by contradiction that the product of two color-invariant operators is in general not well defined.
Consider the case when $\L = 1$, and assume that the operators $\a^{\m}_{\n}(1)$ and $\a^{\da\m}_{\n}(1)$ are 
bosonic.  Let $\g'^a_b = \g'^{11\ld 1}_{11\ld 1}$ and $\ti{f}'^{(a)}_{(b)} = \ti{f}'^{11\ld 1}_{11\ld 1}$, where 
the number 1 shows up $a$ times in the superscript and $b$ times in the subscript of $\g'$.  Moreover, let 
$\Ps'^{(c)} = \Ps'^{(11\ldots 1)}$ and $s'^c = s^{11\ldots 1}$, where the number 1 shows up $c$ times.  

Assume that $\g'^1_1 \g'^1_1 = \sum_{p=1}^r \a_p \g'^p_p + \sum_{q=1}^s \b_q \ti{f}'^{(q)}_{(q)}$, where $\a_1$, 
$\a_2$, \ld, $\a_r$, $\b_1$, $\b_2$, \ld, and $\b_s$ are non-zero complex numbers for some positive integers $r$ 
and $s$.  Then from the equations $\g'^1_1 \g'^1_1 (s'^1) = \g'^1_1 (\g'^1_1 s'^1) = 1^2 s'^1$, $\g'^1_1 \g'^1_1 
(s'^2) = \g'^1_1 (\g'^1_1 s'^1) = 2^2 s'^1$, \ld, and $\g'^1_1 \g'^1_1 (s'^r) = \g'^1_1 (\g'^1_1 s'^1) = r^2 s'^1$, 
we deduce that $\a_1 = 1$ and $\a_2 = \a_3 = \cd = \a_r = 2$.  Hence $\g'^1_1 \g'^1_1 = \g'^1_1 + 2 \sum_{p=2}^r 
\g'^p_p + \sum_{q=1}^s \b_q \ti{f}^{(q)}_{(q)}$.  However, $\g'^1_1 \g'^1_1 (s'^{r+1}) = (r + 1)^2 s'^{r+1}$ and 
$\g'^1_1 + 2 \sum_{p=2}^r \g'^p_p + \sum_{q=1}^s \b_q \ti{f}'^{(q)}_{(q)} (s'^{r+1}) = (r^2 + 2 r - 1) s'^{r+1}$, 
leading to a contradiction\footnote{This proves our assertion at the beginning of Section~\ref{s4.5}, namely that
the product of two operatros of the fourth kind cannot be written as a finite linear combination of operators of
this kind}.  

Thus we assume instead that $\g'^1_1 \g'^1_1 = \sum_{q=1}^s \b_q \ti{f}'^{(q)}_{(q)}$, where the $\b_q$'s are
non-zero complex numbers.  However, $\g'^1_1 \g'^1_1 (\Ps'^{(s+1)}) = (s+1)^2 \Ps'^{(s+1)}$ whereas 
$\sum_{q=1}^s \b_q \ti{f}'^{(q)}_{(q)} \Ps'^{(s+1)} = 0$, leading to a contradiction, too.  Consequently, it is 
impossible to write $\g'^1_1 \g'^1_1$ as a finite linear combination of $\g'$'s and $\tilde{f}$'s.

This proof can be easily generalized to include fermions and to the case $\Lambda > 1$.  Q.E.D. 

\section{Grand String Algebra}
\la{sa5.1}

We would like to show that the binary operations given in Section~\ref{s5.2} are Lie superbrackets.  Thus they 
constitute a Lie superalgebra.

Define the following operators:
\beq
   l'^I_J & \equiv & \g'^I_J - \sum_{i=1}^{\L + 2 \L_F} \g'^{iI}_{iJ} - \ti{f}'^I_J; \nn \\
   r'^I_J & \equiv & \g'^I_J - \sum_{j=1}^{\L + 2 \L_F} \g'^{Ij}_{Jj} - \ti{f}'^I_J; \; \mbox{and} \nn \\
   f'^I_J & \equiv & l'^I_J - \sum_{j=1}^{\L + 2 \L_F} l'^{Ij}_{Jj} \nn \\
	  & = & r'^I_J - \sum_{i=1}^{\L + 2 \L_F} r'^{iI}_{iJ} \nn \\
	  & = & \g'^I_J - \sum_{i=1}^{\L + 2 \L_F} \g'^{iI}_{iJ} - \sum_{j=1}^{\L + 2 \L_F} \g'^{Ij}_{Jj} \nn \\
          & & + \sum_{i,j=1}^{\L + 2 \L_F} \g'^{iIj}_{iIj}
	      - \ti{f}'^I_J + \sum_{j=1}^{\L + 2 \L_F} \ti{f}'^{Ij}_{Jj}.
\la{a5.1.1}
\eeq
The reader can verify that
\beq
   l'^I_J s'^K & = & \sum_{K_1 \dot{K}_2 = K} \d^{K_1}_J s'^{I \dot{K}_2}; \nn \\
   r'^I_J s'^K & = & \sum_{\dot{K}_1 K_2 = K} \d^{K_2}_J s'^{\dot{K}_1 I}; \nn \\
   f'^I_J s'^K & = & \d^K_J s'^I;
\la{a5.1.2}
\eeq
and
\beq
   l'^I_J \Ps'^K = r'^I_J \Ps'^K = f'^I_J \Ps'^k = 0.
\la{a5.1.3}
\eeq

Consider the subspace of the singlet states spanned by all states of the form
\beq
   \bar{\ph}^{\r_1} \otimes s^{\dot{K}} \otimes \ph^{\r_2} \equiv s'^{\r_1 + \L, \dot{K}, \r_2 + \L + \L_F}
   \; \mbox{and} \; \Ps^K \equiv \Ps'^K,
\la{a5.1.4}
\eeq
where any integer in $K$ and $\dot{K}$ is between 1 and $\L$ inclusive, and $1 \leq \r_1, \r_2 \leq \L_F$.
(The justification of the use of the direct products $\otimes$ will be obvious shortly.)  Eq.(\ref{3.2.6}) tells us
that this definition of $\Ps^K$ satisfies Eq.(\ref{2.2.8}):
\beq
   \Ps^K = \Ps^{K_2 K_1}.
\la{a5.1.5}
\eeq
Next, consider a subset of color-invariant operators in the heterix algebra consisting of all finite linear 
combinations of
\beq
   \g^I_J & \equiv & \g'^I_J; \nn \\
   \ti{f}^I_J & \equiv & \ti{f}'^I_J; \nn \\
   \bar{\X}^{\l_1}_{\l_2} \otimes l^{\dot{I}}_{\dot{J}} & \equiv &  
   l'^{\l_1 + \L, \dot{I}}_{\l_2 + \L, \dot{J}}; \nn \\
   r^{\dot{I}}_{\dot{J}} \otimes \X^{\l_3}_{\l_4} & \equiv &
   r'^{\dot{I}, \l_3 + \L + \L_F}_{\dot{J}, \l_4 + \L + \L_F} \mbox{; and} \nn \\
   \bar{\X}^{\l_1}_{\l_2} \otimes f^{\dot{I}}_{\dot{J}} \otimes \X^{\l_3}_{\l_4} & \equiv &
   f'^{\l_1 + \L, \dot{I}, \l_3 + \L + \L_F}_{\l_2 + \L, \dot{J}, \l_4 + \L + \L_F}
\la{a5.1.6}
\eeq
where any integer in $I$, $J$, $\dot{I}$ or $\dot{J}$ is between 1 and $\L$ inclusive, and 
$1 \leq \l_1, \l_2, \l_3 \; \mbox{and} \; \l_4 \leq \L_F$.  (Again it will be obvious shortly why the direct 
products are appropriate.)  It can be shown that this subset of color-invariant operators form a subalgebra of the 
heterix algebra.  {\em A fortiori}, this subset forms a Lie algebra.  Moreover, the subspace of the 
states defined above is a representation space for this Lie algebra, albeit a {\em reducible} one according to
Eq.(\ref{5.3.1}).


\begin{thebibliography}{99}
\bibitem{cteq} R. Brock {\em et al}., Rev. Mod. Phys. {\bf 67}, 157 (1995).
\bibitem{krra} G. S. Krishnaswami and S. G. Rajeev, Phys. Lett. B {\bf 441} 429 (1998).
\bibitem{rajeev99} S. G. Rajeev, e-print hep-th/9905072.
\bibitem{berezin} F. A. Berezin, Commun. Math. Phys. {\bf 63}, 131 (1978).
\bibitem{yaffe} L. Yaffe, Rev. Mod. Phys. {\bf 54}, 407 (1982).
\bibitem{rajeev94} S. G. Rajeev, Int. J. Mod. Phys. A {\bf 9}, 5583 (1994).
\bibitem{arnold} V. I. Arnold, {\em Mathematical Methods of Classical Mechanics}, 2nd ed. (Springer-Verlag, New 
		 York, 1989).
\bibitem{witten78} E. Witten, Nucl. Phys. B {\bf 160}, 57 (1978).
\bibitem{coleman} S. Coleman, {\em Aspects of Symmetry} (Cambridge University Press, Cambridge, 1985).
\bibitem{chpr} V. Chari and A. Pressley, {\em A Guide to Quantum Groups} (Cambridge University Press, Cambridge, 
	       1994).
\bibitem{polchinski} J. Polchinski, {\em String Theory}, Vols. 1 and 2 (Cambridge University Press, Cambridge, 
		     1998).
\bibitem{weba} J. Wess and J. Bagger, {\em Supersymmetry and Supergravity}, 2nd edition (Princeton University Press,
	       Princeton, N.J., 1992).
\bibitem{buku} I. L. Buchbinder and S. M. Kuzenko, {\em Ideas and Methods in Supersymmetry and Supergravity, or a 
	       Walk through Superspace} (Institute of Physics Publishing, Bristol, Philadelphia, 1995).
\bibitem{bfss} T. Banks, W. Fischler, S. H. Shenkar and L. Susskind, Phys. Rev. D {\bf 55}, 5112 (1997).
\bibitem{hubbard} J. Hubbard, Proc. Royal Soc. London, Ser. A {\bf 276}, 238 (1963).
\bibitem{virasoro} M. A. Virasoro, Phys. Rev. D {\bf 1}, 2933 (1970).
\bibitem{gool} P. Goddard and D. Olive, Int. J. Mod. Phys. A {\bf 1}, 303 (1986).
\bibitem{thooft74a} G. t' Hooft, Nucl. Phys. B {\bf 72}, 461 (1974).
\bibitem{thorn79} C. B. Thorn, Phys. Rev. D {\bf 20}, 1435 (1979).
\bibitem{opstal} C.-W. H. Lee and S. G. Rajeev, Nucl. Phys. B {\bf 529}, 656 (1998).
\bibitem{clstal} C.-W. H. Lee and S. G. Rajeev, J. Math. Phys. {\bf 39}, 5199 (1998).
\bibitem{dakl} S. Dalley and I. R. Klebanov, Phys. Rev. D {\bf 47}, 2517 (1993).
\bibitem{kono} S. Kobayashi and K. Nomizu, {\em Foundations of Differential Geometry}, Vol. 1, Wiley Classics
	       Library Edition (Wiley Interscience, 1996).
\bibitem{anda96a} F. Antonuccio and S. Dalley, Nucl. Phys. B {\bf 461}, 275 (1996).
\bibitem{anda96b} F. Antonuccio and S. Dalley, Phys. Lett. B {\bf 376}, 154 (1996).
\bibitem{david} F. David, Nucl. Phys. B {\bf 257}, 45 (1985).
\bibitem{kazakov85} V. A. Kazakov, Phys. Lett. B {\bf 150}, 28 (1985).
\bibitem{frgizi} P. Di Francesco, P. Ginsparg and J. Zinn-Justin, Phys. Rep. {\bf 254}, 1 (1995).
\bibitem{beth} O. Bergman and C. B. Thorn, Phys. Rev. D {\bf 52}, 5980 (1995).
\bibitem{diveve} R. Dijkgraaf, E. Verlinde and H. Verlinde, Nucl. Phys. B {\bf 500}, 43 (1997).
\bibitem{domb} C. Domb, in {\em Ising Model}, edited by C. Domb and M. S. Green, Phase Transitions and Critical
	       Phenomena, Vol. 3 (Academic Press, London, New York, 1974). 
\bibitem{amdujo} J. Ambjorn, B. Durhuus and T. Jonsson, {\em Quantum Geometry --- A Statistical Field Theory
		 Approach}, (Cambridge University Press, Cambridge, 1997).
\bibitem{bethe} H. A. Bethe, Z. Phys. {\bf 71}, 205 (1931).
\bibitem{yang} C. N. Yang, Phys. Rev. Lett. {\bf 19}, 1312 (1967).
\bibitem{baxter} R. J. Baxter, {\em Exactly Solved Models in Statistical Mechanics} (Academic Press, London, 1982).
\bibitem{klsu} I. R. Klebanov and L. Susskind, Nucl. Phys. B {\bf 309}, 175 (1988).
\bibitem{onsager} L. Onsager, Phys. Rev. {\bf 65}, 117 (1944).
\bibitem{frsu} E. Fradkin and L. Susskind, Phys. Rev. D {\bf 17}, 2637 (1978).
\bibitem{kogut} J. B. Kogut, Rev. Mod. Phys. {\bf 51}, 659 (1979).
\bibitem{prl} C.-W. H. Lee and S. G. Rajeev, Phys. Rev. Lett. {\bf 80}, 2285 (1998).
\bibitem{albaba} F. C. Alcaraz, M. N. Barber and M. T. Batchelor, Phys. Rev. Lett. {\bf 58}, 771 (1987).
\bibitem{plb} C.-W. H. Lee and S. G. Rajeev, Phys. Lett. B {\bf 436}, 91 (1998).
\bibitem{britpazu} E. Br\'{e}zin, C. Itzykson, G. Parisi and J. B. Zuber, Comm. Math. Phys. {\bf 59}, 35 (1978).
\bibitem{mehta} M. L. Mehta, Comm. Math. Phys. {\bf 79}, 327 (1981).
\bibitem{douglas} M. R. Douglas, Phys. Lett. B {\bf 238}, 176 (1990).
\bibitem{cuntz} J. Cuntz, Commun. Math. Phys., {\bf 57}, 173 (1977).
\bibitem{evans} D. E. Evans, Publ. RIMS, Kyoto Univ. {\bf 14}, 383 (1980).
\bibitem{zassenhaus} H. Zassenhaus, Hamb. Abh. {\bf 13}, 1 (1939).
\bibitem{chang} H. J. Chang, Hamb. Abh. {\bf 14}, 151 (1941).
\bibitem{seligman} G. B. Seligman, {\em Modular Lie Algebras}, Ergebnisse der Mathematik und Ihrer Grenzgebibete
		   Band 40 (Springer-Verlag, Berlin Heidelberg, 1967).
\bibitem{humphreys} J. E. Humphreys, {\em Introduction to Lie Algebras and Representation Theory} (Springer-Verlag,
                    New York, 1972).
\bibitem{murphy} G. J. Murphy, {\em $C^*$-Algebras and Operator Theory} (Academic Press, San Diego, 1990).
\bibitem{davies90} B. Davies, J. Phys. A: Math. Gen. {\bf 23}, 2245 (1990).
\bibitem{dogr} L. Dolan and M. Grady, Phys. Rev. D {\bf 25}, 1587 (1982).
\bibitem{davies91} B. Davies, J. Math. Phys. {\bf 32}, 2945 (1991).
\bibitem{roan} S.-S. Roan, Max Planck Institute, Bonn, Report No. MPI/91-70 (1991).
\bibitem{honecker} A. Honecker, Ph. D. thesis, hep-th/9503104.
\end{thebibliography}
\end{document}